\documentclass[article,aps, twocolumn,showpacs,amsmath,amssymb,superscriptaddress,nofootinbib,floatfix]{revtex4-2}

\usepackage{graphicx}% Include figure files
\usepackage{dcolumn}% Align table columns on decimal point
\usepackage{bm}% bold math
\usepackage[colorlinks=true,linkcolor=blue,citecolor=blue,urlcolor=blue]{hyperref}

\usepackage{array,booktabs,tabularx}
\usepackage[caption=false]{subfig}
\usepackage[USenglish]{babel}
\usepackage{braket}
\usepackage{xcolor}
\usepackage{amsfonts}
\usepackage{amssymb}
\usepackage[normalem]{ulem}
\usepackage{amsmath}
\definecolor{mycyan}{RGB}{0,255,255}

\begin{document}

\title{Magnetoelastic coupling in intercalated transition metal dichalcogenides}

\author{A. Kar}
\thanks{These authors contributed equally to this work.}
%\email{phys.arunava20@gmail.com}
\affiliation{Donostia International Physics Center (DIPC), San Sebastián, Spain}

\author{R. Basak}
\thanks{These authors contributed equally to this work.}
\affiliation{Department of Physics, University of California San Diego, San Diego, California 92093, USA}

\author{Xue Li}
\thanks{These authors contributed equally to this work.}
\affiliation{Beijing National Laboratory for Condensed Matter Physics and Institute of Physics, Chinese Academy of Sciences, Beijing 100190, China}
\affiliation{University of Chinese Academy of Sciences, Beijing 100049, China}

\author{A. Korshunov}
\thanks{These authors contributed equally to this work.}
\affiliation{Donostia International Physics Center (DIPC), San Sebastián, Spain}

\author{D. Subires}
\affiliation{Donostia International Physics Center (DIPC), San Sebastián, Spain}
\affiliation{Departamento de Física Aplicada I, Universidad del País Vasco UPV/EHU, E-20018 San Sebastián, Spain}

\author{J. Phillips}
\affiliation{Departamento de Física Aplicada, Universidade de Santiago de Compostela, E-15782 Campus Sur s/n, Santiago de Compostela, Spain}
\affiliation{Instituto de Materiais iMATUS, Universidade de Santiago de Compostela, E-15782 Campus Sur s/n, Santiago de Compostela, Spain} 

\author{C.-Y. Lim}
\affiliation{Donostia International Physics Center (DIPC), San Sebastián, Spain}

\author{Feng Zhou}
\affiliation{Beijing National Laboratory for Condensed Matter Physics and Institute of Physics, Chinese Academy of Sciences, Beijing 100190, China}
\affiliation{School of Electronics and Information Engineering, Tiangong University, Tianjin 300387, China}

\author{Linxuan Song}
\affiliation{Beijing National Laboratory for Condensed Matter Physics and Institute of Physics, Chinese Academy of Sciences, Beijing 100190, China}
\affiliation{School of Electronics and Information Engineering, Tiangong University, Tianjin 300387, China}
\author{Wenhong Wang}
\affiliation{School of Electronics and Information Engineering, Tiangong University, Tianjin 300387, China}
\author{Yong-Chang Lau}
\affiliation{Beijing National Laboratory for Condensed Matter Physics and Institute of Physics, Chinese Academy of Sciences, Beijing 100190, China}
\affiliation{University of Chinese Academy of Sciences, Beijing 100049, China}

\author{G. Garbarino}
\affiliation{European Synchrotron Radiation Facility (ESRF), BP 220, F-38043 Grenoble Cedex 9, France}

\author{P. Gargiani}
\affiliation{ALBA Synchrotron Light Source, 08290 Barcelona, Spain}

\author{Y. Zhao}
\affiliation{Deutsches Elektronen-Synchrotron DESY, Notkestr. 85, 22607, Hamburg, Germany}

\author{C. Plueckthun}
\affiliation{Deutsches Elektronen-Synchrotron DESY, Notkestr. 85, 22607, Hamburg, Germany}

\author{S. Francoual}
\affiliation{Deutsches Elektronen-Synchrotron DESY, Notkestr. 85, 22607, Hamburg, Germany}

\author{A. Jana}
\affiliation{CNR-Istituto Officina
dei Materiali (CNR-IOM), Strada Statale 14, km 163.5, 34149 Trieste,
Italy}
\affiliation{International Center for Theoretical Physics (ICTP), 34151 Trieste,
Italy}

\author{I. Vobornik}
\affiliation{CNR-Istituto Officina
dei Materiali (CNR-IOM), Strada Statale 14, km 163.5, 34149 Trieste,
Italy}

\author{T. Valla}
\affiliation{Donostia International Physics Center (DIPC), San Sebastián, Spain}

\author{A. Rajapitamahuni}
\affiliation{National Synchrotron Light Source II, Brookhaven National Laboratory, Upton, New York 11973, USA}

\author{James G. Analytis}
\affiliation{Physics Department, University of California, Berkeley, California 94720, USA}
\affiliation{CIFAR Quantum Materials, CIFAR, Toronto, Ontario M5G 1M1, Canada}

\author{Robert J. Birgeneau}
\affiliation{Materials Science Division, Lawrence Berkeley National Lab, Berkeley, California 94720, USA}
\affiliation{Physics Department, University of California, Berkeley, California 94720, USA}
\affiliation{Department of Materials Science and Engineering, University of California, Berkeley, California 94720, USA}

\author{E. Vescovo}
\affiliation{National Synchrotron Light Source II, Brookhaven National Laboratory, Upton, New York 11973, USA}

\author{A. Bosak}
\affiliation{European Synchrotron Radiation Facility (ESRF), BP 220, F-38043 Grenoble Cedex 9, France}

\author{J. Dai}
\affiliation{ALBA Synchrotron Light Source, 08290 Barcelona, Spain}

\author{M. Tallarida}
\affiliation{ALBA Synchrotron Light Source, 08290 Barcelona, Spain}

\author{A. Frano}
\affiliation{Department of Physics, University of California San Diego, San Diego, California 92093, USA}

\author{V. Pardo}
\affiliation{Departamento de Física Aplicada, Universidade de Santiago de Compostela, E-15782 Campus Sur s/n, Santiago de Compostela, Spain}
\affiliation{Instituto de Materiais iMATUS, Universidade de Santiago de Compostela, E-15782 Campus Sur s/n, Santiago de Compostela, Spain} 

\author{S. Wu}
\affiliation{Department of Physics, Santa Clara University}

\author{S. Blanco-Canosa}
\email{sblanco@dipc.org}
\affiliation{Donostia International Physics Center (DIPC), San Sebastián, Spain}
\affiliation{IKERBASQUE, Basque Foundation for Science, 48013 Bilbao, Spain}

% \footnotetext[1]{phys.arunava20@gmail.com}
% \begingroup
% \renewcommand\thefootnote{}\footnotetext{phys.arunava20@gmail.com}
% \endgroup

\begin{abstract}
The large van der Waals gap in transition metal dichalcogenides (TMDs) offers an avenue to host external metal atoms that modify the ground state of these 2D materials. Here, we experimentally and theoretically address the charge correlations in a family of intercalated TMDs. While short-range charge fluctuations develop in Co$_{1/3}$TaS$_{2}$ and Fe$_{1/3}$TaS$_{2}$, long-range charge order switches-on in Fe$_{1/3}$NbS$_{2}$ driven by the interplay of magnetic order and lattice degrees of freedom. The magnetoelastic coupling is demonstrated in Fe$_{1/3}$NbS$_{2}$ by the enhancement of the charge modulations upon magnetic field below T$_\mathrm{N}$, although Density Functional Perturbation Theory (DFPT) calculations predict negligible electron(spin)-phonon coupling. Furthermore, we show that Co-intercalated TaS$_2$ display a kagome-like Fermi surface, hence opening the path to engineer electronic band structures and study the entanglement of spin, charge, and spin-phonon mechanisms in the large family of intercalated TMDs.

\end{abstract}
\maketitle

A charge density wave (CDW), an exotic collective ground state is routinely found in transition metal dichalcogenides (TMDs) together with topological phases, spin valley, superconductivity, or Moiré excitons  \cite{wilson1975charge,moncton1977neutron,sipos2008mott,guillamon2008superconducting,liu2021monolayer,guo2017modulation,sipos2008mott}. Because of this, TMDs have risen as appealing candidates for applications ranging from nanoelectronics to nanoscale sensing \cite{radisavljevic2011single,splendiani2010emerging,mak2010atomically,manzeli20172d,zhang2020transition,khan2020recent}.
However, effective control of charge modulations remains a challenge as the coupling of the CDW to external stimuli is usually rather weak. In case of the high T$_c$ cuprates, despite the weak (or negligible) coupling of the CDW to the magnetic field, $\overrightarrow{\mathrm{B}}$, the strong competition between superconductivity and CDW allows to tune the intensity and correlation length of the incommensurate charge modulations by $\overrightarrow{\mathrm{B}}$ \cite{chang2016,agterberg2020physics,gerber2015three}. 

Such a collective state of interacting electrons (holes) features a periodic modulation of the electronic density and the accompanying distortion of the lattice, minimizing the system's total energy. Historically, a CDW can be understood from a 1D Peierls chain \cite{peierls1996quantum,pouget2016peierls,smaalen2005peierls}, where the entire Fermi surface can be perfectly nested by a specific wave vector $\mathrm{q}\mathrm{_{CDW}} = 2k_F$, elastically scattering electrons between states at $\pm$\textit{k}$_F$. However, in higher dimensions (2D and 3D), inelastic processes based on electron-phonon interaction (EPI) have to be considered, so that the real and imaginary parts of the electronic susceptibility calculated using the band structure can be compared with the experimental $\mathrm{q}\mathrm{_{CDW}}$
\cite{johannes2008,calandra2009effect,zhu2015classification,gerber2015three}. 
The EPI mechanism has been widely considered a proper description of CDW in 2D transition metal dichalcogenides (TMDs), such as TaS$_2$ \cite{smith1985band,sipos2008mott}, NbSe$_2$ \cite{calandra2009effect,lian2018unveiling,weber2009,valla2004quasiparticle}, TiSe$_2$ \cite{weber2011} or VSe$_2$ \cite{diego2021van,diego2024electronic}, where the nesting mechanism does not work. 

On the other hand, the reduced electric-field screening and the large van der Waals (vdW) gap in 2D TMDs open up new avenues to manipulate their ground state by intercalating transition metal (TM) ions in the vdW gap. Among the TMDs, the intercalted TM$_{1/3}$XS$_2$ (TM= Fe, Co; X=Nb, Ta) class of materials, are perfect candidates to study their frustrated magnetic order and their interplay with the charge modulations of the host lattice. For instance, (Fe, Co)$_{1/3}$TaS$_{2}$ are the only magnetic topological material available today that combines chirality and a triangular lattice, offering an opportunity to explore emergent magnetic and electronic orders. Indeed, Fe$_{1/3}$NbS$_{2}$ shows a coexistence of magnetic and charge correlations at low temperature. Earlier studies have explored unusual magnetism in (Ni, Cr)$_{1/3}$(Nb, Ta)S$_2$ that exhibit a soliton lattice with chiral magnetic ordering \cite{cao2020overview, togawa2012chiral,chapman2014spin}, topological coupling between structural chirality and magnetic order \cite{du2021topological,obeysekera2021magneto}, long-range helical antiferromagnetism and A-type spin configuration \cite{an2023bulk}. On the other hand, large anomalous Hall effect, \textit{Z}$_3$ nematicity and exchange bias effect were reported in non-coplanar antiferromagnetic Co$_{1/3}$NbS$_2$ \cite{ghimire2018large,takagi2023spontaneous,popvcevic2023electronic,tenasini2020giant}, Co$_{1/3}$TaS$_2$ \cite{park2022field,park2023tetrahedral,kim2024electrical,park2024composition} and Fe$_{1/3}$NbS$_2$ \cite{little2020three, nair2020electrical,pan2023fe,liu2022unconventional,mankovsky2016electronic}, respectively. However, despite extensive research on the magneto-transport properties of these intercalated bulk TMDs \cite{lu2020canted,zhang2018electrical}, a detailed understanding of the interplay between magnetism and CDWs has been lacking, while the origin of the periodic lattice modulations remains under debate.

\begin{figure}
	\includegraphics[width=1.0\linewidth]{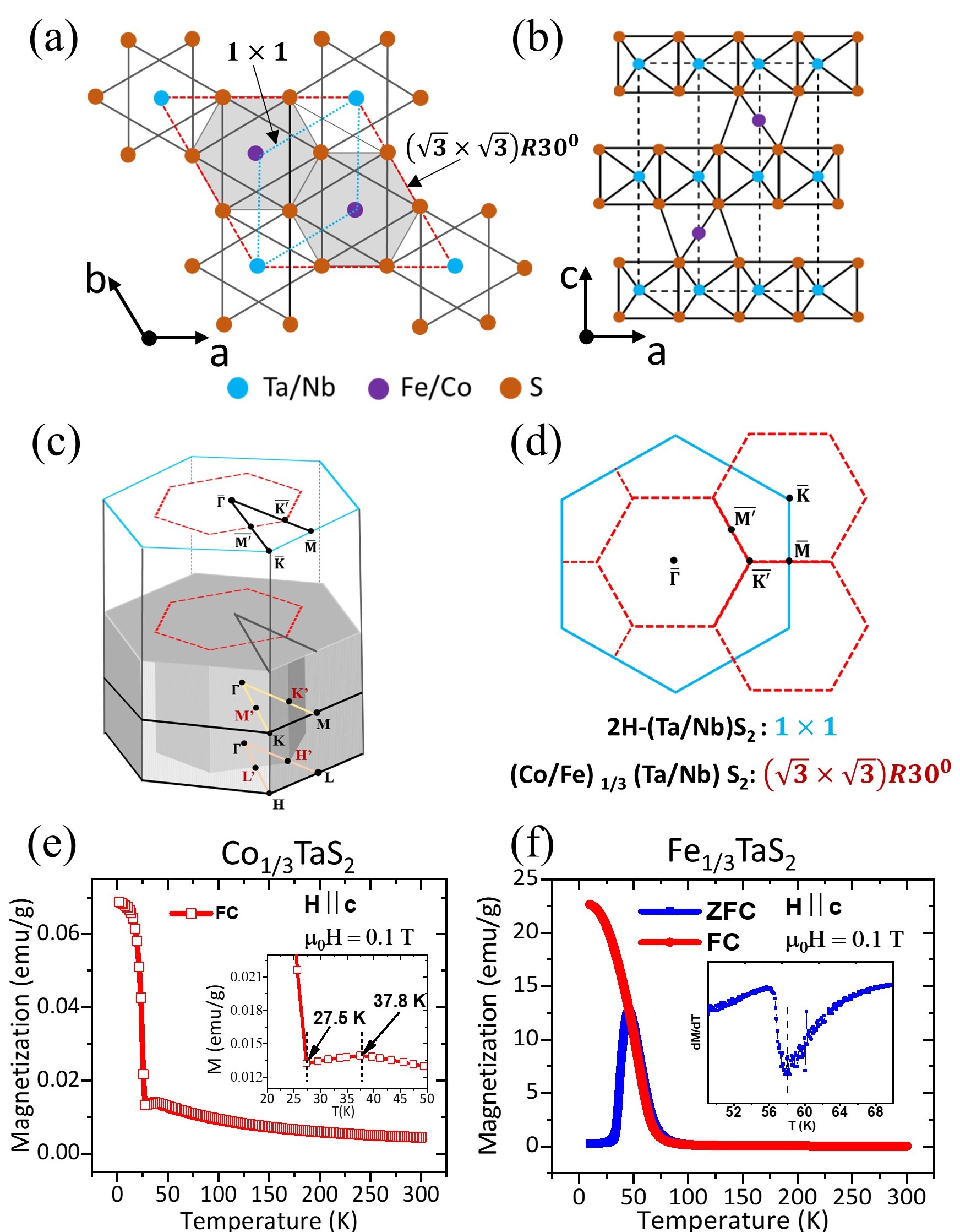}
	\caption{(a)-(b) Crystal structure of (Fe,Co)$_{1/3}$(Nb,Ta)S$_2$ viewed along the \textit{c} and \textit{b} axis, respectively. Intercalated Co/Fe atoms in 2H-TaS$_2$ occupy the octahedral interstitial sites between two TaS$_2$ layers and form a ($\sqrt{3}\times\sqrt{3}$)R30$^\circ$ superstructure.(c)-(d) Bulk and surface BZ of 2H-TaS$_2$ and Co$_{1/3}$TaS$_2$, shown in cyan and red lines, respectively. To indicate the high-symmetry points of the Co$_{1/3}$TaS$_2$ Brillouin Zone (BZ), prime signs (M', K', L', H') are used to distinguish it from the parent BZ. (e)-(f) Temperature dependent magnetization of Co$_{1/3}$TaS$_2$ and Fe$_{1/3}$TaS$_2$, respectively. The inset figures depict the zoomed portion and first derivative of the magnetization curve, respectively, around the magnetic phase transition temperature.}
	\label{fig_1}
\end{figure}

Here, we have used angle-resolved photoemission (ARPES), diffuse scattering (DS), diffraction, and density functional theory (DFT) to study a variety of TM-intercalated TMDs. We show that the intercalated (Co, Fe)$_{1/3}$-TaS$_2$ develop short-range charge modulations with propagation vector ($\frac{1}{2}$\ $\frac{1}{2}$\ 0), while the Fe analog, Fe$_{1/3}$-NbS$_2$, displays charge modulations strongly intertwined with the magnetic order, driven by the magnetoelastic coupling. We also reveal a kagome-like Fermi surface in Co$_{1/3}$TaS$_2$ system. Our results open the possibility for magnetic field sensors and spintronics applications providing a path to advance the knowledge of materials engineering with targeted functionalities.

\begin{figure*}
	\includegraphics[width=1.0
    \linewidth]{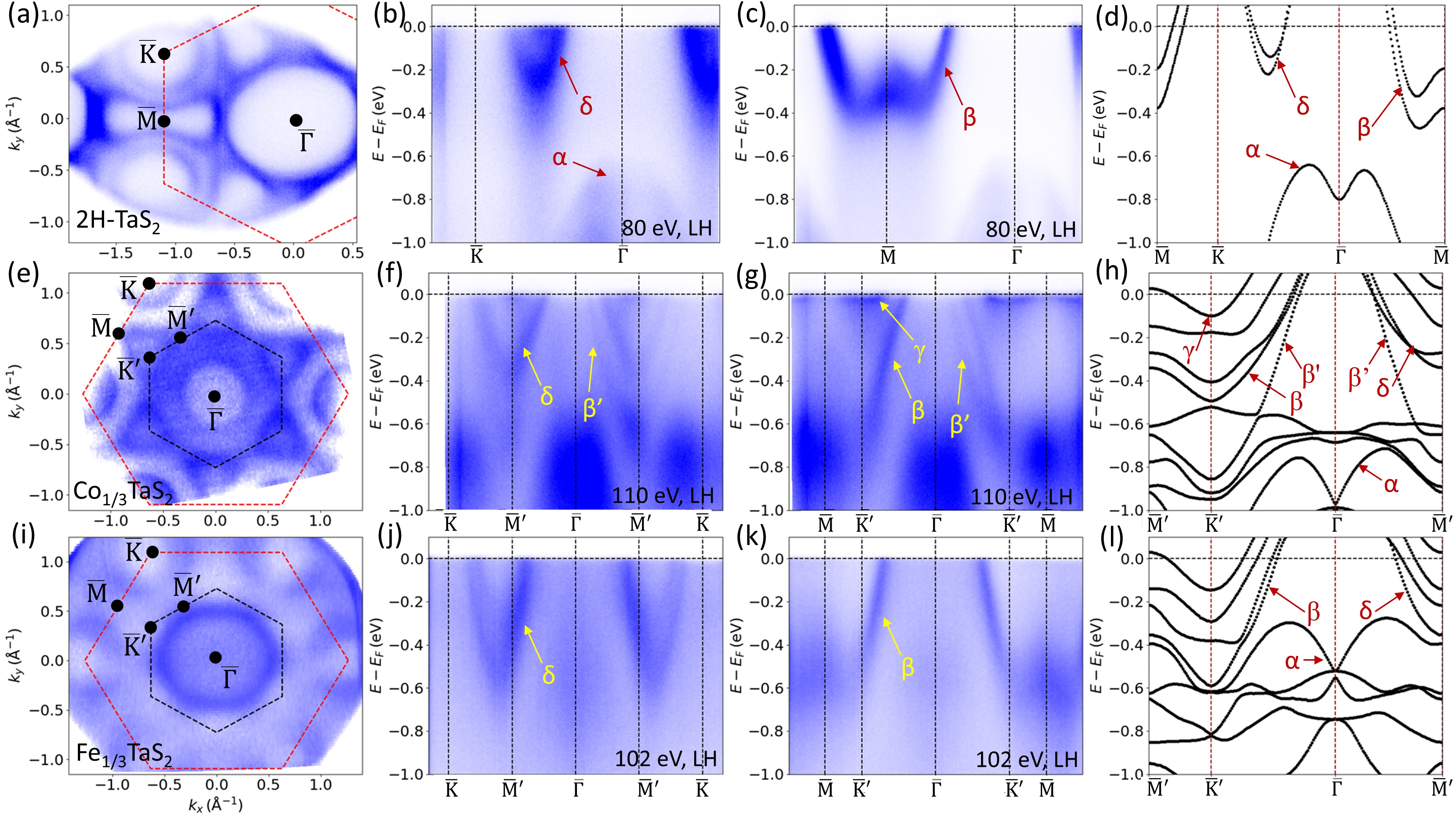}
	\caption{Fermi surface, experimental and DFT computed valence band mapping of the pristine 2H-TaS$_2$ (a)-(d), $\frac{1}{3}$ Co intercalated 2H-TaS$_2$ (e)-(h) and $\frac{1}{3}$ Fe intercalated 2H-TaS$_2$ (i)-(l) samples (T=10 K). The red dotted hexagon in the FS mappings shows the first Brillouin zone of the pristine sample with 1$\times$1 unit cell. The black dotted hexagon shows the first Brillouin zone of the $\frac{1}{3}$ Co/Fe intercalated 2H-TaS$_2$ systems with ($\sqrt{3}\times\sqrt{3}$)R30$^\circ$ superstructure relative to the 1$\times$1 primitive unit cell of 2H-TaS$_2$. The unprimed and primed alphabets indicate the high symmetry point of the pristine and superstructure systems, respectively.}
	\label{fig_2}
\end{figure*}

The bulk crystal structure of (Co,Fe)$_{1/3}$(Ta,Nb)S$_2$ is depicted in Fig.~\ref{fig_1}(a) and 1(b). In the parent TaS$_2$ layer, Ta atoms are surrounded by six S atoms in a trigonal prismatic coordination that follows the 2H\textit{a} stacking order generating octahedral sites inside the van der Waals gaps, Fig.~\ref{fig_1}(b). The TM intercalants occupy the 2\textit{c} Wyckoff positions in between the vdW layers, breaking the space-inversion and mirror symmetries resulting in building a non-centrosymmetric chiral hexagonal structure defined by the space group P6$_3$22 \cite{park2023tetrahedral}. The arrangement of the intercalated TM atoms forms a triangular lattice layer, which can be viewed as ($\sqrt{3}\times\sqrt{3}$)R30$^\circ$ superstructure relative to the 1$\times$1 primitive unit cell of 2H-TaS$_2$, Fig. \ref{fig_1}(a). Figure \ref{fig_1}(e) and (f) display the temperature dependence of the magnetization of Co$_{1/3}$TaS$_2$ and Fe$_{1/3}$TaS$_2$, respectively. The magnetization profile shows that the magnetic ordering of the Co atoms appears below the Néel temperature T$_\mathrm{N1}$ = 38 K, whereas an additional magnetic transition is observed at T$_\mathrm{N2}$ = 27 K (see Supplementary Material (SM) Fig. 1). The magnetization of Fe$_{1/3}$TaS$_2$ shows a magnetic transition at T$_\mathrm{C}\sim$60 K and a splitting between the zero-field cooled (ZFC) and field-cooled (FC) curves below T$_\mathrm{C}$, hinting at an inhomogeneous distribution of magnetic domains. The field dependence of the magnetization shows a saturation magnetization of $\sim$4.2 $\mu_{B}$ (SM Fig. 2), similar to that in Fe$_{1/3}$NbS$_2$ \cite{wu2022highly} and close to the value expected for Fe$^{2+}$ ions in a high spin state. Fitting the inverse susceptibility to a Curie-Weiss law, we retrieve an effective moment for Co of  3.2 $\mu_B$/Co that corresponds to high spin Co$^{2+}$.

A comparative study of the electronic structure of 2H-TaS$_2$, Co$_{1/3}$TaS$_{2}$ and Fe$_{1/3}$TaS$_{2}$ is shown in Fig.~\ref{fig_2}. The electronic structure of Fe$_{1/3}$NbS$_{2}$ has been previously reported in Ref. \cite{wu2023discovery}. The Fermi surface (FS) of 2H-TaS$_2$ (Fig. \ref{fig_2}(a)) consists of multiple quasi-2D cylindrical pockets due to the weak interlayer coupling and is consistent with first-principles calculations and previous studies \cite{PhysRevB.96.125103,PhysRevB.98.035203,dijkstra1989band}. Large hole-like pockets are observed around the $\bar{{\Gamma}}$ and $\bar{\mathrm{K}}$ points in the Brillouin zone and a `dog-bone' shaped electron-pocket is centered at the $\bar{\mathrm{M}}$ point. The double trilayer structure of 2H-TaS$_2$ with two Ta atoms per unit cell corresponds to the double-walled cylinders, originating from Ta 5$d$ orbitals, centered on ${\Gamma}-\mathrm{A}$ and $\mathrm{K-H}$ lines of the Brillouin zone \cite{yan2015structural, PhysRevB.85.224532} and the band sheets are well separated around the $\mathrm{K}$ point. Along the ${\bar{\Gamma}}$-${\bar{\mathrm{K}}}$ direction, the parabolic band ($\delta$) forming the inner hole pockets disperse across the Fermi level with the bottom near 0.35 eV below E$_\mathrm{F}$. While along $\bar{{\Gamma}}$-$\bar{\mathrm{M}}$ direction, the inner carrier pocket, $\beta$ band is extended to 0.4 eV below the Fermi level. Although the Fermi surface of 2H-TaS$_2$ features several nearly parallel regions, their separations do not correspond to the magnitudes of the CDW wave vectors ($\mathrm{q_{CDW}}$ $\sim$ 2/3 ${\Gamma}$M) \cite{thompson1972}, indicating that simple Fermi surface nesting is not sufficient to drive the CDW phase transition in 2H-TaS$_2$ system.

Figures~\ref{fig_2}(e)-(h) and ~\ref{fig_2}(i)-(l) present the experimental and calculated electronic band dispersion for the  1/3 Co- and Fe- intercalated TaS$_{2}$, respectively. The Fermi surfaces for both Fe and Co intercalated systems exhibit circular barrels around the zone center of their BZ. The size of the hole pockets decreases after Co/Fe intercalation due to charge transfer from the intercalated atoms to the host lattice, Fig.~\ref{fig_2}(e) and (i). Moreover, an extra hole pocket ($\beta'$) appears around the $\bar{\Gamma}$ point, and triangular-shaped electron pockets are observed at each corner (${\mathrm{\bar{K}}'}$ points) of the boundary of the small BZ in Co$_{1/3}$TaS$_{2}$ [black dotted line in Fig.~\ref{fig_2}(e)]. Interestingly, the triangular pockets centered at $\mathrm{\bar{K}'}$ points with the corner touching at $\mathrm{\bar{M}'}$ high symmetry points are a characteristic feature of the kagome metals \cite{wilson2024v3sb5,hu2022rich}.

Valence band ARPES measurements along the ${\bar{\Gamma}}$-${\mathrm{\bar{K}}}$ and ${\bar{\Gamma}}$-${\mathrm{\bar{M}}}$ symmetry directions reveal changes in the electronic structure compared to the parent compound. The parabolic band ($\delta$) along ${\bar{\Gamma}}$-${\bar{\mathrm{M'}}}$-${\bar{\mathrm{K}}}$ extends 0.35 eV and 0.5 eV below E$_F$ for the Co and Fe intercalated systems, respectively. In the Co-doped sample, we see new bands crossing E$_\mathrm{F}$, labeled as $\beta$', and $\gamma$ [Fig.~\ref{fig_2}(f)-(g)] that form hole pockets around the zone center, while the $\gamma$ band creates the triangular electron pocket around the $\bar{\mathrm{K'}}$ point. The dispersion and binding energy positions of the $\beta$, $\beta$', $\gamma$, and $\delta$ bands of Co-TaS$_2$ and Fe-TaS$_2$ are well-reproduced by DFT calculations (see details in SM), Fig. \ref{fig_2}(h) and (l). Moreover, the absence of the $\gamma$ pocket in Fig. \ref{fig_2}(g) is a consequence of electron doping that shifts E$_\mathrm{F}$. 

\begin{figure}
\includegraphics[width=1.0\linewidth]{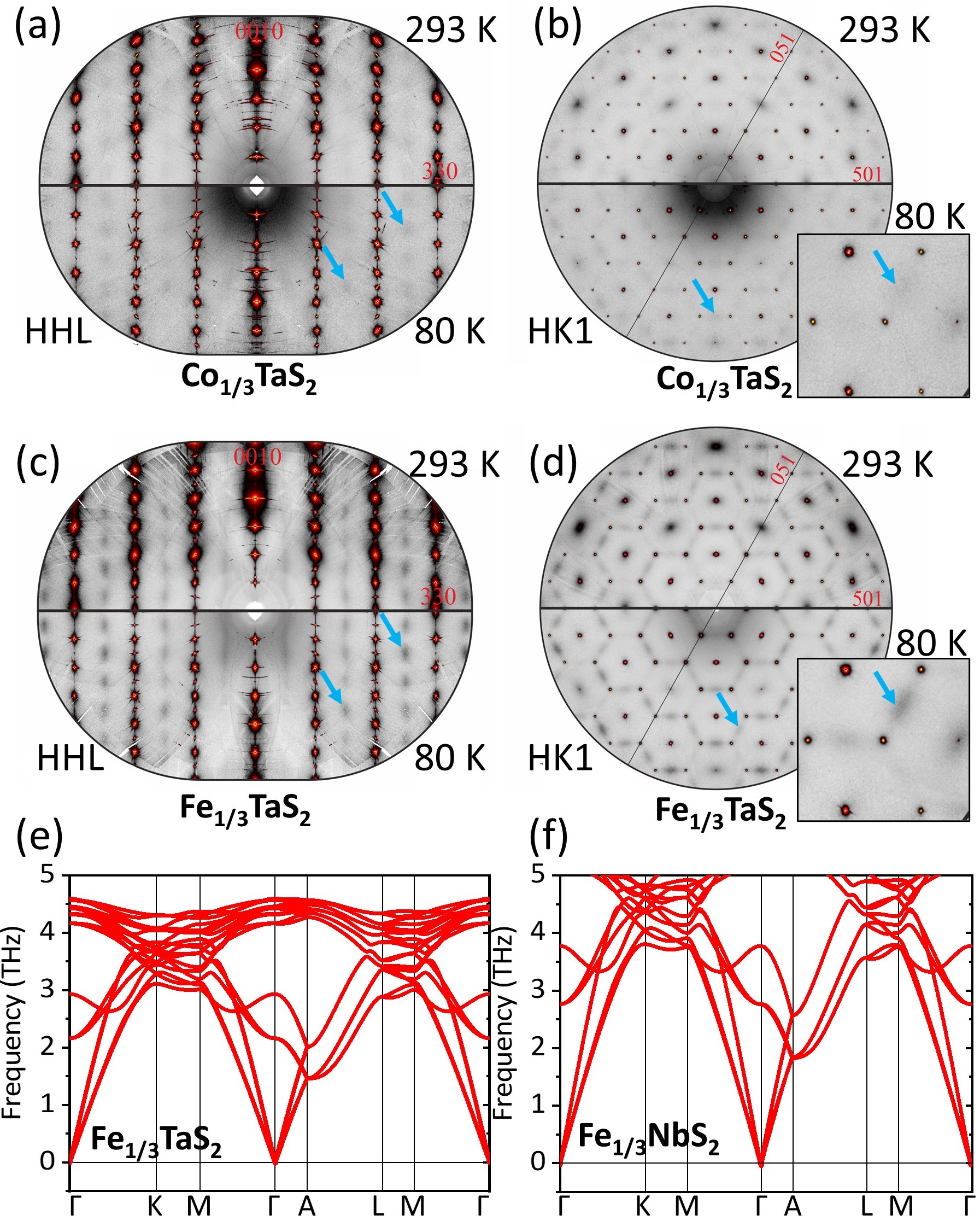}
\caption{Diffuse scattering mapping of (a-b) Co$_{1/3}$TaS$_2$ and (c-d) Fe$_{1/3}$TaS$_2$ systems. (a) and (c) Along the HHL plane at T = 293 K and 80 K, representing data above and below the antiferromagnetic transition, (upper and lower half, respectively). (b) and (d) Mapping along the HK1 plane at T = 293 K and 80 K (upper and lower half, respectively) (A zoomed view with blue arrows are marked to represent the DS positions in the corresponding plane) (e)-(f) DFT-calculated phonon spectrum in the antiferromagnetic phase of  Fe$_{1/3}$TaS$_2$ and Fe$_{1/3}$NbS$_2$ systems, respectively}.
    \label{fig_3}
\end{figure}

Now we proceed with the study of the effects of the intercalation on CDWs and lattice dynamics. Unlike 2H-NbS$_2$ \cite{leroux2012anharmonic}, 2H-TaS$_2$ forms a 3$\times$3 commensurate CDW at low temperature \cite{thompson1972}. Figure \ref{fig_3} displays the diffuse scattering (DS) maps of the Co and Fe intercalated 2H-TaS$_2$. While Co$_{1/3}$TaS$_2$ shows a blurred diffuse charge precursor, highlighted with blue arrows in Fig. \ref{fig_3}(a) and (b), Fe doped TaS$_2$ develops an anisotropic diffuse signal characteristic of a CDW, Fig.~\ref{fig_3} (c) and (d), concentrated around the wavevector q$_\mathrm{CDW}$= ($\frac{1}{2}$, 0, L). The DS develops a slightly larger spectral weight around the reciprocal lattice vector $\mathrm{\textbf{G}}_{300}$, see blue arrows in Fig. \ref{fig_3}(d), characteristic of displacement disorder, thereby demonstrating the prominent role of phonon fluctuations and differs from the substitutional disorder-driven diffuse signal with a honeycomb structure in Cr$_{1/3}$NbS$_2$ \cite{dyadkhin2015}. The interlayer correlations of Co and Fe are manifested in the form of modulated diffuse clouds, elongated along the \textit{c} direction, Fig. \ref{fig_3}(a) and (c). Such correlated disorder can influence the magnetic ordering and the transition temperature \cite{dyadkhin2015}. The diffuse intensity remains constant with temperature and does not condense into CDW reflections at low-temperature, remaining weakly correlated in the in- and out-of-plane directions (11 \r{A} along L, ~17 \r{A} and ~69 \r{A} in the HK plane, parallel and perpendicular to the anisotropic DS, respectively). Furthermore, the propagation vector differs from the parent 2H-TaS$_2$, q$_\mathrm{CDW}$= ($\frac{1}{3}$, 0, L), but matches the long-range charge order observed in Fe$_{1/3}$NbS$_2$ \cite{wu2023discovery}. Intriguingly, our harmonic phonon calculations do not show imaginary modes at q$_\mathrm{CDW}$, Fig. \ref{fig_3}(e) and (f), hence discarding Fermi surface nesting and electron-phonon interaction mechanisms as the origin of the charge modulations. This is further corroborated by the absence of nesting vectors and CDW gaps in the Fermi surface, Fig. \ref{fig_2}. Importantly, magnetic DFT calculations do not reproduce imaginary phonons in Fe$_{1/3}$NbS$_2$, well known for its long-range charge modulations \cite{wu2023discovery}, over the whole Brillouin zone center, Fig. \ref{fig_3}(e-f). However, the interaction between magnetic moments and the lattice distortions associated with the CDW intertwines the CDW amplitude and the lattice strain with the magnetic order parameter in Fe$_{1/3}$NbS$_2$. Such magnetoelastic effect strongly influences the CDW state by modifying the lattice distortions via magnetostriction, enabling external control via magnetic fields or mechanical forces.

\begin{figure}
    \includegraphics[width=1.0\linewidth,trim={1cm 1cm 1cm 1cm}, clip]{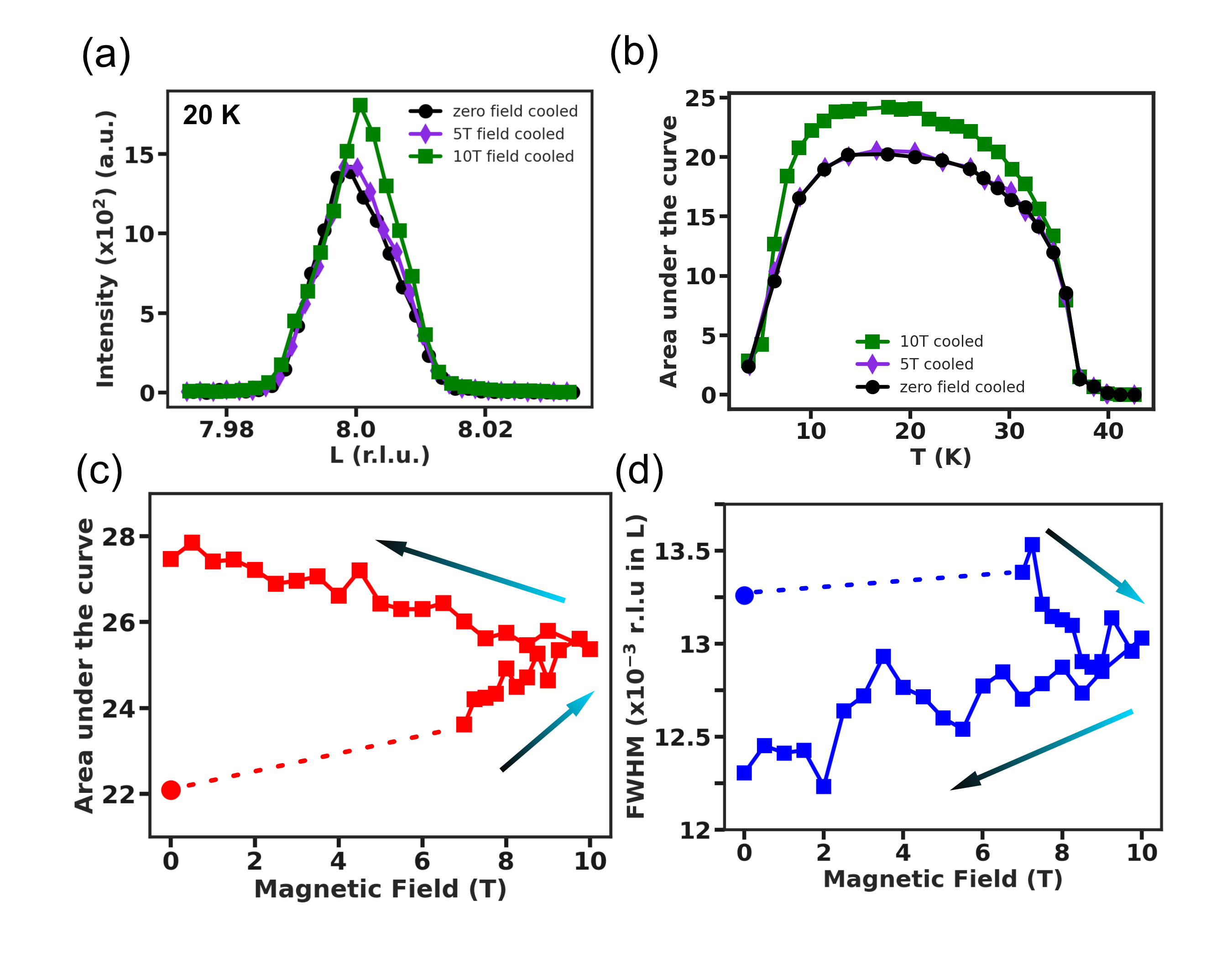}
	\caption{Magnetic field and temperature dependent charge ordering behavior in Fe$_{0.35}$NbS$_{2}$ system. (a) Intensity of the charge order peak at \textbf{Q} = (0.5, -0.5, 8) with direct field cooled conditions at 20 K. (b)  The temperature-dependent evolution of the area of the CDW peak under several magnetic field cooled conditions, measured at \textbf{Q} = (0.5, -0.5, 8). (c) and (d) Evolution of the charge-ordered peak characteristics, including the peak area and full width at half maximum (FWHM), as a function of the applied external magnetic field.}
	\label{magneto-elastic}
\end{figure}

Indeed, the long-range charge order below T$_\mathrm{N}$ in the collinear antiferromagnet Fe-NbS$_2$, hints at a possible magnetoelastic origin of the charge modulations. 
Figure \ref{magneto-elastic} details the magnetic field dependence of the charge modulations in Fe intercalated NbS$_2$, which shows a prominent super-lattice peak at (0.5, 0, L), where L is any integer \cite{wu2023discovery}, below T$_\mathrm{CDW}$=T$_\mathrm{N}$. \color{black} To investigate the field dependence,\color{black} 
we first cooled the sample under a magnetic field and then collected radial scans on the CDW peak while warming up the sample under that external magnetic field (see Methods for details). 
First, we report a clear observable field dependence of the charge order peak intensity affirming the expectation. 
Below T$_\mathrm{CDW}$ and down to 15 K, Fe$_{0.35}$NbS$_2$ exhibits a field-independent charge order behavior up to 5 T, both in intensity and linewidth, while a pronounced enhancement in the peak intensity is observed beyond this threshold field, Fig. \ref{magneto-elastic}(a,c), demonstrating a coupling of the electronic modulations to the magnetic field.
We propose that the magneto-elastic coupling arises from the interaction between magnetism and lattice potentially manifesting more prominently in Fe-NbS$_2$ due to larger magnetic moments on Fe sites in this compound compared to other compounds while retaining the lattice symmetry.
In contrast, the transition temperature for the charge order (T$_\mathrm{CDW}$) remains independent of the applied field. 
The weak interaction between antiferromagnetic order and the charge order may be considered as the Landau free energy term in the form of $\gamma \rho m^2$, where $\gamma$ is the coupling strength, $\rho$ is the CDW order parameter, and $m$ is the antiferromagnetic order parameter \cite{wu2023discovery}. 
Applying an in-plane ($\perp$ c-axis) magnetic field $\overrightarrow{\mathrm{B}}$ remarkably facilitates the CDW in material as evidenced in Fig. \ref{magneto-elastic}(c) where the magnetic field was ramped at a fixed temperature of 20 K\color{black}. 
Theoretical models demonstrated that within the Hartree-Fock approximation, CDW coexist with ferromagnetic order \cite{Balseiro1980}, although the model assumes a propensity to Fermi surface nesting. Furthermore, the non-centrosymmetric P6$_3$22 space group may allow for, besides the Heisenberg symmetric exchange, the antisymmetric Dzyaloshinskii Moriya (DM) interaction, further enhancing the magnetostriction coupling \cite{weber2022}.

We note a couple of interesting observations which point towards further complexity of the magneto-elastic coupling in these materials demanding future work from both theory and experiments. Firstly, the charge order exhibits a non-monotonic temperature evolution below 15 K, independent of the external field. The integrated peak intensity decreases below 15 K, indicating a re-entrant behavior—contrasting with prior work that reported a monotonic trend \cite{wu2023discovery}, see supp. inf. Fig. S9. At low-temperature the decrease in intensity of the CDW peak occurs regardless of the applied field, suggesting the involvement of the structural defects. 
Indeed structural defects, such as vacancies and interstitials, affect charge and antiferromagnetic orders differently. While defects disrupt antiferromagnetic exchange interactions, they act as a random field for charge order \cite{dyadkhin2015,schuessler2021,korshunov2024}. The resulting charge configurations evolve slowly at low temperatures, preventing the system from reaching equilibrium within the short interaction time with the X-rays. 
Hence at 20 K, where the temperature evolution of the FC data shows maximum contrast with the ZFC data, we vary the external field to observe a continuous increase in CDW peak intensity with increasing magnetic field,
directly evidenced in Fig. \ref{magneto-elastic}(c-d). We consider such a response of the CDW intensity to the magneto-elastic coupling of the material.

Finally, we would like to comment on the increase in the CDW peak after the magnetic field is removed. This suggests a complex free energy landscape where multiple degrees of freedom interact within each domain, and multi-$\mathrm{\textbf{q}}$ domains influence each other through long wavelength strain field \cite{gursoy2025dark}.
We speculate that the reorientation of twin domains leads to greater lattice distortion or strain, which helps stabilize the CDW. The simultaneous increase in CDW peak intensity and correlation length under varying magnetic fields deep within the ordered phase (at 20 K) supports this idea.
When the field is reduced, the system retains a ``memory" of the high-field state due to long ranged strain interactions between charge-ordered domains. As the field decreases, the system releases stored energy by reducing the number of less favorable domains.
The persistent lattice distortions from this memory effect may further enhance charge localization, strengthening the CDW even without an external magnetic field.

In summary, we have presented a combination of experimental data and \textit{ab initio} calculations to comprehensively study the formation of the CDW in intercalated TMDs. We have demonstrated that the enhanced magnetostriction phenomena result in a coupling of magnetism and lattice that drives the formation of charge modulations, strongly intertwined with the magnetic order, despite the negligible spin-phonon coupling observed in DFT. Moreover, the intercalation of magnetic ions allows for precise engineering of the spin and lattice degrees of freedom for efficient manipulation of magnetic states through external perturbations, such as strain, proximity effect, and pressure.

\section{Acknowledgements} 
This work was supported by the MINECO of Spain, projects PID2021-122609NB-C21 and PID2021-122609NB-C22. A.K. and S.B-C. acknowledge financial support by the European Union Next Generation EU/PRTR-C17.I1, as well as by IKUR Strategy under the collaboration agreement between IKERBASQUE Foundation and DIPC on behalf of the Department of Education of the Basque Government. C-Y.L. was supported by the European Research Council (ERC) under the European Union’s Horizon 2020 research and innovation program (Grant Agreement No. 101020833). J. P. thanks MECD for the financial support received through the 'Ayudas para contratos predoctorales para la formación de doctores' grant PRE2019-087338. We acknowledge DESY (Hamburg, Germany), a member of
the Helmholtz Association HGF, for the provision of experimental
facilities. Parts of this research were carried out at beamline P09 at PETRA III at DESY. Beamtime was allocated for proposal I-20230669. We acknowledge the computational resources provided by the Galician Supercomputing Center (CESGA). X.L., F.Z., L.S., W.W., and Y.-C.L were supported by the Strategic Priority Research Program (B) of the Chinese Academy of Sciences (CAS) (XDB33000000). Work at UC San Diego was supported by the National Science Foundation under Grant No. DMR-2145080. A portion of this work was carried out at the Synergetic Extreme Condition User Facility (SECUF). Work at the university of California, Berkeley, and Lawrence Berkeley National Laboratory was funded by the U.S. DOE, Ofrials of Science, Office of Basic Energy Sciences, Material Sciences and Engineering Division under contract No. DE-AC02-05CH11231 (Quantum Materials Program KC2202). This work has been partly performed in
the framework of the nanoscience foundry and fine analysis (NFFA-MUR Italy Progetti Internazionali) facility. This research used resources of the ESM beamline of the National Synchrotron Light Source II, a U.S. Department of Energy (DOE) Office of Science User Facility operated for the DOE Office of Science by Brookhaven National Laboratory under Contract No. DE-SC0012704.

\section{Experimental details}
Diffuse scattering measurements were performed at the ID28 beamline of the European Synchrotron Radiation Facility (ESRF) using an incident energy of 17.8 keV and a Dectris PILATUS3 1M X area detector. The orientation matrix refinement was performed using the CrysAlis software package. Reciprocal space maps were reconstructed using the ID28 software ProjectN and visualized with Albula. 

Angle-Resolved Photoemission Spectroscopy (ARPES) measurements were carried out at multiple synchrotron facilities, including the LOREA beamline at ALBA (using MBS electron analyzer with a base pressure of 10$^{-10}$, angular resolution of 0.2$^{\circ}$, energy resolution 10 meV), the APE-LE beamline at ELETTRA (equipped with a DA30 electron analyzer, base pressure 10$^{-11}$ mbar, angular resolution of 0.2$^{\circ}$, and energy resolution of 25 meV) and NSLS II synchrotron (a Scienta DA30 electron energy analyzer, energy, and angular resolutions were better than 15 meV and 0.2 \r{A}$^{-1}$, respectively) research facilities. During measurement, the sample temperature was kept at 10 K below the magnetic transition transition temperature for both the intercalated TMDC samples.

Magnetic field dependent hard X-ray scattering measurements were done at beamline P09 of Deutsches Elektronen-Synchrotron (DESY) in a horizontal scattering geometry. The lattice parameters used: $a=5.774$ \AA, $b=5.774$ \AA, $c=12.197$ \AA, $ \alpha = 90^{\circ}, \beta = 90^{\circ}, \gamma = 120^{\circ}$. The incoming beam energy was set to 7.248 keV. The beamspot at the sample after focusing was 70 $\mu m \times$ 200 $\mu m$. An avalanche photodiode was used as a detector. The diffractometer was tuned to center on the (0.5 -0.5 8) charge order peak. We verified the sample position to ensure consistency throughout the measurements. L-scans were collected on warming after field cooling the sample down to the base temperature of 3.7 K, at different magnetic field strengths. The magnetic field was applied vertically, perpendicular to the scattering plane, and parallel to the sample surface. The maximum contrast between different field cooled diffraction measurement was found around 20K where we set the sample temperature and ramped the field up to 10 T and down to 0 T while collecting the longitudinal scans. To extract area under the curve, we fit a Gaussian to the intensity distribution collected from each scans. 

\section{computational details}
The electronic structure calculations, band structures, Fermi surfaces and density of states
were obtained by means of calculations based on the density functional theory\cite{dft}, as
implemented in the {\sc wien2k} code\cite{wien2k, Blaha2020wien2k}. Phonon band
structures were obtained through the {\sc phonopy} code \cite{phonopy, phonopy-phono3py-JPSJ} with forces and total energies obtained using the {\sc VASP} code\cite{kresse1993ab,kresse1996efficiency,kresse1996efficient}.
In particular, when computing with {\sc wien2k}, all results where obtained using a value of
R$_{mt}$K$_{max}$=7.0. The R$_{mt}$ values used in a.u. used were the following: 2.46 a.u.
for Ta, 2.06 for S, 2.32 for Fe and 2.32 a.u. for Co. All atomic positions were fully relaxed within
space group no. 149 (P$_{312}$), that stems from the experimental space group no. 182
(P$_{6322}$) by making the two Fe (Co) atoms in the unit cell inequivalent.
When computing with {\sc VASP}, all results where obtained with the following cutoffs: The
cut-off energy of the plane wave representation of the augmentation charges was ENAUG =
500 eV for the Co compound and 550 eV for the Fe one. The cutoff energy for the plane-wave-
basis set used was ENCUT = 300 eV. The converged mesh used in all compounds was a
13$\times$13$\times$5 k-mesh using the Monkhorst-Pack scheme\cite{monkhorst_pack}.
The harmonic spectrum was computed using the real-space supercell approach\cite{ phonopy,
phonopy-phono3py-JPSJ}. We used for all compounds a 2$\times$2$\times$2 supercell, and a
converged 7$\times$7$\times$3 k-mesh for all non-magnetic calculations and a converged
6$\times$6$\times$2 k-mesh for all magnetic calculations.

\bibliography{reference}

\begin{thebibliography}{75}%
\makeatletter
\providecommand \@ifxundefined [1]{%
 \@ifx{#1\undefined}
}%
\providecommand \@ifnum [1]{%
 \ifnum #1\expandafter \@firstoftwo
 \else \expandafter \@secondoftwo
 \fi
}%
\providecommand \@ifx [1]{%
 \ifx #1\expandafter \@firstoftwo
 \else \expandafter \@secondoftwo
 \fi
}%
\providecommand \natexlab [1]{#1}%
\providecommand \enquote  [1]{``#1''}%
\providecommand \bibnamefont  [1]{#1}%
\providecommand \bibfnamefont [1]{#1}%
\providecommand \citenamefont [1]{#1}%
\providecommand \href@noop [0]{\@secondoftwo}%
\providecommand \href [0]{\begingroup \@sanitize@url \@href}%
\providecommand \@href[1]{\@@startlink{#1}\@@href}%
\providecommand \@@href[1]{\endgroup#1\@@endlink}%
\providecommand \@sanitize@url [0]{\catcode `\\12\catcode `\$12\catcode
  `\&12\catcode `\#12\catcode `\^12\catcode `\_12\catcode `\%12\relax}%
\providecommand \@@startlink[1]{}%
\providecommand \@@endlink[0]{}%
\providecommand \url  [0]{\begingroup\@sanitize@url \@url }%
\providecommand \@url [1]{\endgroup\@href {#1}{\urlprefix }}%
\providecommand \urlprefix  [0]{URL }%
\providecommand \Eprint [0]{\href }%
\providecommand \doibase [0]{https://doi.org/}%
\providecommand \selectlanguage [0]{\@gobble}%
\providecommand \bibinfo  [0]{\@secondoftwo}%
\providecommand \bibfield  [0]{\@secondoftwo}%
\providecommand \translation [1]{[#1]}%
\providecommand \BibitemOpen [0]{}%
\providecommand \bibitemStop [0]{}%
\providecommand \bibitemNoStop [0]{.\EOS\space}%
\providecommand \EOS [0]{\spacefactor3000\relax}%
\providecommand \BibitemShut  [1]{\csname bibitem#1\endcsname}%
\let\auto@bib@innerbib\@empty
%</preamble>
\bibitem [{\citenamefont {Wilson}\ \emph {et~al.}(1975)\citenamefont {Wilson},
  \citenamefont {Di~Salvo},\ and\ \citenamefont {Mahajan}}]{wilson1975charge}%
  \BibitemOpen
  \bibfield  {author} {\bibinfo {author} {\bibfnamefont {J.~A.}\ \bibnamefont
  {Wilson}}, \bibinfo {author} {\bibfnamefont {F.}~\bibnamefont {Di~Salvo}},\
  and\ \bibinfo {author} {\bibfnamefont {S.}~\bibnamefont {Mahajan}},\
  }\bibfield  {title} {\bibinfo {title} {Charge-density waves and superlattices
  in the metallic layered transition metal dichalcogenides},\ }\href@noop {}
  {\bibfield  {journal} {\bibinfo  {journal} {Advances in Physics}\ }\textbf
  {\bibinfo {volume} {24}},\ \bibinfo {pages} {117} (\bibinfo {year}
  {1975})}\BibitemShut {NoStop}%
\bibitem [{\citenamefont {Moncton}\ \emph {et~al.}(1977)\citenamefont
  {Moncton}, \citenamefont {Axe},\ and\ \citenamefont
  {DiSalvo}}]{moncton1977neutron}%
  \BibitemOpen
  \bibfield  {author} {\bibinfo {author} {\bibfnamefont {D.~E.}\ \bibnamefont
  {Moncton}}, \bibinfo {author} {\bibfnamefont {J.}~\bibnamefont {Axe}},\ and\
  \bibinfo {author} {\bibfnamefont {F.}~\bibnamefont {DiSalvo}},\ }\bibfield
  {title} {\bibinfo {title} {Neutron scattering study of the charge-density
  wave transitions in 2 h- ta se 2 and 2 h- nb se 2},\ }\href@noop {}
  {\bibfield  {journal} {\bibinfo  {journal} {Physical Review B}\ }\textbf
  {\bibinfo {volume} {16}},\ \bibinfo {pages} {801} (\bibinfo {year}
  {1977})}\BibitemShut {NoStop}%
\bibitem [{\citenamefont {Sipos}\ \emph {et~al.}(2008)\citenamefont {Sipos},
  \citenamefont {Kusmartseva}, \citenamefont {Akrap}, \citenamefont {Berger},
  \citenamefont {Forr{\'o}},\ and\ \citenamefont
  {Tuti{\v{s}}}}]{sipos2008mott}%
  \BibitemOpen
  \bibfield  {author} {\bibinfo {author} {\bibfnamefont {B.}~\bibnamefont
  {Sipos}}, \bibinfo {author} {\bibfnamefont {A.~F.}\ \bibnamefont
  {Kusmartseva}}, \bibinfo {author} {\bibfnamefont {A.}~\bibnamefont {Akrap}},
  \bibinfo {author} {\bibfnamefont {H.}~\bibnamefont {Berger}}, \bibinfo
  {author} {\bibfnamefont {L.}~\bibnamefont {Forr{\'o}}},\ and\ \bibinfo
  {author} {\bibfnamefont {E.}~\bibnamefont {Tuti{\v{s}}}},\ }\bibfield
  {title} {\bibinfo {title} {From mott state to superconductivity in 1t-tas2},\
  }\href@noop {} {\bibfield  {journal} {\bibinfo  {journal} {Nature materials}\
  }\textbf {\bibinfo {volume} {7}},\ \bibinfo {pages} {960} (\bibinfo {year}
  {2008})}\BibitemShut {NoStop}%
\bibitem [{\citenamefont {Guillam{\'o}n}\ \emph {et~al.}(2008)\citenamefont
  {Guillam{\'o}n}, \citenamefont {Suderow}, \citenamefont {Vieira},
  \citenamefont {Cario}, \citenamefont {Diener},\ and\ \citenamefont
  {Rodiere}}]{guillamon2008superconducting}%
  \BibitemOpen
  \bibfield  {author} {\bibinfo {author} {\bibfnamefont {I.}~\bibnamefont
  {Guillam{\'o}n}}, \bibinfo {author} {\bibfnamefont {H.}~\bibnamefont
  {Suderow}}, \bibinfo {author} {\bibfnamefont {S.}~\bibnamefont {Vieira}},
  \bibinfo {author} {\bibfnamefont {L.}~\bibnamefont {Cario}}, \bibinfo
  {author} {\bibfnamefont {P.}~\bibnamefont {Diener}},\ and\ \bibinfo {author}
  {\bibfnamefont {P.}~\bibnamefont {Rodiere}},\ }\bibfield  {title} {\bibinfo
  {title} {Superconducting density of states and vortex cores of 2h-nbs 2},\
  }\href@noop {} {\bibfield  {journal} {\bibinfo  {journal} {Physical review
  letters}\ }\textbf {\bibinfo {volume} {101}},\ \bibinfo {pages} {166407}
  (\bibinfo {year} {2008})}\BibitemShut {NoStop}%
\bibitem [{\citenamefont {Liu}\ \emph {et~al.}(2021)\citenamefont {Liu},
  \citenamefont {Leveillee}, \citenamefont {Lu}, \citenamefont {Yu},
  \citenamefont {Kim}, \citenamefont {Tian}, \citenamefont {Shi}, \citenamefont
  {Lai}, \citenamefont {Zhang}, \citenamefont {Giustino} \emph
  {et~al.}}]{liu2021monolayer}%
  \BibitemOpen
  \bibfield  {author} {\bibinfo {author} {\bibfnamefont {M.}~\bibnamefont
  {Liu}}, \bibinfo {author} {\bibfnamefont {J.}~\bibnamefont {Leveillee}},
  \bibinfo {author} {\bibfnamefont {S.}~\bibnamefont {Lu}}, \bibinfo {author}
  {\bibfnamefont {J.}~\bibnamefont {Yu}}, \bibinfo {author} {\bibfnamefont
  {H.}~\bibnamefont {Kim}}, \bibinfo {author} {\bibfnamefont {C.}~\bibnamefont
  {Tian}}, \bibinfo {author} {\bibfnamefont {Y.}~\bibnamefont {Shi}}, \bibinfo
  {author} {\bibfnamefont {K.}~\bibnamefont {Lai}}, \bibinfo {author}
  {\bibfnamefont {C.}~\bibnamefont {Zhang}}, \bibinfo {author} {\bibfnamefont
  {F.}~\bibnamefont {Giustino}}, \emph {et~al.},\ }\bibfield  {title} {\bibinfo
  {title} {Monolayer 1t-nbse2 as a 2d-correlated magnetic insulator},\
  }\href@noop {} {\bibfield  {journal} {\bibinfo  {journal} {Science advances}\
  }\textbf {\bibinfo {volume} {7}},\ \bibinfo {pages} {eabi6339} (\bibinfo
  {year} {2021})}\BibitemShut {NoStop}%
\bibitem [{\citenamefont {Guo}\ \emph {et~al.}(2017)\citenamefont {Guo},
  \citenamefont {Deng}, \citenamefont {Sun}, \citenamefont {Li}, \citenamefont
  {Zhao}, \citenamefont {Wu}, \citenamefont {Chu}, \citenamefont {Zhang},
  \citenamefont {Pan}, \citenamefont {Zheng} \emph
  {et~al.}}]{guo2017modulation}%
  \BibitemOpen
  \bibfield  {author} {\bibinfo {author} {\bibfnamefont {Y.}~\bibnamefont
  {Guo}}, \bibinfo {author} {\bibfnamefont {H.}~\bibnamefont {Deng}}, \bibinfo
  {author} {\bibfnamefont {X.}~\bibnamefont {Sun}}, \bibinfo {author}
  {\bibfnamefont {X.}~\bibnamefont {Li}}, \bibinfo {author} {\bibfnamefont
  {J.}~\bibnamefont {Zhao}}, \bibinfo {author} {\bibfnamefont {J.}~\bibnamefont
  {Wu}}, \bibinfo {author} {\bibfnamefont {W.}~\bibnamefont {Chu}}, \bibinfo
  {author} {\bibfnamefont {S.}~\bibnamefont {Zhang}}, \bibinfo {author}
  {\bibfnamefont {H.}~\bibnamefont {Pan}}, \bibinfo {author} {\bibfnamefont
  {X.}~\bibnamefont {Zheng}}, \emph {et~al.},\ }\bibfield  {title} {\bibinfo
  {title} {Modulation of metal and insulator states in 2d ferromagnetic vs2 by
  van der waals interaction engineering},\ }\href@noop {} {\bibfield  {journal}
  {\bibinfo  {journal} {Advanced Materials}\ }\textbf {\bibinfo {volume}
  {29}},\ \bibinfo {pages} {1700715} (\bibinfo {year} {2017})}\BibitemShut
  {NoStop}%
\bibitem [{\citenamefont {Radisavljevic}\ \emph {et~al.}(2011)\citenamefont
  {Radisavljevic}, \citenamefont {Radenovic}, \citenamefont {Brivio},
  \citenamefont {Giacometti},\ and\ \citenamefont
  {Kis}}]{radisavljevic2011single}%
  \BibitemOpen
  \bibfield  {author} {\bibinfo {author} {\bibfnamefont {B.}~\bibnamefont
  {Radisavljevic}}, \bibinfo {author} {\bibfnamefont {A.}~\bibnamefont
  {Radenovic}}, \bibinfo {author} {\bibfnamefont {J.}~\bibnamefont {Brivio}},
  \bibinfo {author} {\bibfnamefont {V.}~\bibnamefont {Giacometti}},\ and\
  \bibinfo {author} {\bibfnamefont {A.}~\bibnamefont {Kis}},\ }\bibfield
  {title} {\bibinfo {title} {Single-layer mos2 transistors},\ }\href@noop {}
  {\bibfield  {journal} {\bibinfo  {journal} {Nature nanotechnology}\ }\textbf
  {\bibinfo {volume} {6}},\ \bibinfo {pages} {147} (\bibinfo {year}
  {2011})}\BibitemShut {NoStop}%
\bibitem [{\citenamefont {Splendiani}\ \emph {et~al.}(2010)\citenamefont
  {Splendiani}, \citenamefont {Sun}, \citenamefont {Zhang}, \citenamefont {Li},
  \citenamefont {Kim}, \citenamefont {Chim}, \citenamefont {Galli},\ and\
  \citenamefont {Wang}}]{splendiani2010emerging}%
  \BibitemOpen
  \bibfield  {author} {\bibinfo {author} {\bibfnamefont {A.}~\bibnamefont
  {Splendiani}}, \bibinfo {author} {\bibfnamefont {L.}~\bibnamefont {Sun}},
  \bibinfo {author} {\bibfnamefont {Y.}~\bibnamefont {Zhang}}, \bibinfo
  {author} {\bibfnamefont {T.}~\bibnamefont {Li}}, \bibinfo {author}
  {\bibfnamefont {J.}~\bibnamefont {Kim}}, \bibinfo {author} {\bibfnamefont
  {C.-Y.}\ \bibnamefont {Chim}}, \bibinfo {author} {\bibfnamefont
  {G.}~\bibnamefont {Galli}},\ and\ \bibinfo {author} {\bibfnamefont
  {F.}~\bibnamefont {Wang}},\ }\bibfield  {title} {\bibinfo {title} {Emerging
  photoluminescence in monolayer mos2},\ }\href@noop {} {\bibfield  {journal}
  {\bibinfo  {journal} {Nano letters}\ }\textbf {\bibinfo {volume} {10}},\
  \bibinfo {pages} {1271} (\bibinfo {year} {2010})}\BibitemShut {NoStop}%
\bibitem [{\citenamefont {Mak}\ \emph {et~al.}(2010)\citenamefont {Mak},
  \citenamefont {Lee}, \citenamefont {Hone}, \citenamefont {Shan},\ and\
  \citenamefont {Heinz}}]{mak2010atomically}%
  \BibitemOpen
  \bibfield  {author} {\bibinfo {author} {\bibfnamefont {K.~F.}\ \bibnamefont
  {Mak}}, \bibinfo {author} {\bibfnamefont {C.}~\bibnamefont {Lee}}, \bibinfo
  {author} {\bibfnamefont {J.}~\bibnamefont {Hone}}, \bibinfo {author}
  {\bibfnamefont {J.}~\bibnamefont {Shan}},\ and\ \bibinfo {author}
  {\bibfnamefont {T.~F.}\ \bibnamefont {Heinz}},\ }\bibfield  {title} {\bibinfo
  {title} {Atomically thin mos 2: a new direct-gap semiconductor},\ }\href@noop
  {} {\bibfield  {journal} {\bibinfo  {journal} {Physical review letters}\
  }\textbf {\bibinfo {volume} {105}},\ \bibinfo {pages} {136805} (\bibinfo
  {year} {2010})}\BibitemShut {NoStop}%
\bibitem [{\citenamefont {Manzeli}\ \emph {et~al.}(2017)\citenamefont
  {Manzeli}, \citenamefont {Ovchinnikov}, \citenamefont {Pasquier},
  \citenamefont {Yazyev},\ and\ \citenamefont {Kis}}]{manzeli20172d}%
  \BibitemOpen
  \bibfield  {author} {\bibinfo {author} {\bibfnamefont {S.}~\bibnamefont
  {Manzeli}}, \bibinfo {author} {\bibfnamefont {D.}~\bibnamefont
  {Ovchinnikov}}, \bibinfo {author} {\bibfnamefont {D.}~\bibnamefont
  {Pasquier}}, \bibinfo {author} {\bibfnamefont {O.~V.}\ \bibnamefont
  {Yazyev}},\ and\ \bibinfo {author} {\bibfnamefont {A.}~\bibnamefont {Kis}},\
  }\bibfield  {title} {\bibinfo {title} {2d transition metal dichalcogenides},\
  }\href@noop {} {\bibfield  {journal} {\bibinfo  {journal} {Nature Reviews
  Materials}\ }\textbf {\bibinfo {volume} {2}},\ \bibinfo {pages} {1} (\bibinfo
  {year} {2017})}\BibitemShut {NoStop}%
\bibitem [{\citenamefont {Zhang}\ \emph {et~al.}(2020)\citenamefont {Zhang},
  \citenamefont {Teng}, \citenamefont {Loy}, \citenamefont {How}, \citenamefont
  {Leong},\ and\ \citenamefont {Tao}}]{zhang2020transition}%
  \BibitemOpen
  \bibfield  {author} {\bibinfo {author} {\bibfnamefont {X.}~\bibnamefont
  {Zhang}}, \bibinfo {author} {\bibfnamefont {S.~Y.}\ \bibnamefont {Teng}},
  \bibinfo {author} {\bibfnamefont {A.~C.~M.}\ \bibnamefont {Loy}}, \bibinfo
  {author} {\bibfnamefont {B.~S.}\ \bibnamefont {How}}, \bibinfo {author}
  {\bibfnamefont {W.~D.}\ \bibnamefont {Leong}},\ and\ \bibinfo {author}
  {\bibfnamefont {X.}~\bibnamefont {Tao}},\ }\bibfield  {title} {\bibinfo
  {title} {Transition metal dichalcogenides for the application of pollution
  reduction: A review},\ }\href@noop {} {\bibfield  {journal} {\bibinfo
  {journal} {Nanomaterials}\ }\textbf {\bibinfo {volume} {10}},\ \bibinfo
  {pages} {1012} (\bibinfo {year} {2020})}\BibitemShut {NoStop}%
\bibitem [{\citenamefont {Khan}\ \emph {et~al.}(2020)\citenamefont {Khan},
  \citenamefont {Tareen}, \citenamefont {Aslam}, \citenamefont {Wang},
  \citenamefont {Zhang}, \citenamefont {Mahmood}, \citenamefont {Ouyang},
  \citenamefont {Zhang},\ and\ \citenamefont {Guo}}]{khan2020recent}%
  \BibitemOpen
  \bibfield  {author} {\bibinfo {author} {\bibfnamefont {K.}~\bibnamefont
  {Khan}}, \bibinfo {author} {\bibfnamefont {A.~K.}\ \bibnamefont {Tareen}},
  \bibinfo {author} {\bibfnamefont {M.}~\bibnamefont {Aslam}}, \bibinfo
  {author} {\bibfnamefont {R.}~\bibnamefont {Wang}}, \bibinfo {author}
  {\bibfnamefont {Y.}~\bibnamefont {Zhang}}, \bibinfo {author} {\bibfnamefont
  {A.}~\bibnamefont {Mahmood}}, \bibinfo {author} {\bibfnamefont
  {Z.}~\bibnamefont {Ouyang}}, \bibinfo {author} {\bibfnamefont
  {H.}~\bibnamefont {Zhang}},\ and\ \bibinfo {author} {\bibfnamefont
  {Z.}~\bibnamefont {Guo}},\ }\bibfield  {title} {\bibinfo {title} {Recent
  developments in emerging two-dimensional materials and their applications},\
  }\href@noop {} {\bibfield  {journal} {\bibinfo  {journal} {Journal of
  Materials Chemistry C}\ }\textbf {\bibinfo {volume} {8}},\ \bibinfo {pages}
  {387} (\bibinfo {year} {2020})}\BibitemShut {NoStop}%
\bibitem [{\citenamefont {Chang}\ \emph {et~al.}(2016)\citenamefont {Chang},
  \citenamefont {Blackburn}, \citenamefont {Ivashko}, \citenamefont {Holmes},
  \citenamefont {Christensen}, \citenamefont {H{\"u}cker}, \citenamefont
  {Liang}, \citenamefont {Bonn}, \citenamefont {Hardy}, \citenamefont
  {R{\"u}tt}, \citenamefont {Zimmermann}, \citenamefont {Forgan},\ and\
  \citenamefont {Hayden}}]{chang2016}%
  \BibitemOpen
  \bibfield  {author} {\bibinfo {author} {\bibfnamefont {J.}~\bibnamefont
  {Chang}}, \bibinfo {author} {\bibfnamefont {E.}~\bibnamefont {Blackburn}},
  \bibinfo {author} {\bibfnamefont {O.}~\bibnamefont {Ivashko}}, \bibinfo
  {author} {\bibfnamefont {A.~T.}\ \bibnamefont {Holmes}}, \bibinfo {author}
  {\bibfnamefont {N.~B.}\ \bibnamefont {Christensen}}, \bibinfo {author}
  {\bibfnamefont {M.}~\bibnamefont {H{\"u}cker}}, \bibinfo {author}
  {\bibfnamefont {R.}~\bibnamefont {Liang}}, \bibinfo {author} {\bibfnamefont
  {D.~A.}\ \bibnamefont {Bonn}}, \bibinfo {author} {\bibfnamefont {W.~N.}\
  \bibnamefont {Hardy}}, \bibinfo {author} {\bibfnamefont {U.}~\bibnamefont
  {R{\"u}tt}}, \bibinfo {author} {\bibfnamefont {M.~v.}\ \bibnamefont
  {Zimmermann}}, \bibinfo {author} {\bibfnamefont {E.~M.}\ \bibnamefont
  {Forgan}},\ and\ \bibinfo {author} {\bibfnamefont {S.~M.}\ \bibnamefont
  {Hayden}},\ }\bibfield  {title} {\bibinfo {title} {Magnetic field controlled
  charge density wave coupling in underdoped yba2cu3o6+x},\ }\href
  {https://doi.org/10.1038/ncomms11494} {\bibfield  {journal} {\bibinfo
  {journal} {Nature Communications}\ }\textbf {\bibinfo {volume} {7}},\
  \bibinfo {pages} {11494} (\bibinfo {year} {2016})}\BibitemShut {NoStop}%
\bibitem [{\citenamefont {Agterberg}\ \emph {et~al.}(2020)\citenamefont
  {Agterberg}, \citenamefont {Davis}, \citenamefont {Edkins}, \citenamefont
  {Fradkin}, \citenamefont {Van~Harlingen}, \citenamefont {Kivelson},
  \citenamefont {Lee}, \citenamefont {Radzihovsky}, \citenamefont {Tranquada},\
  and\ \citenamefont {Wang}}]{agterberg2020physics}%
  \BibitemOpen
  \bibfield  {author} {\bibinfo {author} {\bibfnamefont {D.~F.}\ \bibnamefont
  {Agterberg}}, \bibinfo {author} {\bibfnamefont {J.~S.}\ \bibnamefont
  {Davis}}, \bibinfo {author} {\bibfnamefont {S.~D.}\ \bibnamefont {Edkins}},
  \bibinfo {author} {\bibfnamefont {E.}~\bibnamefont {Fradkin}}, \bibinfo
  {author} {\bibfnamefont {D.~J.}\ \bibnamefont {Van~Harlingen}}, \bibinfo
  {author} {\bibfnamefont {S.~A.}\ \bibnamefont {Kivelson}}, \bibinfo {author}
  {\bibfnamefont {P.~A.}\ \bibnamefont {Lee}}, \bibinfo {author} {\bibfnamefont
  {L.}~\bibnamefont {Radzihovsky}}, \bibinfo {author} {\bibfnamefont {J.~M.}\
  \bibnamefont {Tranquada}},\ and\ \bibinfo {author} {\bibfnamefont
  {Y.}~\bibnamefont {Wang}},\ }\bibfield  {title} {\bibinfo {title} {The
  physics of pair-density waves: cuprate superconductors and beyond},\
  }\href@noop {} {\bibfield  {journal} {\bibinfo  {journal} {Annual Review of
  Condensed Matter Physics}\ }\textbf {\bibinfo {volume} {11}},\ \bibinfo
  {pages} {231} (\bibinfo {year} {2020})}\BibitemShut {NoStop}%
\bibitem [{\citenamefont {Gerber}\ \emph {et~al.}(2015)\citenamefont {Gerber},
  \citenamefont {Jang}, \citenamefont {Nojiri}, \citenamefont {Matsuzawa},
  \citenamefont {Yasumura}, \citenamefont {Bonn}, \citenamefont {Liang},
  \citenamefont {Hardy}, \citenamefont {Islam}, \citenamefont {Mehta} \emph
  {et~al.}}]{gerber2015three}%
  \BibitemOpen
  \bibfield  {author} {\bibinfo {author} {\bibfnamefont {S.}~\bibnamefont
  {Gerber}}, \bibinfo {author} {\bibfnamefont {H.}~\bibnamefont {Jang}},
  \bibinfo {author} {\bibfnamefont {H.}~\bibnamefont {Nojiri}}, \bibinfo
  {author} {\bibfnamefont {S.}~\bibnamefont {Matsuzawa}}, \bibinfo {author}
  {\bibfnamefont {H.}~\bibnamefont {Yasumura}}, \bibinfo {author}
  {\bibfnamefont {D.}~\bibnamefont {Bonn}}, \bibinfo {author} {\bibfnamefont
  {R.}~\bibnamefont {Liang}}, \bibinfo {author} {\bibfnamefont
  {W.}~\bibnamefont {Hardy}}, \bibinfo {author} {\bibfnamefont
  {Z.}~\bibnamefont {Islam}}, \bibinfo {author} {\bibfnamefont
  {A.}~\bibnamefont {Mehta}}, \emph {et~al.},\ }\bibfield  {title} {\bibinfo
  {title} {Three-dimensional charge density wave order in yba2cu3o6. 67 at high
  magnetic fields},\ }\href@noop {} {\bibfield  {journal} {\bibinfo  {journal}
  {Science}\ }\textbf {\bibinfo {volume} {350}},\ \bibinfo {pages} {949}
  (\bibinfo {year} {2015})}\BibitemShut {NoStop}%
\bibitem [{\citenamefont {Peierls}(1996)}]{peierls1996quantum}%
  \BibitemOpen
  \bibfield  {author} {\bibinfo {author} {\bibfnamefont {R.~E.}\ \bibnamefont
  {Peierls}},\ }\href@noop {} {\emph {\bibinfo {title} {Quantum theory of
  solids}}}\ (\bibinfo  {publisher} {Clarendon Press},\ \bibinfo {year}
  {1996})\BibitemShut {NoStop}%
\bibitem [{\citenamefont {Pouget}(2016)}]{pouget2016peierls}%
  \BibitemOpen
  \bibfield  {author} {\bibinfo {author} {\bibfnamefont {J.-P.}\ \bibnamefont
  {Pouget}},\ }\bibfield  {title} {\bibinfo {title} {The peierls instability
  and charge density wave in one-dimensional electronic conductors},\
  }\href@noop {} {\bibfield  {journal} {\bibinfo  {journal} {Comptes Rendus
  Physique}\ }\textbf {\bibinfo {volume} {17}},\ \bibinfo {pages} {332}
  (\bibinfo {year} {2016})}\BibitemShut {NoStop}%
\bibitem [{\citenamefont {Smaalen}(2005)}]{smaalen2005peierls}%
  \BibitemOpen
  \bibfield  {author} {\bibinfo {author} {\bibfnamefont {S.~v.}\ \bibnamefont
  {Smaalen}},\ }\bibfield  {title} {\bibinfo {title} {The peierls transition in
  low-dimensional electronic crystals},\ }\href@noop {} {\bibfield  {journal}
  {\bibinfo  {journal} {Acta Crystallographica Section A: Foundations of
  Crystallography}\ }\textbf {\bibinfo {volume} {61}},\ \bibinfo {pages} {51}
  (\bibinfo {year} {2005})}\BibitemShut {NoStop}%
\bibitem [{\citenamefont {Johannes}\ and\ \citenamefont
  {Mazin}(2008)}]{johannes2008}%
  \BibitemOpen
  \bibfield  {author} {\bibinfo {author} {\bibfnamefont {M.~D.}\ \bibnamefont
  {Johannes}}\ and\ \bibinfo {author} {\bibfnamefont {I.~I.}\ \bibnamefont
  {Mazin}},\ }\bibfield  {title} {\bibinfo {title} {Fermi surface nesting and
  the origin of charge density waves in metals},\ }\href
  {https://doi.org/10.1103/PhysRevB.77.165135} {\bibfield  {journal} {\bibinfo
  {journal} {Phys. Rev. B}\ }\textbf {\bibinfo {volume} {77}},\ \bibinfo
  {pages} {165135} (\bibinfo {year} {2008})}\BibitemShut {NoStop}%
\bibitem [{\citenamefont {Calandra}\ \emph {et~al.}(2009)\citenamefont
  {Calandra}, \citenamefont {Mazin},\ and\ \citenamefont
  {Mauri}}]{calandra2009effect}%
  \BibitemOpen
  \bibfield  {author} {\bibinfo {author} {\bibfnamefont {M.}~\bibnamefont
  {Calandra}}, \bibinfo {author} {\bibfnamefont {I.}~\bibnamefont {Mazin}},\
  and\ \bibinfo {author} {\bibfnamefont {F.}~\bibnamefont {Mauri}},\ }\bibfield
   {title} {\bibinfo {title} {Effect of dimensionality on the charge-density
  wave in few-layer 2 h-nbse 2},\ }\href@noop {} {\bibfield  {journal}
  {\bibinfo  {journal} {Physical Review B—Condensed Matter and Materials
  Physics}\ }\textbf {\bibinfo {volume} {80}},\ \bibinfo {pages} {241108}
  (\bibinfo {year} {2009})}\BibitemShut {NoStop}%
\bibitem [{\citenamefont {Zhu}\ \emph {et~al.}(2015)\citenamefont {Zhu},
  \citenamefont {Cao}, \citenamefont {Zhang}, \citenamefont {Plummer},\ and\
  \citenamefont {Guo}}]{zhu2015classification}%
  \BibitemOpen
  \bibfield  {author} {\bibinfo {author} {\bibfnamefont {X.}~\bibnamefont
  {Zhu}}, \bibinfo {author} {\bibfnamefont {Y.}~\bibnamefont {Cao}}, \bibinfo
  {author} {\bibfnamefont {J.}~\bibnamefont {Zhang}}, \bibinfo {author}
  {\bibfnamefont {E.}~\bibnamefont {Plummer}},\ and\ \bibinfo {author}
  {\bibfnamefont {J.}~\bibnamefont {Guo}},\ }\bibfield  {title} {\bibinfo
  {title} {Classification of charge density waves based on their nature},\
  }\href@noop {} {\bibfield  {journal} {\bibinfo  {journal} {Proceedings of the
  National Academy of Sciences}\ }\textbf {\bibinfo {volume} {112}},\ \bibinfo
  {pages} {2367} (\bibinfo {year} {2015})}\BibitemShut {NoStop}%
\bibitem [{\citenamefont {Smith}\ \emph {et~al.}(1985)\citenamefont {Smith},
  \citenamefont {Kevan},\ and\ \citenamefont {DiSalvo}}]{smith1985band}%
  \BibitemOpen
  \bibfield  {author} {\bibinfo {author} {\bibfnamefont {N.}~\bibnamefont
  {Smith}}, \bibinfo {author} {\bibfnamefont {S.}~\bibnamefont {Kevan}},\ and\
  \bibinfo {author} {\bibfnamefont {F.}~\bibnamefont {DiSalvo}},\ }\bibfield
  {title} {\bibinfo {title} {Band structures of the layer compounds 1t-tas2 and
  2h-tase2 in the presence of commensurate charge-density waves},\ }\href@noop
  {} {\bibfield  {journal} {\bibinfo  {journal} {Journal of Physics C: Solid
  State Physics}\ }\textbf {\bibinfo {volume} {18}},\ \bibinfo {pages} {3175}
  (\bibinfo {year} {1985})}\BibitemShut {NoStop}%
\bibitem [{\citenamefont {Lian}\ \emph {et~al.}(2018)\citenamefont {Lian},
  \citenamefont {Si},\ and\ \citenamefont {Duan}}]{lian2018unveiling}%
  \BibitemOpen
  \bibfield  {author} {\bibinfo {author} {\bibfnamefont {C.-S.}\ \bibnamefont
  {Lian}}, \bibinfo {author} {\bibfnamefont {C.}~\bibnamefont {Si}},\ and\
  \bibinfo {author} {\bibfnamefont {W.}~\bibnamefont {Duan}},\ }\bibfield
  {title} {\bibinfo {title} {Unveiling charge-density wave, superconductivity,
  and their competitive nature in two-dimensional nbse2},\ }\href@noop {}
  {\bibfield  {journal} {\bibinfo  {journal} {Nano letters}\ }\textbf {\bibinfo
  {volume} {18}},\ \bibinfo {pages} {2924} (\bibinfo {year}
  {2018})}\BibitemShut {NoStop}%
\bibitem [{\citenamefont {Weber}\ \emph
  {et~al.}(2011{\natexlab{a}})\citenamefont {Weber}, \citenamefont
  {Rosenkranz}, \citenamefont {Castellan}, \citenamefont {Osborn},
  \citenamefont {Hott}, \citenamefont {Heid}, \citenamefont {Bohnen},
  \citenamefont {Egami}, \citenamefont {Said},\ and\ \citenamefont
  {Reznik}}]{weber2009}%
  \BibitemOpen
  \bibfield  {author} {\bibinfo {author} {\bibfnamefont {F.}~\bibnamefont
  {Weber}}, \bibinfo {author} {\bibfnamefont {S.}~\bibnamefont {Rosenkranz}},
  \bibinfo {author} {\bibfnamefont {J.-P.}\ \bibnamefont {Castellan}}, \bibinfo
  {author} {\bibfnamefont {R.}~\bibnamefont {Osborn}}, \bibinfo {author}
  {\bibfnamefont {R.}~\bibnamefont {Hott}}, \bibinfo {author} {\bibfnamefont
  {R.}~\bibnamefont {Heid}}, \bibinfo {author} {\bibfnamefont {K.-P.}\
  \bibnamefont {Bohnen}}, \bibinfo {author} {\bibfnamefont {T.}~\bibnamefont
  {Egami}}, \bibinfo {author} {\bibfnamefont {A.~H.}\ \bibnamefont {Said}},\
  and\ \bibinfo {author} {\bibfnamefont {D.}~\bibnamefont {Reznik}},\
  }\bibfield  {title} {\bibinfo {title} {Extended phonon collapse and the
  origin of the charge-density wave in
  $2h\mathrm{\text{\ensuremath{-}}}{\mathrm{nbse}}_{2}$},\ }\href
  {https://doi.org/10.1103/PhysRevLett.107.107403} {\bibfield  {journal}
  {\bibinfo  {journal} {Phys. Rev. Lett.}\ }\textbf {\bibinfo {volume} {107}},\
  \bibinfo {pages} {107403} (\bibinfo {year} {2011}{\natexlab{a}})}\BibitemShut
  {NoStop}%
\bibitem [{\citenamefont {Valla}\ \emph {et~al.}(2004)\citenamefont {Valla},
  \citenamefont {Fedorov}, \citenamefont {Johnson}, \citenamefont {Glans},
  \citenamefont {McGuinness}, \citenamefont {Smith}, \citenamefont {Andrei},\
  and\ \citenamefont {Berger}}]{valla2004quasiparticle}%
  \BibitemOpen
  \bibfield  {author} {\bibinfo {author} {\bibfnamefont {T.}~\bibnamefont
  {Valla}}, \bibinfo {author} {\bibfnamefont {A.}~\bibnamefont {Fedorov}},
  \bibinfo {author} {\bibfnamefont {P.}~\bibnamefont {Johnson}}, \bibinfo
  {author} {\bibfnamefont {P.}~\bibnamefont {Glans}}, \bibinfo {author}
  {\bibfnamefont {C.}~\bibnamefont {McGuinness}}, \bibinfo {author}
  {\bibfnamefont {K.}~\bibnamefont {Smith}}, \bibinfo {author} {\bibfnamefont
  {E.}~\bibnamefont {Andrei}},\ and\ \bibinfo {author} {\bibfnamefont
  {H.}~\bibnamefont {Berger}},\ }\bibfield  {title} {\bibinfo {title}
  {Quasiparticle spectra, charge-density waves, superconductivity,<? format?>
  and electron-phonon coupling in 2 h-n b s e 2},\ }\href@noop {} {\bibfield
  {journal} {\bibinfo  {journal} {Physical review letters}\ }\textbf {\bibinfo
  {volume} {92}},\ \bibinfo {pages} {086401} (\bibinfo {year}
  {2004})}\BibitemShut {NoStop}%
\bibitem [{\citenamefont {Weber}\ \emph
  {et~al.}(2011{\natexlab{b}})\citenamefont {Weber}, \citenamefont
  {Rosenkranz}, \citenamefont {Castellan}, \citenamefont {Osborn},
  \citenamefont {Karapetrov}, \citenamefont {Hott}, \citenamefont {Heid},
  \citenamefont {Bohnen},\ and\ \citenamefont {Alatas}}]{weber2011}%
  \BibitemOpen
  \bibfield  {author} {\bibinfo {author} {\bibfnamefont {F.}~\bibnamefont
  {Weber}}, \bibinfo {author} {\bibfnamefont {S.}~\bibnamefont {Rosenkranz}},
  \bibinfo {author} {\bibfnamefont {J.-P.}\ \bibnamefont {Castellan}}, \bibinfo
  {author} {\bibfnamefont {R.}~\bibnamefont {Osborn}}, \bibinfo {author}
  {\bibfnamefont {G.}~\bibnamefont {Karapetrov}}, \bibinfo {author}
  {\bibfnamefont {R.}~\bibnamefont {Hott}}, \bibinfo {author} {\bibfnamefont
  {R.}~\bibnamefont {Heid}}, \bibinfo {author} {\bibfnamefont {K.-P.}\
  \bibnamefont {Bohnen}},\ and\ \bibinfo {author} {\bibfnamefont
  {A.}~\bibnamefont {Alatas}},\ }\bibfield  {title} {\bibinfo {title}
  {Electron-phonon coupling and the soft phonon mode in
  ${\mathrm{tise}}_{2}$},\ }\href
  {https://doi.org/10.1103/PhysRevLett.107.266401} {\bibfield  {journal}
  {\bibinfo  {journal} {Phys. Rev. Lett.}\ }\textbf {\bibinfo {volume} {107}},\
  \bibinfo {pages} {266401} (\bibinfo {year} {2011}{\natexlab{b}})}\BibitemShut
  {NoStop}%
\bibitem [{\citenamefont {Diego}\ \emph {et~al.}(2021)\citenamefont {Diego},
  \citenamefont {Said}, \citenamefont {Mahatha}, \citenamefont {Bianco},
  \citenamefont {Monacelli}, \citenamefont {Calandra}, \citenamefont {Mauri},
  \citenamefont {Rossnagel}, \citenamefont {Errea},\ and\ \citenamefont
  {Blanco-Canosa}}]{diego2021van}%
  \BibitemOpen
  \bibfield  {author} {\bibinfo {author} {\bibfnamefont {J.}~\bibnamefont
  {Diego}}, \bibinfo {author} {\bibfnamefont {A.}~\bibnamefont {Said}},
  \bibinfo {author} {\bibfnamefont {S.~K.}\ \bibnamefont {Mahatha}}, \bibinfo
  {author} {\bibfnamefont {R.}~\bibnamefont {Bianco}}, \bibinfo {author}
  {\bibfnamefont {L.}~\bibnamefont {Monacelli}}, \bibinfo {author}
  {\bibfnamefont {M.}~\bibnamefont {Calandra}}, \bibinfo {author}
  {\bibfnamefont {F.}~\bibnamefont {Mauri}}, \bibinfo {author} {\bibfnamefont
  {K.}~\bibnamefont {Rossnagel}}, \bibinfo {author} {\bibfnamefont
  {I.}~\bibnamefont {Errea}},\ and\ \bibinfo {author} {\bibfnamefont
  {S.}~\bibnamefont {Blanco-Canosa}},\ }\bibfield  {title} {\bibinfo {title}
  {van der waals driven anharmonic melting of the 3d charge density wave in
  vse2},\ }\href@noop {} {\bibfield  {journal} {\bibinfo  {journal} {Nature
  communications}\ }\textbf {\bibinfo {volume} {12}},\ \bibinfo {pages} {598}
  (\bibinfo {year} {2021})}\BibitemShut {NoStop}%
\bibitem [{\citenamefont {Diego}\ \emph {et~al.}(2024)\citenamefont {Diego},
  \citenamefont {Subires}, \citenamefont {Said}, \citenamefont {Chaney},
  \citenamefont {Korshunov}, \citenamefont {Garbarino}, \citenamefont
  {Diekmann}, \citenamefont {Mahatha}, \citenamefont {Pardo}, \citenamefont
  {Wilkinson} \emph {et~al.}}]{diego2024electronic}%
  \BibitemOpen
  \bibfield  {author} {\bibinfo {author} {\bibfnamefont {J.}~\bibnamefont
  {Diego}}, \bibinfo {author} {\bibfnamefont {D.}~\bibnamefont {Subires}},
  \bibinfo {author} {\bibfnamefont {A.}~\bibnamefont {Said}}, \bibinfo {author}
  {\bibfnamefont {D.}~\bibnamefont {Chaney}}, \bibinfo {author} {\bibfnamefont
  {A.}~\bibnamefont {Korshunov}}, \bibinfo {author} {\bibfnamefont
  {G.}~\bibnamefont {Garbarino}}, \bibinfo {author} {\bibfnamefont
  {F.}~\bibnamefont {Diekmann}}, \bibinfo {author} {\bibfnamefont {S.~K.}\
  \bibnamefont {Mahatha}}, \bibinfo {author} {\bibfnamefont {V.}~\bibnamefont
  {Pardo}}, \bibinfo {author} {\bibfnamefont {J.}~\bibnamefont {Wilkinson}},
  \emph {et~al.},\ }\bibfield  {title} {\bibinfo {title} {Electronic structure
  and lattice dynamics of 1 t-vse 2: Origin of the three-dimensional charge
  density wave},\ }\href@noop {} {\bibfield  {journal} {\bibinfo  {journal}
  {Physical Review B}\ }\textbf {\bibinfo {volume} {109}},\ \bibinfo {pages}
  {035133} (\bibinfo {year} {2024})}\BibitemShut {NoStop}%
\bibitem [{\citenamefont {Cao}\ \emph {et~al.}(2020)\citenamefont {Cao},
  \citenamefont {Huang}, \citenamefont {Yin}, \citenamefont {Xie},
  \citenamefont {Liu}, \citenamefont {Wang}, \citenamefont {Zhu}, \citenamefont
  {Mandrus}, \citenamefont {Wang},\ and\ \citenamefont
  {Huang}}]{cao2020overview}%
  \BibitemOpen
  \bibfield  {author} {\bibinfo {author} {\bibfnamefont {Y.}~\bibnamefont
  {Cao}}, \bibinfo {author} {\bibfnamefont {Z.}~\bibnamefont {Huang}}, \bibinfo
  {author} {\bibfnamefont {Y.}~\bibnamefont {Yin}}, \bibinfo {author}
  {\bibfnamefont {H.}~\bibnamefont {Xie}}, \bibinfo {author} {\bibfnamefont
  {B.}~\bibnamefont {Liu}}, \bibinfo {author} {\bibfnamefont {W.}~\bibnamefont
  {Wang}}, \bibinfo {author} {\bibfnamefont {C.}~\bibnamefont {Zhu}}, \bibinfo
  {author} {\bibfnamefont {D.}~\bibnamefont {Mandrus}}, \bibinfo {author}
  {\bibfnamefont {L.}~\bibnamefont {Wang}},\ and\ \bibinfo {author}
  {\bibfnamefont {W.}~\bibnamefont {Huang}},\ }\bibfield  {title} {\bibinfo
  {title} {Overview and advances in a layered chiral helimagnet cr1/3nbs2},\
  }\href@noop {} {\bibfield  {journal} {\bibinfo  {journal} {Materials Today
  Advances}\ }\textbf {\bibinfo {volume} {7}},\ \bibinfo {pages} {100080}
  (\bibinfo {year} {2020})}\BibitemShut {NoStop}%
\bibitem [{\citenamefont {Togawa}\ \emph {et~al.}(2012)\citenamefont {Togawa},
  \citenamefont {Koyama}, \citenamefont {Takayanagi}, \citenamefont {Mori},
  \citenamefont {Kousaka}, \citenamefont {Akimitsu}, \citenamefont {Nishihara},
  \citenamefont {Inoue}, \citenamefont {Ovchinnikov},\ and\ \citenamefont
  {Kishine}}]{togawa2012chiral}%
  \BibitemOpen
  \bibfield  {author} {\bibinfo {author} {\bibfnamefont {Y.}~\bibnamefont
  {Togawa}}, \bibinfo {author} {\bibfnamefont {T.}~\bibnamefont {Koyama}},
  \bibinfo {author} {\bibfnamefont {K.}~\bibnamefont {Takayanagi}}, \bibinfo
  {author} {\bibfnamefont {S.}~\bibnamefont {Mori}}, \bibinfo {author}
  {\bibfnamefont {Y.}~\bibnamefont {Kousaka}}, \bibinfo {author} {\bibfnamefont
  {J.}~\bibnamefont {Akimitsu}}, \bibinfo {author} {\bibfnamefont
  {S.}~\bibnamefont {Nishihara}}, \bibinfo {author} {\bibfnamefont
  {K.}~\bibnamefont {Inoue}}, \bibinfo {author} {\bibfnamefont
  {A.}~\bibnamefont {Ovchinnikov}},\ and\ \bibinfo {author} {\bibfnamefont
  {J.-i.}\ \bibnamefont {Kishine}},\ }\bibfield  {title} {\bibinfo {title}
  {Chiral magnetic soliton lattice on a chiral helimagnet},\ }\href@noop {}
  {\bibfield  {journal} {\bibinfo  {journal} {Physical review letters}\
  }\textbf {\bibinfo {volume} {108}},\ \bibinfo {pages} {107202} (\bibinfo
  {year} {2012})}\BibitemShut {NoStop}%
\bibitem [{\citenamefont {Chapman}\ \emph {et~al.}(2014)\citenamefont
  {Chapman}, \citenamefont {Bornstein}, \citenamefont {Ghimire}, \citenamefont
  {Mandrus},\ and\ \citenamefont {Lee}}]{chapman2014spin}%
  \BibitemOpen
  \bibfield  {author} {\bibinfo {author} {\bibfnamefont {B.~J.}\ \bibnamefont
  {Chapman}}, \bibinfo {author} {\bibfnamefont {A.~C.}\ \bibnamefont
  {Bornstein}}, \bibinfo {author} {\bibfnamefont {N.~J.}\ \bibnamefont
  {Ghimire}}, \bibinfo {author} {\bibfnamefont {D.}~\bibnamefont {Mandrus}},\
  and\ \bibinfo {author} {\bibfnamefont {M.}~\bibnamefont {Lee}},\ }\bibfield
  {title} {\bibinfo {title} {Spin structure of the anisotropic helimagnet cr1/
  3nbs2 in a magnetic field},\ }\href@noop {} {\bibfield  {journal} {\bibinfo
  {journal} {Applied Physics Letters}\ }\textbf {\bibinfo {volume} {105}}
  (\bibinfo {year} {2014})}\BibitemShut {NoStop}%
\bibitem [{\citenamefont {Du}\ \emph {et~al.}(2021)\citenamefont {Du},
  \citenamefont {Huang}, \citenamefont {Kim}, \citenamefont {Lim},
  \citenamefont {Gamage}, \citenamefont {Yang}, \citenamefont {Mostovoy},
  \citenamefont {Garlow}, \citenamefont {Han}, \citenamefont {Zhu} \emph
  {et~al.}}]{du2021topological}%
  \BibitemOpen
  \bibfield  {author} {\bibinfo {author} {\bibfnamefont {K.}~\bibnamefont
  {Du}}, \bibinfo {author} {\bibfnamefont {F.-T.}\ \bibnamefont {Huang}},
  \bibinfo {author} {\bibfnamefont {J.}~\bibnamefont {Kim}}, \bibinfo {author}
  {\bibfnamefont {S.~J.}\ \bibnamefont {Lim}}, \bibinfo {author} {\bibfnamefont
  {K.}~\bibnamefont {Gamage}}, \bibinfo {author} {\bibfnamefont
  {J.}~\bibnamefont {Yang}}, \bibinfo {author} {\bibfnamefont {M.}~\bibnamefont
  {Mostovoy}}, \bibinfo {author} {\bibfnamefont {J.}~\bibnamefont {Garlow}},
  \bibinfo {author} {\bibfnamefont {M.-G.}\ \bibnamefont {Han}}, \bibinfo
  {author} {\bibfnamefont {Y.}~\bibnamefont {Zhu}}, \emph {et~al.},\ }\bibfield
   {title} {\bibinfo {title} {Topological spin/structure couplings in layered
  chiral magnet cr1/3tas2: The discovery of spiral magnetic superstructure},\
  }\href@noop {} {\bibfield  {journal} {\bibinfo  {journal} {Proceedings of the
  National Academy of Sciences}\ }\textbf {\bibinfo {volume} {118}},\ \bibinfo
  {pages} {e2023337118} (\bibinfo {year} {2021})}\BibitemShut {NoStop}%
\bibitem [{\citenamefont {Obeysekera}\ \emph {et~al.}(2021)\citenamefont
  {Obeysekera}, \citenamefont {Gamage}, \citenamefont {Gao}, \citenamefont
  {Cheong},\ and\ \citenamefont {Yang}}]{obeysekera2021magneto}%
  \BibitemOpen
  \bibfield  {author} {\bibinfo {author} {\bibfnamefont {D.}~\bibnamefont
  {Obeysekera}}, \bibinfo {author} {\bibfnamefont {K.}~\bibnamefont {Gamage}},
  \bibinfo {author} {\bibfnamefont {Y.}~\bibnamefont {Gao}}, \bibinfo {author}
  {\bibfnamefont {S.-w.}\ \bibnamefont {Cheong}},\ and\ \bibinfo {author}
  {\bibfnamefont {J.}~\bibnamefont {Yang}},\ }\bibfield  {title} {\bibinfo
  {title} {The magneto-transport properties of cr1/3tas2 with chiral magnetic
  solitons},\ }\href@noop {} {\bibfield  {journal} {\bibinfo  {journal}
  {Advanced Electronic Materials}\ }\textbf {\bibinfo {volume} {7}},\ \bibinfo
  {pages} {2100424} (\bibinfo {year} {2021})}\BibitemShut {NoStop}%
\bibitem [{\citenamefont {An}\ \emph {et~al.}(2023)\citenamefont {An},
  \citenamefont {Park}, \citenamefont {Kim}, \citenamefont {Zhang},
  \citenamefont {Kim}, \citenamefont {Avdeev}, \citenamefont {Kim},
  \citenamefont {Han}, \citenamefont {Noh}, \citenamefont {Seong} \emph
  {et~al.}}]{an2023bulk}%
  \BibitemOpen
  \bibfield  {author} {\bibinfo {author} {\bibfnamefont {Y.}~\bibnamefont
  {An}}, \bibinfo {author} {\bibfnamefont {P.}~\bibnamefont {Park}}, \bibinfo
  {author} {\bibfnamefont {C.}~\bibnamefont {Kim}}, \bibinfo {author}
  {\bibfnamefont {K.}~\bibnamefont {Zhang}}, \bibinfo {author} {\bibfnamefont
  {H.}~\bibnamefont {Kim}}, \bibinfo {author} {\bibfnamefont {M.}~\bibnamefont
  {Avdeev}}, \bibinfo {author} {\bibfnamefont {J.}~\bibnamefont {Kim}},
  \bibinfo {author} {\bibfnamefont {M.-J.}\ \bibnamefont {Han}}, \bibinfo
  {author} {\bibfnamefont {H.-J.}\ \bibnamefont {Noh}}, \bibinfo {author}
  {\bibfnamefont {S.}~\bibnamefont {Seong}}, \emph {et~al.},\ }\bibfield
  {title} {\bibinfo {title} {Bulk properties of the chiral metallic triangular
  antiferromagnets ni 1/3 nb s 2 and ni 1/3 ta s 2},\ }\href@noop {} {\bibfield
   {journal} {\bibinfo  {journal} {Physical Review B}\ }\textbf {\bibinfo
  {volume} {108}},\ \bibinfo {pages} {054418} (\bibinfo {year}
  {2023})}\BibitemShut {NoStop}%
\bibitem [{\citenamefont {Ghimire}\ \emph {et~al.}(2018)\citenamefont
  {Ghimire}, \citenamefont {Botana}, \citenamefont {Jiang}, \citenamefont
  {Zhang}, \citenamefont {Chen},\ and\ \citenamefont
  {Mitchell}}]{ghimire2018large}%
  \BibitemOpen
  \bibfield  {author} {\bibinfo {author} {\bibfnamefont {N.~J.}\ \bibnamefont
  {Ghimire}}, \bibinfo {author} {\bibfnamefont {A.}~\bibnamefont {Botana}},
  \bibinfo {author} {\bibfnamefont {J.}~\bibnamefont {Jiang}}, \bibinfo
  {author} {\bibfnamefont {J.}~\bibnamefont {Zhang}}, \bibinfo {author}
  {\bibfnamefont {Y.-S.}\ \bibnamefont {Chen}},\ and\ \bibinfo {author}
  {\bibfnamefont {J.}~\bibnamefont {Mitchell}},\ }\bibfield  {title} {\bibinfo
  {title} {Large anomalous hall effect in the chiral-lattice antiferromagnet
  conb3s6},\ }\href@noop {} {\bibfield  {journal} {\bibinfo  {journal} {Nature
  communications}\ }\textbf {\bibinfo {volume} {9}},\ \bibinfo {pages} {3280}
  (\bibinfo {year} {2018})}\BibitemShut {NoStop}%
\bibitem [{\citenamefont {Takagi}\ \emph {et~al.}(2023)\citenamefont {Takagi},
  \citenamefont {Takagi}, \citenamefont {Minami}, \citenamefont {Nomoto},
  \citenamefont {Ohishi}, \citenamefont {Suzuki}, \citenamefont {Yanagi},
  \citenamefont {Hirayama}, \citenamefont {Khanh}, \citenamefont {Karube} \emph
  {et~al.}}]{takagi2023spontaneous}%
  \BibitemOpen
  \bibfield  {author} {\bibinfo {author} {\bibfnamefont {H.}~\bibnamefont
  {Takagi}}, \bibinfo {author} {\bibfnamefont {R.}~\bibnamefont {Takagi}},
  \bibinfo {author} {\bibfnamefont {S.}~\bibnamefont {Minami}}, \bibinfo
  {author} {\bibfnamefont {T.}~\bibnamefont {Nomoto}}, \bibinfo {author}
  {\bibfnamefont {K.}~\bibnamefont {Ohishi}}, \bibinfo {author} {\bibfnamefont
  {M.-T.}\ \bibnamefont {Suzuki}}, \bibinfo {author} {\bibfnamefont
  {Y.}~\bibnamefont {Yanagi}}, \bibinfo {author} {\bibfnamefont
  {M.}~\bibnamefont {Hirayama}}, \bibinfo {author} {\bibfnamefont
  {N.}~\bibnamefont {Khanh}}, \bibinfo {author} {\bibfnamefont
  {K.}~\bibnamefont {Karube}}, \emph {et~al.},\ }\bibfield  {title} {\bibinfo
  {title} {Spontaneous topological hall effect induced by non-coplanar
  antiferromagnetic order in intercalated van der waals materials},\
  }\href@noop {} {\bibfield  {journal} {\bibinfo  {journal} {Nature Physics}\
  }\textbf {\bibinfo {volume} {19}},\ \bibinfo {pages} {961} (\bibinfo {year}
  {2023})}\BibitemShut {NoStop}%
\bibitem [{\citenamefont {Pop{\v{c}}evi{\'c}}\ \emph
  {et~al.}(2023)\citenamefont {Pop{\v{c}}evi{\'c}}, \citenamefont
  {Batisti{\'c}}, \citenamefont {Smontara}, \citenamefont {Velebit},
  \citenamefont {Ja{\'c}imovi{\'c}}, \citenamefont {{\v{Z}}ivkovi{\'c}},
  \citenamefont {Tsyrulin}, \citenamefont {Piatek}, \citenamefont {Berger},
  \citenamefont {Sidorenko} \emph {et~al.}}]{popvcevic2023electronic}%
  \BibitemOpen
  \bibfield  {author} {\bibinfo {author} {\bibfnamefont {P.}~\bibnamefont
  {Pop{\v{c}}evi{\'c}}}, \bibinfo {author} {\bibfnamefont {I.}~\bibnamefont
  {Batisti{\'c}}}, \bibinfo {author} {\bibfnamefont {A.}~\bibnamefont
  {Smontara}}, \bibinfo {author} {\bibfnamefont {K.}~\bibnamefont {Velebit}},
  \bibinfo {author} {\bibfnamefont {J.}~\bibnamefont {Ja{\'c}imovi{\'c}}},
  \bibinfo {author} {\bibfnamefont {I.}~\bibnamefont {{\v{Z}}ivkovi{\'c}}},
  \bibinfo {author} {\bibfnamefont {N.}~\bibnamefont {Tsyrulin}}, \bibinfo
  {author} {\bibfnamefont {J.}~\bibnamefont {Piatek}}, \bibinfo {author}
  {\bibfnamefont {H.}~\bibnamefont {Berger}}, \bibinfo {author} {\bibfnamefont
  {A.}~\bibnamefont {Sidorenko}}, \emph {et~al.},\ }\bibfield  {title}
  {\bibinfo {title} {Electronic transport and magnetism in the alternating
  stack of metallic and highly frustrated magnetic layers in co 1/3 nb s 2},\
  }\href@noop {} {\bibfield  {journal} {\bibinfo  {journal} {Physical Review
  B}\ }\textbf {\bibinfo {volume} {107}},\ \bibinfo {pages} {235149} (\bibinfo
  {year} {2023})}\BibitemShut {NoStop}%
\bibitem [{\citenamefont {Tenasini}\ \emph {et~al.}(2020)\citenamefont
  {Tenasini}, \citenamefont {Martino}, \citenamefont {Ubrig}, \citenamefont
  {Ghimire}, \citenamefont {Berger}, \citenamefont {Zaharko}, \citenamefont
  {Wu}, \citenamefont {Mitchell}, \citenamefont {Martin}, \citenamefont
  {Forr{\'o}} \emph {et~al.}}]{tenasini2020giant}%
  \BibitemOpen
  \bibfield  {author} {\bibinfo {author} {\bibfnamefont {G.}~\bibnamefont
  {Tenasini}}, \bibinfo {author} {\bibfnamefont {E.}~\bibnamefont {Martino}},
  \bibinfo {author} {\bibfnamefont {N.}~\bibnamefont {Ubrig}}, \bibinfo
  {author} {\bibfnamefont {N.~J.}\ \bibnamefont {Ghimire}}, \bibinfo {author}
  {\bibfnamefont {H.}~\bibnamefont {Berger}}, \bibinfo {author} {\bibfnamefont
  {O.}~\bibnamefont {Zaharko}}, \bibinfo {author} {\bibfnamefont
  {F.}~\bibnamefont {Wu}}, \bibinfo {author} {\bibfnamefont {J.}~\bibnamefont
  {Mitchell}}, \bibinfo {author} {\bibfnamefont {I.}~\bibnamefont {Martin}},
  \bibinfo {author} {\bibfnamefont {L.}~\bibnamefont {Forr{\'o}}}, \emph
  {et~al.},\ }\bibfield  {title} {\bibinfo {title} {Giant anomalous hall effect
  in quasi-two-dimensional layered antiferromagnet co 1/3 nbs 2},\ }\href@noop
  {} {\bibfield  {journal} {\bibinfo  {journal} {Physical Review Research}\
  }\textbf {\bibinfo {volume} {2}},\ \bibinfo {pages} {023051} (\bibinfo {year}
  {2020})}\BibitemShut {NoStop}%
\bibitem [{\citenamefont {Park}\ \emph {et~al.}(2022)\citenamefont {Park},
  \citenamefont {Kang}, \citenamefont {Kim}, \citenamefont {Lee}, \citenamefont
  {Noh}, \citenamefont {Han},\ and\ \citenamefont {Park}}]{park2022field}%
  \BibitemOpen
  \bibfield  {author} {\bibinfo {author} {\bibfnamefont {P.}~\bibnamefont
  {Park}}, \bibinfo {author} {\bibfnamefont {Y.-G.}\ \bibnamefont {Kang}},
  \bibinfo {author} {\bibfnamefont {J.}~\bibnamefont {Kim}}, \bibinfo {author}
  {\bibfnamefont {K.~H.}\ \bibnamefont {Lee}}, \bibinfo {author} {\bibfnamefont
  {H.-J.}\ \bibnamefont {Noh}}, \bibinfo {author} {\bibfnamefont {M.~J.}\
  \bibnamefont {Han}},\ and\ \bibinfo {author} {\bibfnamefont {J.-G.}\
  \bibnamefont {Park}},\ }\bibfield  {title} {\bibinfo {title} {Field-tunable
  toroidal moment and anomalous hall effect in noncollinear antiferromagnetic
  weyl semimetal co1/3tas2},\ }\href@noop {} {\bibfield  {journal} {\bibinfo
  {journal} {npj Quantum Materials}\ }\textbf {\bibinfo {volume} {7}},\
  \bibinfo {pages} {42} (\bibinfo {year} {2022})}\BibitemShut {NoStop}%
\bibitem [{\citenamefont {Park}\ \emph {et~al.}(2023)\citenamefont {Park},
  \citenamefont {Cho}, \citenamefont {Kim}, \citenamefont {An}, \citenamefont
  {Kang}, \citenamefont {Avdeev}, \citenamefont {Sibille}, \citenamefont
  {Iida}, \citenamefont {Kajimoto}, \citenamefont {Lee} \emph
  {et~al.}}]{park2023tetrahedral}%
  \BibitemOpen
  \bibfield  {author} {\bibinfo {author} {\bibfnamefont {P.}~\bibnamefont
  {Park}}, \bibinfo {author} {\bibfnamefont {W.}~\bibnamefont {Cho}}, \bibinfo
  {author} {\bibfnamefont {C.}~\bibnamefont {Kim}}, \bibinfo {author}
  {\bibfnamefont {Y.}~\bibnamefont {An}}, \bibinfo {author} {\bibfnamefont
  {Y.-G.}\ \bibnamefont {Kang}}, \bibinfo {author} {\bibfnamefont
  {M.}~\bibnamefont {Avdeev}}, \bibinfo {author} {\bibfnamefont
  {R.}~\bibnamefont {Sibille}}, \bibinfo {author} {\bibfnamefont
  {K.}~\bibnamefont {Iida}}, \bibinfo {author} {\bibfnamefont {R.}~\bibnamefont
  {Kajimoto}}, \bibinfo {author} {\bibfnamefont {K.~H.}\ \bibnamefont {Lee}},
  \emph {et~al.},\ }\bibfield  {title} {\bibinfo {title} {Tetrahedral triple-q
  magnetic ordering and large spontaneous hall conductivity in the metallic
  triangular antiferromagnet co1/3tas2},\ }\href@noop {} {\bibfield  {journal}
  {\bibinfo  {journal} {Nature Communications}\ }\textbf {\bibinfo {volume}
  {14}},\ \bibinfo {pages} {8346} (\bibinfo {year} {2023})}\BibitemShut
  {NoStop}%
\bibitem [{\citenamefont {Kim}\ \emph {et~al.}(2024)\citenamefont {Kim},
  \citenamefont {Zhang}, \citenamefont {Park}, \citenamefont {Cho},
  \citenamefont {Kim},\ and\ \citenamefont {Park}}]{kim2024electrical}%
  \BibitemOpen
  \bibfield  {author} {\bibinfo {author} {\bibfnamefont {J.}~\bibnamefont
  {Kim}}, \bibinfo {author} {\bibfnamefont {K.}~\bibnamefont {Zhang}}, \bibinfo
  {author} {\bibfnamefont {P.}~\bibnamefont {Park}}, \bibinfo {author}
  {\bibfnamefont {W.}~\bibnamefont {Cho}}, \bibinfo {author} {\bibfnamefont
  {H.}~\bibnamefont {Kim}},\ and\ \bibinfo {author} {\bibfnamefont {J.-G.}\
  \bibnamefont {Park}},\ }\bibfield  {title} {\bibinfo {title} {Electrical
  control of topological 3q state in an intercalated van der waals
  antiferromagnet},\ }\href@noop {} {\bibfield  {journal} {\bibinfo  {journal}
  {arXiv preprint arXiv:2409.02710}\ } (\bibinfo {year} {2024})}\BibitemShut
  {NoStop}%
\bibitem [{\citenamefont {Park}\ \emph {et~al.}(2024)\citenamefont {Park},
  \citenamefont {Cho}, \citenamefont {Kim}, \citenamefont {An}, \citenamefont
  {Avdeev}, \citenamefont {Iida}, \citenamefont {Kajimoto},\ and\ \citenamefont
  {Park}}]{park2024composition}%
  \BibitemOpen
  \bibfield  {author} {\bibinfo {author} {\bibfnamefont {P.}~\bibnamefont
  {Park}}, \bibinfo {author} {\bibfnamefont {W.}~\bibnamefont {Cho}}, \bibinfo
  {author} {\bibfnamefont {C.}~\bibnamefont {Kim}}, \bibinfo {author}
  {\bibfnamefont {Y.}~\bibnamefont {An}}, \bibinfo {author} {\bibfnamefont
  {M.}~\bibnamefont {Avdeev}}, \bibinfo {author} {\bibfnamefont
  {K.}~\bibnamefont {Iida}}, \bibinfo {author} {\bibfnamefont {R.}~\bibnamefont
  {Kajimoto}},\ and\ \bibinfo {author} {\bibfnamefont {J.-G.}\ \bibnamefont
  {Park}},\ }\bibfield  {title} {\bibinfo {title} {Composition dependence of
  bulk properties in the co-intercalated transition metal dichalcogenide co 1/3
  ta s 2},\ }\href@noop {} {\bibfield  {journal} {\bibinfo  {journal} {Physical
  Review B}\ }\textbf {\bibinfo {volume} {109}},\ \bibinfo {pages} {L060403}
  (\bibinfo {year} {2024})}\BibitemShut {NoStop}%
\bibitem [{\citenamefont {Little}\ \emph {et~al.}(2020)\citenamefont {Little},
  \citenamefont {Lee}, \citenamefont {John}, \citenamefont {Doyle},
  \citenamefont {Maniv}, \citenamefont {Nair}, \citenamefont {Chen},
  \citenamefont {Rees}, \citenamefont {Venderbos}, \citenamefont {Fernandes}
  \emph {et~al.}}]{little2020three}%
  \BibitemOpen
  \bibfield  {author} {\bibinfo {author} {\bibfnamefont {A.}~\bibnamefont
  {Little}}, \bibinfo {author} {\bibfnamefont {C.}~\bibnamefont {Lee}},
  \bibinfo {author} {\bibfnamefont {C.}~\bibnamefont {John}}, \bibinfo {author}
  {\bibfnamefont {S.}~\bibnamefont {Doyle}}, \bibinfo {author} {\bibfnamefont
  {E.}~\bibnamefont {Maniv}}, \bibinfo {author} {\bibfnamefont {N.~L.}\
  \bibnamefont {Nair}}, \bibinfo {author} {\bibfnamefont {W.}~\bibnamefont
  {Chen}}, \bibinfo {author} {\bibfnamefont {D.}~\bibnamefont {Rees}}, \bibinfo
  {author} {\bibfnamefont {J.~W.}\ \bibnamefont {Venderbos}}, \bibinfo {author}
  {\bibfnamefont {R.~M.}\ \bibnamefont {Fernandes}}, \emph {et~al.},\
  }\bibfield  {title} {\bibinfo {title} {Three-state nematicity in the
  triangular lattice antiferromagnet fe1/3nbs2},\ }\href@noop {} {\bibfield
  {journal} {\bibinfo  {journal} {Nature materials}\ }\textbf {\bibinfo
  {volume} {19}},\ \bibinfo {pages} {1062} (\bibinfo {year}
  {2020})}\BibitemShut {NoStop}%
\bibitem [{\citenamefont {Nair}\ \emph {et~al.}(2020)\citenamefont {Nair},
  \citenamefont {Maniv}, \citenamefont {John}, \citenamefont {Doyle},
  \citenamefont {Orenstein},\ and\ \citenamefont
  {Analytis}}]{nair2020electrical}%
  \BibitemOpen
  \bibfield  {author} {\bibinfo {author} {\bibfnamefont {N.~L.}\ \bibnamefont
  {Nair}}, \bibinfo {author} {\bibfnamefont {E.}~\bibnamefont {Maniv}},
  \bibinfo {author} {\bibfnamefont {C.}~\bibnamefont {John}}, \bibinfo {author}
  {\bibfnamefont {S.}~\bibnamefont {Doyle}}, \bibinfo {author} {\bibfnamefont
  {J.}~\bibnamefont {Orenstein}},\ and\ \bibinfo {author} {\bibfnamefont
  {J.~G.}\ \bibnamefont {Analytis}},\ }\bibfield  {title} {\bibinfo {title}
  {Electrical switching in a magnetically intercalated transition metal
  dichalcogenide},\ }\href@noop {} {\bibfield  {journal} {\bibinfo  {journal}
  {Nature materials}\ }\textbf {\bibinfo {volume} {19}},\ \bibinfo {pages}
  {153} (\bibinfo {year} {2020})}\BibitemShut {NoStop}%
\bibitem [{\citenamefont {Pan}\ \emph {et~al.}(2023)\citenamefont {Pan},
  \citenamefont {Tang}, \citenamefont {Bai}, \citenamefont {Tang},
  \citenamefont {Zhang}, \citenamefont {Chen}, \citenamefont {Guo},
  \citenamefont {Zhu}, \citenamefont {Xu},\ and\ \citenamefont
  {Xu}}]{pan2023fe}%
  \BibitemOpen
  \bibfield  {author} {\bibinfo {author} {\bibfnamefont {S.}~\bibnamefont
  {Pan}}, \bibinfo {author} {\bibfnamefont {L.}~\bibnamefont {Tang}}, \bibinfo
  {author} {\bibfnamefont {Y.}~\bibnamefont {Bai}}, \bibinfo {author}
  {\bibfnamefont {J.}~\bibnamefont {Tang}}, \bibinfo {author} {\bibfnamefont
  {Z.}~\bibnamefont {Zhang}}, \bibinfo {author} {\bibfnamefont
  {B.}~\bibnamefont {Chen}}, \bibinfo {author} {\bibfnamefont {Y.}~\bibnamefont
  {Guo}}, \bibinfo {author} {\bibfnamefont {J.}~\bibnamefont {Zhu}}, \bibinfo
  {author} {\bibfnamefont {G.}~\bibnamefont {Xu}},\ and\ \bibinfo {author}
  {\bibfnamefont {F.}~\bibnamefont {Xu}},\ }\bibfield  {title} {\bibinfo
  {title} {Fe and cr co-intercalation in 2 h-nb s 2 single crystals for
  realization of perpendicular magnetic anisotropy and large anomalous hall
  effect},\ }\href@noop {} {\bibfield  {journal} {\bibinfo  {journal} {Physical
  Review Materials}\ }\textbf {\bibinfo {volume} {7}},\ \bibinfo {pages}
  {034407} (\bibinfo {year} {2023})}\BibitemShut {NoStop}%
\bibitem [{\citenamefont {Liu}\ \emph {et~al.}(2022)\citenamefont {Liu},
  \citenamefont {Zhu}, \citenamefont {Wu}, \citenamefont {Lu},\ and\
  \citenamefont {Pu}}]{liu2022unconventional}%
  \BibitemOpen
  \bibfield  {author} {\bibinfo {author} {\bibfnamefont {P.}~\bibnamefont
  {Liu}}, \bibinfo {author} {\bibfnamefont {H.}~\bibnamefont {Zhu}}, \bibinfo
  {author} {\bibfnamefont {Q.}~\bibnamefont {Wu}}, \bibinfo {author}
  {\bibfnamefont {Y.}~\bibnamefont {Lu}},\ and\ \bibinfo {author}
  {\bibfnamefont {Y.}~\bibnamefont {Pu}},\ }\bibfield  {title} {\bibinfo
  {title} {Unconventional magneto-transport properties of the layered
  antiferromagnet fe1/3nbs2},\ }\href@noop {} {\bibfield  {journal} {\bibinfo
  {journal} {Applied Physics Letters}\ }\textbf {\bibinfo {volume} {121}}
  (\bibinfo {year} {2022})}\BibitemShut {NoStop}%
\bibitem [{\citenamefont {Mankovsky}\ \emph {et~al.}(2016)\citenamefont
  {Mankovsky}, \citenamefont {Polesya}, \citenamefont {Ebert},\ and\
  \citenamefont {Bensch}}]{mankovsky2016electronic}%
  \BibitemOpen
  \bibfield  {author} {\bibinfo {author} {\bibfnamefont {S.}~\bibnamefont
  {Mankovsky}}, \bibinfo {author} {\bibfnamefont {S.}~\bibnamefont {Polesya}},
  \bibinfo {author} {\bibfnamefont {H.}~\bibnamefont {Ebert}},\ and\ \bibinfo
  {author} {\bibfnamefont {W.}~\bibnamefont {Bensch}},\ }\bibfield  {title}
  {\bibinfo {title} {Electronic and magnetic properties of 2 h- nbs 2
  intercalated by 3 d transition metals},\ }\href@noop {} {\bibfield  {journal}
  {\bibinfo  {journal} {Physical Review B}\ }\textbf {\bibinfo {volume} {94}},\
  \bibinfo {pages} {184430} (\bibinfo {year} {2016})}\BibitemShut {NoStop}%
\bibitem [{\citenamefont {Lu}\ \emph {et~al.}(2020)\citenamefont {Lu},
  \citenamefont {Sapkota}, \citenamefont {DeBeer-Schmitt}, \citenamefont {Wu},
  \citenamefont {Cao}, \citenamefont {Mannella}, \citenamefont {Mandrus},
  \citenamefont {Aczel},\ and\ \citenamefont {MacDougall}}]{lu2020canted}%
  \BibitemOpen
  \bibfield  {author} {\bibinfo {author} {\bibfnamefont {K.}~\bibnamefont
  {Lu}}, \bibinfo {author} {\bibfnamefont {D.}~\bibnamefont {Sapkota}},
  \bibinfo {author} {\bibfnamefont {L.}~\bibnamefont {DeBeer-Schmitt}},
  \bibinfo {author} {\bibfnamefont {Y.}~\bibnamefont {Wu}}, \bibinfo {author}
  {\bibfnamefont {H.}~\bibnamefont {Cao}}, \bibinfo {author} {\bibfnamefont
  {N.}~\bibnamefont {Mannella}}, \bibinfo {author} {\bibfnamefont
  {D.}~\bibnamefont {Mandrus}}, \bibinfo {author} {\bibfnamefont {A.~A.}\
  \bibnamefont {Aczel}},\ and\ \bibinfo {author} {\bibfnamefont {G.~J.}\
  \bibnamefont {MacDougall}},\ }\bibfield  {title} {\bibinfo {title} {Canted
  antiferromagnetic order in the monoaxial chiral magnets v 1/3 tas 2 and v 1/3
  nbs 2},\ }\href@noop {} {\bibfield  {journal} {\bibinfo  {journal} {Physical
  Review Materials}\ }\textbf {\bibinfo {volume} {4}},\ \bibinfo {pages}
  {054416} (\bibinfo {year} {2020})}\BibitemShut {NoStop}%
\bibitem [{\citenamefont {Zhang}\ \emph {et~al.}(2018)\citenamefont {Zhang},
  \citenamefont {Wei}, \citenamefont {Zheng}, \citenamefont {Lu}, \citenamefont
  {Wu}, \citenamefont {Zhu}, \citenamefont {Tang}, \citenamefont {Ning},
  \citenamefont {Han}, \citenamefont {Ling} \emph
  {et~al.}}]{zhang2018electrical}%
  \BibitemOpen
  \bibfield  {author} {\bibinfo {author} {\bibfnamefont {H.}~\bibnamefont
  {Zhang}}, \bibinfo {author} {\bibfnamefont {W.}~\bibnamefont {Wei}}, \bibinfo
  {author} {\bibfnamefont {G.}~\bibnamefont {Zheng}}, \bibinfo {author}
  {\bibfnamefont {J.}~\bibnamefont {Lu}}, \bibinfo {author} {\bibfnamefont
  {M.}~\bibnamefont {Wu}}, \bibinfo {author} {\bibfnamefont {X.}~\bibnamefont
  {Zhu}}, \bibinfo {author} {\bibfnamefont {J.}~\bibnamefont {Tang}}, \bibinfo
  {author} {\bibfnamefont {W.}~\bibnamefont {Ning}}, \bibinfo {author}
  {\bibfnamefont {Y.}~\bibnamefont {Han}}, \bibinfo {author} {\bibfnamefont
  {L.}~\bibnamefont {Ling}}, \emph {et~al.},\ }\bibfield  {title} {\bibinfo
  {title} {Electrical and anisotropic magnetic properties in layered mn1/3tas2
  crystals},\ }\href@noop {} {\bibfield  {journal} {\bibinfo  {journal}
  {Applied Physics Letters}\ }\textbf {\bibinfo {volume} {113}} (\bibinfo
  {year} {2018})}\BibitemShut {NoStop}%
\bibitem [{\citenamefont {Wu}\ \emph {et~al.}(2022)\citenamefont {Wu},
  \citenamefont {Xu}, \citenamefont {Haley}, \citenamefont {Weber},
  \citenamefont {Acharya}, \citenamefont {Maniv}, \citenamefont {Qiu},
  \citenamefont {Aczel}, \citenamefont {Settineri}, \citenamefont {Neaton}
  \emph {et~al.}}]{wu2022highly}%
  \BibitemOpen
  \bibfield  {author} {\bibinfo {author} {\bibfnamefont {S.}~\bibnamefont
  {Wu}}, \bibinfo {author} {\bibfnamefont {Z.}~\bibnamefont {Xu}}, \bibinfo
  {author} {\bibfnamefont {S.~C.}\ \bibnamefont {Haley}}, \bibinfo {author}
  {\bibfnamefont {S.~F.}\ \bibnamefont {Weber}}, \bibinfo {author}
  {\bibfnamefont {A.}~\bibnamefont {Acharya}}, \bibinfo {author} {\bibfnamefont
  {E.}~\bibnamefont {Maniv}}, \bibinfo {author} {\bibfnamefont
  {Y.}~\bibnamefont {Qiu}}, \bibinfo {author} {\bibfnamefont {A.}~\bibnamefont
  {Aczel}}, \bibinfo {author} {\bibfnamefont {N.~S.}\ \bibnamefont
  {Settineri}}, \bibinfo {author} {\bibfnamefont {J.~B.}\ \bibnamefont
  {Neaton}}, \emph {et~al.},\ }\bibfield  {title} {\bibinfo {title} {Highly
  tunable magnetic phases in transition-metal dichalcogenide fe 1/3+ $\delta$
  nbs 2},\ }\href@noop {} {\bibfield  {journal} {\bibinfo  {journal} {Physical
  Review X}\ }\textbf {\bibinfo {volume} {12}},\ \bibinfo {pages} {021003}
  (\bibinfo {year} {2022})}\BibitemShut {NoStop}%
\bibitem [{\citenamefont {Wu}\ \emph {et~al.}(2023)\citenamefont {Wu},
  \citenamefont {Basak}, \citenamefont {Li}, \citenamefont {Kim}, \citenamefont
  {Ryan}, \citenamefont {Lu}, \citenamefont {Hashimoto}, \citenamefont
  {Nelson}, \citenamefont {Acevedo-Esteves}, \citenamefont {Haley} \emph
  {et~al.}}]{wu2023discovery}%
  \BibitemOpen
  \bibfield  {author} {\bibinfo {author} {\bibfnamefont {S.}~\bibnamefont
  {Wu}}, \bibinfo {author} {\bibfnamefont {R.}~\bibnamefont {Basak}}, \bibinfo
  {author} {\bibfnamefont {W.}~\bibnamefont {Li}}, \bibinfo {author}
  {\bibfnamefont {J.-W.}\ \bibnamefont {Kim}}, \bibinfo {author} {\bibfnamefont
  {P.~J.}\ \bibnamefont {Ryan}}, \bibinfo {author} {\bibfnamefont
  {D.}~\bibnamefont {Lu}}, \bibinfo {author} {\bibfnamefont {M.}~\bibnamefont
  {Hashimoto}}, \bibinfo {author} {\bibfnamefont {C.}~\bibnamefont {Nelson}},
  \bibinfo {author} {\bibfnamefont {R.}~\bibnamefont {Acevedo-Esteves}},
  \bibinfo {author} {\bibfnamefont {S.~C.}\ \bibnamefont {Haley}}, \emph
  {et~al.},\ }\bibfield  {title} {\bibinfo {title} {Discovery of charge order
  in the transition metal dichalcogenide fe x nbs 2},\ }\href@noop {}
  {\bibfield  {journal} {\bibinfo  {journal} {Physical Review Letters}\
  }\textbf {\bibinfo {volume} {131}},\ \bibinfo {pages} {186701} (\bibinfo
  {year} {2023})}\BibitemShut {NoStop}%
\bibitem [{\citenamefont {Zhao}\ \emph {et~al.}(2017)\citenamefont {Zhao},
  \citenamefont {Wijayaratne}, \citenamefont {Butler}, \citenamefont {Yang},
  \citenamefont {Malliakas}, \citenamefont {Chung}, \citenamefont {Louca},
  \citenamefont {Kanatzidis}, \citenamefont {van Wezel},\ and\ \citenamefont
  {Chatterjee}}]{PhysRevB.96.125103}%
  \BibitemOpen
  \bibfield  {author} {\bibinfo {author} {\bibfnamefont {J.}~\bibnamefont
  {Zhao}}, \bibinfo {author} {\bibfnamefont {K.}~\bibnamefont {Wijayaratne}},
  \bibinfo {author} {\bibfnamefont {A.}~\bibnamefont {Butler}}, \bibinfo
  {author} {\bibfnamefont {J.}~\bibnamefont {Yang}}, \bibinfo {author}
  {\bibfnamefont {C.~D.}\ \bibnamefont {Malliakas}}, \bibinfo {author}
  {\bibfnamefont {D.~Y.}\ \bibnamefont {Chung}}, \bibinfo {author}
  {\bibfnamefont {D.}~\bibnamefont {Louca}}, \bibinfo {author} {\bibfnamefont
  {M.~G.}\ \bibnamefont {Kanatzidis}}, \bibinfo {author} {\bibfnamefont
  {J.}~\bibnamefont {van Wezel}},\ and\ \bibinfo {author} {\bibfnamefont
  {U.}~\bibnamefont {Chatterjee}},\ }\bibfield  {title} {\bibinfo {title}
  {Orbital selectivity causing anisotropy and particle-hole asymmetry in the
  charge density wave gap of $2h\text{\ensuremath{-}}{\mathrm{tas}}_{2}$},\
  }\href {https://doi.org/10.1103/PhysRevB.96.125103} {\bibfield  {journal}
  {\bibinfo  {journal} {Phys. Rev. B}\ }\textbf {\bibinfo {volume} {96}},\
  \bibinfo {pages} {125103} (\bibinfo {year} {2017})}\BibitemShut {NoStop}%
\bibitem [{\citenamefont {Yang}\ \emph {et~al.}(2018)\citenamefont {Yang},
  \citenamefont {Fang}, \citenamefont {Fatemi}, \citenamefont {Ruhman},
  \citenamefont {Navarro-Moratalla}, \citenamefont {Watanabe}, \citenamefont
  {Taniguchi}, \citenamefont {Kaxiras},\ and\ \citenamefont
  {Jarillo-Herrero}}]{PhysRevB.98.035203}%
  \BibitemOpen
  \bibfield  {author} {\bibinfo {author} {\bibfnamefont {Y.}~\bibnamefont
  {Yang}}, \bibinfo {author} {\bibfnamefont {S.}~\bibnamefont {Fang}}, \bibinfo
  {author} {\bibfnamefont {V.}~\bibnamefont {Fatemi}}, \bibinfo {author}
  {\bibfnamefont {J.}~\bibnamefont {Ruhman}}, \bibinfo {author} {\bibfnamefont
  {E.}~\bibnamefont {Navarro-Moratalla}}, \bibinfo {author} {\bibfnamefont
  {K.}~\bibnamefont {Watanabe}}, \bibinfo {author} {\bibfnamefont
  {T.}~\bibnamefont {Taniguchi}}, \bibinfo {author} {\bibfnamefont
  {E.}~\bibnamefont {Kaxiras}},\ and\ \bibinfo {author} {\bibfnamefont
  {P.}~\bibnamefont {Jarillo-Herrero}},\ }\bibfield  {title} {\bibinfo {title}
  {Enhanced superconductivity upon weakening of charge density wave transport
  in $2h{\text{-tas}}_{2}$ in the two-dimensional limit},\ }\href
  {https://doi.org/10.1103/PhysRevB.98.035203} {\bibfield  {journal} {\bibinfo
  {journal} {Phys. Rev. B}\ }\textbf {\bibinfo {volume} {98}},\ \bibinfo
  {pages} {035203} (\bibinfo {year} {2018})}\BibitemShut {NoStop}%
\bibitem [{\citenamefont {Dijkstra}\ \emph {et~al.}(1989)\citenamefont
  {Dijkstra}, \citenamefont {Zijlema}, \citenamefont {Van~Bruggen},
  \citenamefont {Haas},\ and\ \citenamefont {De~Groot}}]{dijkstra1989band}%
  \BibitemOpen
  \bibfield  {author} {\bibinfo {author} {\bibfnamefont {J.}~\bibnamefont
  {Dijkstra}}, \bibinfo {author} {\bibfnamefont {P.}~\bibnamefont {Zijlema}},
  \bibinfo {author} {\bibfnamefont {C.}~\bibnamefont {Van~Bruggen}}, \bibinfo
  {author} {\bibfnamefont {C.}~\bibnamefont {Haas}},\ and\ \bibinfo {author}
  {\bibfnamefont {R.}~\bibnamefont {De~Groot}},\ }\bibfield  {title} {\bibinfo
  {title} {Band-structure calculations of fe1/3tas2 and mn1/3tas2, and
  transport and magnetic properties of fe0. 28tas2},\ }\href@noop {} {\bibfield
   {journal} {\bibinfo  {journal} {Journal of Physics: Condensed Matter}\
  }\textbf {\bibinfo {volume} {1}},\ \bibinfo {pages} {6363} (\bibinfo {year}
  {1989})}\BibitemShut {NoStop}%
\bibitem [{\citenamefont {Yan}\ \emph {et~al.}(2015)\citenamefont {Yan},
  \citenamefont {Cruz}, \citenamefont {Cook},\ and\ \citenamefont
  {Varga}}]{yan2015structural}%
  \BibitemOpen
  \bibfield  {author} {\bibinfo {author} {\bibfnamefont {J.-A.}\ \bibnamefont
  {Yan}}, \bibinfo {author} {\bibfnamefont {M.~A.~D.}\ \bibnamefont {Cruz}},
  \bibinfo {author} {\bibfnamefont {B.}~\bibnamefont {Cook}},\ and\ \bibinfo
  {author} {\bibfnamefont {K.}~\bibnamefont {Varga}},\ }\bibfield  {title}
  {\bibinfo {title} {Structural, electronic and vibrational properties of
  few-layer 2h-and 1t-tase2},\ }\href@noop {} {\bibfield  {journal} {\bibinfo
  {journal} {Scientific reports}\ }\textbf {\bibinfo {volume} {5}},\ \bibinfo
  {pages} {1} (\bibinfo {year} {2015})}\BibitemShut {NoStop}%
\bibitem [{\citenamefont {Rahn}\ \emph {et~al.}(2012)\citenamefont {Rahn},
  \citenamefont {Hellmann}, \citenamefont {Kall\"ane}, \citenamefont {Sohrt},
  \citenamefont {Kim}, \citenamefont {Kipp},\ and\ \citenamefont
  {Rossnagel}}]{PhysRevB.85.224532}%
  \BibitemOpen
  \bibfield  {author} {\bibinfo {author} {\bibfnamefont {D.~J.}\ \bibnamefont
  {Rahn}}, \bibinfo {author} {\bibfnamefont {S.}~\bibnamefont {Hellmann}},
  \bibinfo {author} {\bibfnamefont {M.}~\bibnamefont {Kall\"ane}}, \bibinfo
  {author} {\bibfnamefont {C.}~\bibnamefont {Sohrt}}, \bibinfo {author}
  {\bibfnamefont {T.~K.}\ \bibnamefont {Kim}}, \bibinfo {author} {\bibfnamefont
  {L.}~\bibnamefont {Kipp}},\ and\ \bibinfo {author} {\bibfnamefont
  {K.}~\bibnamefont {Rossnagel}},\ }\bibfield  {title} {\bibinfo {title} {Gaps
  and kinks in the electronic structure of the superconductor $2h$-nbse${}_{2}$
  from angle-resolved photoemission at 1 k},\ }\href
  {https://doi.org/10.1103/PhysRevB.85.224532} {\bibfield  {journal} {\bibinfo
  {journal} {Phys. Rev. B}\ }\textbf {\bibinfo {volume} {85}},\ \bibinfo
  {pages} {224532} (\bibinfo {year} {2012})}\BibitemShut {NoStop}%
\bibitem [{\citenamefont {Thompson}\ \emph {et~al.}(1972)\citenamefont
  {Thompson}, \citenamefont {Gamble},\ and\ \citenamefont
  {Koehler}}]{thompson1972}%
  \BibitemOpen
  \bibfield  {author} {\bibinfo {author} {\bibfnamefont {A.~H.}\ \bibnamefont
  {Thompson}}, \bibinfo {author} {\bibfnamefont {F.~R.}\ \bibnamefont
  {Gamble}},\ and\ \bibinfo {author} {\bibfnamefont {R.~F.}\ \bibnamefont
  {Koehler}},\ }\bibfield  {title} {\bibinfo {title} {Effects of intercalation
  on electron transport in tantalum disulfide},\ }\href
  {https://doi.org/10.1103/PhysRevB.5.2811} {\bibfield  {journal} {\bibinfo
  {journal} {Phys. Rev. B}\ }\textbf {\bibinfo {volume} {5}},\ \bibinfo {pages}
  {2811} (\bibinfo {year} {1972})}\BibitemShut {NoStop}%
\bibitem [{\citenamefont {Wilson}\ and\ \citenamefont
  {Ortiz}(2024)}]{wilson2024v3sb5}%
  \BibitemOpen
  \bibfield  {author} {\bibinfo {author} {\bibfnamefont {S.~D.}\ \bibnamefont
  {Wilson}}\ and\ \bibinfo {author} {\bibfnamefont {B.~R.}\ \bibnamefont
  {Ortiz}},\ }\bibfield  {title} {\bibinfo {title} {A v3sb5 kagome
  superconductors},\ }\href@noop {} {\bibfield  {journal} {\bibinfo  {journal}
  {Nature Reviews Materials}\ ,\ \bibinfo {pages} {1}} (\bibinfo {year}
  {2024})}\BibitemShut {NoStop}%
\bibitem [{\citenamefont {Hu}\ \emph {et~al.}(2022)\citenamefont {Hu},
  \citenamefont {Wu}, \citenamefont {Ortiz}, \citenamefont {Ju}, \citenamefont
  {Han}, \citenamefont {Ma}, \citenamefont {Plumb}, \citenamefont {Radovic},
  \citenamefont {Thomale}, \citenamefont {Wilson} \emph {et~al.}}]{hu2022rich}%
  \BibitemOpen
  \bibfield  {author} {\bibinfo {author} {\bibfnamefont {Y.}~\bibnamefont
  {Hu}}, \bibinfo {author} {\bibfnamefont {X.}~\bibnamefont {Wu}}, \bibinfo
  {author} {\bibfnamefont {B.~R.}\ \bibnamefont {Ortiz}}, \bibinfo {author}
  {\bibfnamefont {S.}~\bibnamefont {Ju}}, \bibinfo {author} {\bibfnamefont
  {X.}~\bibnamefont {Han}}, \bibinfo {author} {\bibfnamefont {J.}~\bibnamefont
  {Ma}}, \bibinfo {author} {\bibfnamefont {N.~C.}\ \bibnamefont {Plumb}},
  \bibinfo {author} {\bibfnamefont {M.}~\bibnamefont {Radovic}}, \bibinfo
  {author} {\bibfnamefont {R.}~\bibnamefont {Thomale}}, \bibinfo {author}
  {\bibfnamefont {S.~D.}\ \bibnamefont {Wilson}}, \emph {et~al.},\ }\bibfield
  {title} {\bibinfo {title} {Rich nature of van hove singularities in kagome
  superconductor csv3sb5},\ }\href@noop {} {\bibfield  {journal} {\bibinfo
  {journal} {Nature Communications}\ }\textbf {\bibinfo {volume} {13}},\
  \bibinfo {pages} {2220} (\bibinfo {year} {2022})}\BibitemShut {NoStop}%
\bibitem [{\citenamefont {Leroux}\ \emph {et~al.}(2012)\citenamefont {Leroux},
  \citenamefont {Le~Tacon}, \citenamefont {Calandra}, \citenamefont {Cario},
  \citenamefont {Measson}, \citenamefont {Diener}, \citenamefont {Borrissenko},
  \citenamefont {Bosak},\ and\ \citenamefont {Rodiere}}]{leroux2012anharmonic}%
  \BibitemOpen
  \bibfield  {author} {\bibinfo {author} {\bibfnamefont {M.}~\bibnamefont
  {Leroux}}, \bibinfo {author} {\bibfnamefont {M.}~\bibnamefont {Le~Tacon}},
  \bibinfo {author} {\bibfnamefont {M.}~\bibnamefont {Calandra}}, \bibinfo
  {author} {\bibfnamefont {L.}~\bibnamefont {Cario}}, \bibinfo {author}
  {\bibfnamefont {M.~A.}\ \bibnamefont {Measson}}, \bibinfo {author}
  {\bibfnamefont {P.}~\bibnamefont {Diener}}, \bibinfo {author} {\bibfnamefont
  {E.}~\bibnamefont {Borrissenko}}, \bibinfo {author} {\bibfnamefont
  {A.}~\bibnamefont {Bosak}},\ and\ \bibinfo {author} {\bibfnamefont
  {P.}~\bibnamefont {Rodiere}},\ }\bibfield  {title} {\bibinfo {title}
  {Anharmonic suppression of charge density waves in 2 h-nbs 2},\ }\href@noop
  {} {\bibfield  {journal} {\bibinfo  {journal} {Physical Review B—Condensed
  Matter and Materials Physics}\ }\textbf {\bibinfo {volume} {86}},\ \bibinfo
  {pages} {155125} (\bibinfo {year} {2012})}\BibitemShut {NoStop}%
\bibitem [{\citenamefont {Dyadkin}\ \emph {et~al.}(2015)\citenamefont
  {Dyadkin}, \citenamefont {Mushenok}, \citenamefont {Bosak}, \citenamefont
  {Menzel}, \citenamefont {Grigoriev}, \citenamefont {Pattison},\ and\
  \citenamefont {Chernyshov}}]{dyadkhin2015}%
  \BibitemOpen
  \bibfield  {author} {\bibinfo {author} {\bibfnamefont {V.}~\bibnamefont
  {Dyadkin}}, \bibinfo {author} {\bibfnamefont {F.}~\bibnamefont {Mushenok}},
  \bibinfo {author} {\bibfnamefont {A.}~\bibnamefont {Bosak}}, \bibinfo
  {author} {\bibfnamefont {D.}~\bibnamefont {Menzel}}, \bibinfo {author}
  {\bibfnamefont {S.}~\bibnamefont {Grigoriev}}, \bibinfo {author}
  {\bibfnamefont {P.}~\bibnamefont {Pattison}},\ and\ \bibinfo {author}
  {\bibfnamefont {D.}~\bibnamefont {Chernyshov}},\ }\bibfield  {title}
  {\bibinfo {title} {Structural disorder versus chiral magnetism in
  ${\mathrm{cr}}_{1/3}{\mathrm{nbs}}_{2}$},\ }\href
  {https://doi.org/10.1103/PhysRevB.91.184205} {\bibfield  {journal} {\bibinfo
  {journal} {Phys. Rev. B}\ }\textbf {\bibinfo {volume} {91}},\ \bibinfo
  {pages} {184205} (\bibinfo {year} {2015})}\BibitemShut {NoStop}%
\bibitem [{\citenamefont {Balseiro}\ \emph {et~al.}(1980)\citenamefont
  {Balseiro}, \citenamefont {Schlottmann},\ and\ \citenamefont
  {Yndurain}}]{Balseiro1980}%
  \BibitemOpen
  \bibfield  {author} {\bibinfo {author} {\bibfnamefont {C.~A.}\ \bibnamefont
  {Balseiro}}, \bibinfo {author} {\bibfnamefont {P.}~\bibnamefont
  {Schlottmann}},\ and\ \bibinfo {author} {\bibfnamefont {F.}~\bibnamefont
  {Yndurain}},\ }\bibfield  {title} {\bibinfo {title} {Coexistence of
  charge-density waves and magnetic order},\ }\href
  {https://doi.org/10.1103/PhysRevB.21.5267} {\bibfield  {journal} {\bibinfo
  {journal} {Phys. Rev. B}\ }\textbf {\bibinfo {volume} {21}},\ \bibinfo
  {pages} {5267} (\bibinfo {year} {1980})}\BibitemShut {NoStop}%
\bibitem [{\citenamefont {Weber}\ \emph {et~al.}(2022)\citenamefont {Weber},
  \citenamefont {Guennou}, \citenamefont {Evans}, \citenamefont {Toulouse},
  \citenamefont {Simonov}, \citenamefont {Kholina}, \citenamefont {Ma},
  \citenamefont {Ren}, \citenamefont {Cao}, \citenamefont {Carpenter},
  \citenamefont {Dkhil}, \citenamefont {Fiebig},\ and\ \citenamefont
  {Kreisel}}]{weber2022}%
  \BibitemOpen
  \bibfield  {author} {\bibinfo {author} {\bibfnamefont {M.~C.}\ \bibnamefont
  {Weber}}, \bibinfo {author} {\bibfnamefont {M.}~\bibnamefont {Guennou}},
  \bibinfo {author} {\bibfnamefont {D.~M.}\ \bibnamefont {Evans}}, \bibinfo
  {author} {\bibfnamefont {C.}~\bibnamefont {Toulouse}}, \bibinfo {author}
  {\bibfnamefont {A.}~\bibnamefont {Simonov}}, \bibinfo {author} {\bibfnamefont
  {Y.}~\bibnamefont {Kholina}}, \bibinfo {author} {\bibfnamefont
  {X.}~\bibnamefont {Ma}}, \bibinfo {author} {\bibfnamefont {W.}~\bibnamefont
  {Ren}}, \bibinfo {author} {\bibfnamefont {S.}~\bibnamefont {Cao}}, \bibinfo
  {author} {\bibfnamefont {M.~A.}\ \bibnamefont {Carpenter}}, \bibinfo {author}
  {\bibfnamefont {B.}~\bibnamefont {Dkhil}}, \bibinfo {author} {\bibfnamefont
  {M.}~\bibnamefont {Fiebig}},\ and\ \bibinfo {author} {\bibfnamefont
  {J.}~\bibnamefont {Kreisel}},\ }\bibfield  {title} {\bibinfo {title}
  {Emerging spin--phonon coupling through cross-talk of two magnetic
  sublattices},\ }\href {https://doi.org/10.1038/s41467-021-27267-8} {\bibfield
   {journal} {\bibinfo  {journal} {Nature Communications}\ }\textbf {\bibinfo
  {volume} {13}},\ \bibinfo {pages} {443} (\bibinfo {year} {2022})}\BibitemShut
  {NoStop}%
\bibitem [{\citenamefont {Ricci}\ \emph {et~al.}(2021)\citenamefont {Ricci},
  \citenamefont {Poccia}, \citenamefont {Campi}, \citenamefont {Mishra},
  \citenamefont {M\"uller}, \citenamefont {Joseph}, \citenamefont {Shi},
  \citenamefont {Zozulya}, \citenamefont {Buchholz}, \citenamefont {Trabant},
  \citenamefont {Lee}, \citenamefont {Viefhaus}, \citenamefont {Goedkoop},
  \citenamefont {Nugroho}, \citenamefont {Braden}, \citenamefont {Roy},
  \citenamefont {Sprung},\ and\ \citenamefont
  {Sch\"u\ss{}ler-Langeheine}}]{schuessler2021}%
  \BibitemOpen
  \bibfield  {author} {\bibinfo {author} {\bibfnamefont {A.}~\bibnamefont
  {Ricci}}, \bibinfo {author} {\bibfnamefont {N.}~\bibnamefont {Poccia}},
  \bibinfo {author} {\bibfnamefont {G.}~\bibnamefont {Campi}}, \bibinfo
  {author} {\bibfnamefont {S.}~\bibnamefont {Mishra}}, \bibinfo {author}
  {\bibfnamefont {L.}~\bibnamefont {M\"uller}}, \bibinfo {author}
  {\bibfnamefont {B.}~\bibnamefont {Joseph}}, \bibinfo {author} {\bibfnamefont
  {B.}~\bibnamefont {Shi}}, \bibinfo {author} {\bibfnamefont {A.}~\bibnamefont
  {Zozulya}}, \bibinfo {author} {\bibfnamefont {M.}~\bibnamefont {Buchholz}},
  \bibinfo {author} {\bibfnamefont {C.}~\bibnamefont {Trabant}}, \bibinfo
  {author} {\bibfnamefont {J.~C.~T.}\ \bibnamefont {Lee}}, \bibinfo {author}
  {\bibfnamefont {J.}~\bibnamefont {Viefhaus}}, \bibinfo {author}
  {\bibfnamefont {J.~B.}\ \bibnamefont {Goedkoop}}, \bibinfo {author}
  {\bibfnamefont {A.~A.}\ \bibnamefont {Nugroho}}, \bibinfo {author}
  {\bibfnamefont {M.}~\bibnamefont {Braden}}, \bibinfo {author} {\bibfnamefont
  {S.}~\bibnamefont {Roy}}, \bibinfo {author} {\bibfnamefont {M.}~\bibnamefont
  {Sprung}},\ and\ \bibinfo {author} {\bibfnamefont {C.}~\bibnamefont
  {Sch\"u\ss{}ler-Langeheine}},\ }\bibfield  {title} {\bibinfo {title}
  {Measurement of spin dynamics in a layered nickelate using x-ray photon
  correlation spectroscopy: Evidence for intrinsic destabilization of
  incommensurate stripes at low temperatures},\ }\href
  {https://doi.org/10.1103/PhysRevLett.127.057001} {\bibfield  {journal}
  {\bibinfo  {journal} {Phys. Rev. Lett.}\ }\textbf {\bibinfo {volume} {127}},\
  \bibinfo {pages} {057001} (\bibinfo {year} {2021})}\BibitemShut {NoStop}%
\bibitem [{\citenamefont {Korshunov}\ \emph {et~al.}(2024)\citenamefont
  {Korshunov}, \citenamefont {Kar}, \citenamefont {Lim}, \citenamefont
  {Subires}, \citenamefont {Deng}, \citenamefont {Jiang}, \citenamefont {Hu},
  \citenamefont {Călugăru}, \citenamefont {Yi}, \citenamefont {Roychowdhury},
  \citenamefont {Shekhar}, \citenamefont {Garbarino}, \citenamefont {Törmä},
  \citenamefont {Felser}, \citenamefont {Bernevig},\ and\ \citenamefont
  {Blanco-Canosa}}]{korshunov2024}%
  \BibitemOpen
  \bibfield  {author} {\bibinfo {author} {\bibfnamefont {A.}~\bibnamefont
  {Korshunov}}, \bibinfo {author} {\bibfnamefont {A.}~\bibnamefont {Kar}},
  \bibinfo {author} {\bibfnamefont {C.~Y.}\ \bibnamefont {Lim}}, \bibinfo
  {author} {\bibfnamefont {D.}~\bibnamefont {Subires}}, \bibinfo {author}
  {\bibfnamefont {J.}~\bibnamefont {Deng}}, \bibinfo {author} {\bibfnamefont
  {Y.}~\bibnamefont {Jiang}}, \bibinfo {author} {\bibfnamefont
  {H.}~\bibnamefont {Hu}}, \bibinfo {author} {\bibfnamefont {D.}~\bibnamefont
  {Călugăru}}, \bibinfo {author} {\bibfnamefont {C.}~\bibnamefont {Yi}},
  \bibinfo {author} {\bibfnamefont {S.}~\bibnamefont {Roychowdhury}}, \bibinfo
  {author} {\bibfnamefont {C.}~\bibnamefont {Shekhar}}, \bibinfo {author}
  {\bibfnamefont {G.}~\bibnamefont {Garbarino}}, \bibinfo {author}
  {\bibfnamefont {P.}~\bibnamefont {Törmä}}, \bibinfo {author} {\bibfnamefont
  {C.}~\bibnamefont {Felser}}, \bibinfo {author} {\bibfnamefont {B.~A.}\
  \bibnamefont {Bernevig}},\ and\ \bibinfo {author} {\bibfnamefont
  {S.}~\bibnamefont {Blanco-Canosa}},\ }\bibfield  {title} {\bibinfo {title}
  {Pressure induced quasi-long-range $\sqrt{3} \times \sqrt{3}$ charge density
  wave and competing orders in the kagome metal fege},\ }\href
  {https://arxiv.org/abs/2409.04325} {\  (\bibinfo {year} {2024})},\ \Eprint
  {https://arxiv.org/abs/2409.04325} {arXiv:2409.04325 [cond-mat.str-el]}
  \BibitemShut {NoStop}%
\bibitem [{\citenamefont {G{\"u}rsoy}\ \emph {et~al.}(2025)\citenamefont
  {G{\"u}rsoy}, \citenamefont {Yay}, \citenamefont {Kisiel}, \citenamefont
  {Wojcik}, \citenamefont {Sheyfer}, \citenamefont {Last}, \citenamefont
  {Highland}, \citenamefont {Fisher}, \citenamefont {Hruszkewycz},\ and\
  \citenamefont {Islam}}]{gursoy2025dark}%
  \BibitemOpen
  \bibfield  {author} {\bibinfo {author} {\bibfnamefont {D.}~\bibnamefont
  {G{\"u}rsoy}}, \bibinfo {author} {\bibfnamefont {K.~A.}\ \bibnamefont {Yay}},
  \bibinfo {author} {\bibfnamefont {E.}~\bibnamefont {Kisiel}}, \bibinfo
  {author} {\bibfnamefont {M.}~\bibnamefont {Wojcik}}, \bibinfo {author}
  {\bibfnamefont {D.}~\bibnamefont {Sheyfer}}, \bibinfo {author} {\bibfnamefont
  {A.}~\bibnamefont {Last}}, \bibinfo {author} {\bibfnamefont {M.}~\bibnamefont
  {Highland}}, \bibinfo {author} {\bibfnamefont {I.~R.}\ \bibnamefont
  {Fisher}}, \bibinfo {author} {\bibfnamefont {S.}~\bibnamefont
  {Hruszkewycz}},\ and\ \bibinfo {author} {\bibfnamefont {Z.}~\bibnamefont
  {Islam}},\ }\bibfield  {title} {\bibinfo {title} {Dark-field x-ray microscopy
  with structured illumination for three-dimensional imaging},\ }\href@noop {}
  {\bibfield  {journal} {\bibinfo  {journal} {Communications Physics}\ }\textbf
  {\bibinfo {volume} {8}},\ \bibinfo {pages} {34} (\bibinfo {year}
  {2025})}\BibitemShut {NoStop}%
\bibitem [{\citenamefont {Hohenberg}\ and\ \citenamefont {Kohn}(1964)}]{dft}%
  \BibitemOpen
  \bibfield  {author} {\bibinfo {author} {\bibfnamefont {P.}~\bibnamefont
  {Hohenberg}}\ and\ \bibinfo {author} {\bibfnamefont {W.}~\bibnamefont
  {Kohn}},\ }\bibfield  {title} {\bibinfo {title} {{Inhomogeneous Electron
  Gas}},\ }\href {https://doi.org/10.1103/PhysRev.136.B864} {\bibfield
  {journal} {\bibinfo  {journal} {Phys. Rev.}\ }\textbf {\bibinfo {volume}
  {136}},\ \bibinfo {pages} {B864} (\bibinfo {year} {1964})}\BibitemShut
  {NoStop}%
\bibitem [{\citenamefont {Schwarz}\ and\ \citenamefont {Blaha}(2003)}]{wien2k}%
  \BibitemOpen
  \bibfield  {author} {\bibinfo {author} {\bibfnamefont {K.}~\bibnamefont
  {Schwarz}}\ and\ \bibinfo {author} {\bibfnamefont {P.}~\bibnamefont
  {Blaha}},\ }\bibfield  {title} {\bibinfo {title} {Solid state calculations
  using {W}{I}{E}{N}2k},\ }\href@noop {} {\bibfield  {journal} {\bibinfo
  {journal} {Comp. Mater. Sci.}\ }\textbf {\bibinfo {volume} {28}},\ \bibinfo
  {pages} {259} (\bibinfo {year} {2003})}\BibitemShut {NoStop}%
\bibitem [{\citenamefont {Blaha}\ \emph {et~al.}(2020)\citenamefont {Blaha},
  \citenamefont {Schwarz}, \citenamefont {Tran}, \citenamefont {Laskowski},
  \citenamefont {Madsen},\ and\ \citenamefont {Marks}}]{Blaha2020wien2k}%
  \BibitemOpen
  \bibfield  {author} {\bibinfo {author} {\bibfnamefont {P.}~\bibnamefont
  {Blaha}}, \bibinfo {author} {\bibfnamefont {K.}~\bibnamefont {Schwarz}},
  \bibinfo {author} {\bibfnamefont {F.}~\bibnamefont {Tran}}, \bibinfo {author}
  {\bibfnamefont {R.}~\bibnamefont {Laskowski}}, \bibinfo {author}
  {\bibfnamefont {G.~K.~H.}\ \bibnamefont {Madsen}},\ and\ \bibinfo {author}
  {\bibfnamefont {L.~D.}\ \bibnamefont {Marks}},\ }\bibfield  {title} {\bibinfo
  {title} {Wien2k: An apw+lo program for calculating the properties of
  solids},\ }\href {https://doi.org/10.1063/1.5143061} {\bibfield  {journal}
  {\bibinfo  {journal} {The Journal of Chemical Physics}\ }\textbf {\bibinfo
  {volume} {152}},\ \bibinfo {pages} {074101} (\bibinfo {year}
  {2020})}\BibitemShut {NoStop}%
\bibitem [{\citenamefont {Togo}\ and\ \citenamefont {Tanaka}(2015)}]{phonopy}%
  \BibitemOpen
  \bibfield  {author} {\bibinfo {author} {\bibfnamefont {A.}~\bibnamefont
  {Togo}}\ and\ \bibinfo {author} {\bibfnamefont {I.}~\bibnamefont {Tanaka}},\
  }\bibfield  {title} {\bibinfo {title} {First principles phonon calculations
  in materials science},\ }\href@noop {} {\bibfield  {journal} {\bibinfo
  {journal} {Scr. Mater.}\ }\textbf {\bibinfo {volume} {108}},\ \bibinfo
  {pages} {1} (\bibinfo {year} {2015})}\BibitemShut {NoStop}%
\bibitem [{\citenamefont {Togo}(2023)}]{phonopy-phono3py-JPSJ}%
  \BibitemOpen
  \bibfield  {author} {\bibinfo {author} {\bibfnamefont {A.}~\bibnamefont
  {Togo}},\ }\bibfield  {title} {\bibinfo {title} {First-principles phonon
  calculations with phonopy and phono3py},\ }\href
  {https://doi.org/10.7566/JPSJ.92.012001} {\bibfield  {journal} {\bibinfo
  {journal} {J. Phys. Soc. Jpn.}\ }\textbf {\bibinfo {volume} {92}},\ \bibinfo
  {pages} {012001} (\bibinfo {year} {2023})}\BibitemShut {NoStop}%
\bibitem [{\citenamefont {Kresse}\ and\ \citenamefont
  {Hafner}(1993)}]{kresse1993ab}%
  \BibitemOpen
  \bibfield  {author} {\bibinfo {author} {\bibfnamefont {G.}~\bibnamefont
  {Kresse}}\ and\ \bibinfo {author} {\bibfnamefont {J.}~\bibnamefont
  {Hafner}},\ }\bibfield  {title} {\bibinfo {title} {Ab initio molecular
  dynamics for liquid metals},\ }\href@noop {} {\bibfield  {journal} {\bibinfo
  {journal} {Physical review B}\ }\textbf {\bibinfo {volume} {47}},\ \bibinfo
  {pages} {558} (\bibinfo {year} {1993})}\BibitemShut {NoStop}%
\bibitem [{\citenamefont {Kresse}\ and\ \citenamefont
  {Furthm{\"u}ller}(1996{\natexlab{a}})}]{kresse1996efficiency}%
  \BibitemOpen
  \bibfield  {author} {\bibinfo {author} {\bibfnamefont {G.}~\bibnamefont
  {Kresse}}\ and\ \bibinfo {author} {\bibfnamefont {J.}~\bibnamefont
  {Furthm{\"u}ller}},\ }\bibfield  {title} {\bibinfo {title} {Efficiency of
  ab-initio total energy calculations for metals and semiconductors using a
  plane-wave basis set},\ }\href@noop {} {\bibfield  {journal} {\bibinfo
  {journal} {Computational materials science}\ }\textbf {\bibinfo {volume}
  {6}},\ \bibinfo {pages} {15} (\bibinfo {year}
  {1996}{\natexlab{a}})}\BibitemShut {NoStop}%
\bibitem [{\citenamefont {Kresse}\ and\ \citenamefont
  {Furthm{\"u}ller}(1996{\natexlab{b}})}]{kresse1996efficient}%
  \BibitemOpen
  \bibfield  {author} {\bibinfo {author} {\bibfnamefont {G.}~\bibnamefont
  {Kresse}}\ and\ \bibinfo {author} {\bibfnamefont {J.}~\bibnamefont
  {Furthm{\"u}ller}},\ }\bibfield  {title} {\bibinfo {title} {Efficient
  iterative schemes for ab initio total-energy calculations using a plane-wave
  basis set},\ }\href@noop {} {\bibfield  {journal} {\bibinfo  {journal}
  {Physical review B}\ }\textbf {\bibinfo {volume} {54}},\ \bibinfo {pages}
  {11169} (\bibinfo {year} {1996}{\natexlab{b}})}\BibitemShut {NoStop}%
\bibitem [{\citenamefont {Monkhorst}\ and\ \citenamefont
  {Pack}(1976)}]{monkhorst_pack}%
  \BibitemOpen
  \bibfield  {author} {\bibinfo {author} {\bibfnamefont {H.~J.}\ \bibnamefont
  {Monkhorst}}\ and\ \bibinfo {author} {\bibfnamefont {J.~D.}\ \bibnamefont
  {Pack}},\ }\bibfield  {title} {\bibinfo {title} {Special points for
  brillouin-zone integrations},\ }\href
  {https://doi.org/10.1103/PhysRevB.13.5188} {\bibfield  {journal} {\bibinfo
  {journal} {Phys. Rev. B}\ }\textbf {\bibinfo {volume} {13}},\ \bibinfo
  {pages} {5188} (\bibinfo {year} {1976})}\BibitemShut {NoStop}%
\end{thebibliography}%


\begin{thebibliography}{2}%
\makeatletter
\providecommand \@ifxundefined [1]{%
 \@ifx{#1\undefined}
}%
\providecommand \@ifnum [1]{%
 \ifnum #1\expandafter \@firstoftwo
 \else \expandafter \@secondoftwo
 \fi
}%
\providecommand \@ifx [1]{%
 \ifx #1\expandafter \@firstoftwo
 \else \expandafter \@secondoftwo
 \fi
}%
\providecommand \natexlab [1]{#1}%
\providecommand \enquote  [1]{``#1''}%
\providecommand \bibnamefont  [1]{#1}%
\providecommand \bibfnamefont [1]{#1}%
\providecommand \citenamefont [1]{#1}%
\providecommand \href@noop [0]{\@secondoftwo}%
\providecommand \href [0]{\begingroup \@sanitize@url \@href}%
\providecommand \@href[1]{\@@startlink{#1}\@@href}%
\providecommand \@@href[1]{\endgroup#1\@@endlink}%
\providecommand \@sanitize@url [0]{\catcode `\\12\catcode `\$12\catcode
  `\&12\catcode `\#12\catcode `\^12\catcode `\_12\catcode `\%12\relax}%
\providecommand \@@startlink[1]{}%
\providecommand \@@endlink[0]{}%
\providecommand \url  [0]{\begingroup\@sanitize@url \@url }%
\providecommand \@url [1]{\endgroup\@href {#1}{\urlprefix }}%
\providecommand \urlprefix  [0]{URL }%
\providecommand \Eprint [0]{\href }%
\providecommand \doibase [0]{https://doi.org/}%
\providecommand \selectlanguage [0]{\@gobble}%
\providecommand \bibinfo  [0]{\@secondoftwo}%
\providecommand \bibfield  [0]{\@secondoftwo}%
\providecommand \translation [1]{[#1]}%
\providecommand \BibitemOpen [0]{}%
\providecommand \bibitemStop [0]{}%
\providecommand \bibitemNoStop [0]{.\EOS\space}%
\providecommand \EOS [0]{\spacefactor3000\relax}%
\providecommand \BibitemShut  [1]{\csname bibitem#1\endcsname}%
\let\auto@bib@innerbib\@empty
%</preamble>
\bibitem [{\citenamefont {Wu}\ \emph {et~al.}(2022)\citenamefont {Wu},
  \citenamefont {Xu}, \citenamefont {Haley}, \citenamefont {Weber},
  \citenamefont {Acharya}, \citenamefont {Maniv}, \citenamefont {Qiu},
  \citenamefont {Aczel}, \citenamefont {Settineri}, \citenamefont {Neaton}
  \emph {et~al.}}]{wu2022highly}%
  \BibitemOpen
  \bibfield  {author} {\bibinfo {author} {\bibfnamefont {S.}~\bibnamefont
  {Wu}}, \bibinfo {author} {\bibfnamefont {Z.}~\bibnamefont {Xu}}, \bibinfo
  {author} {\bibfnamefont {S.~C.}\ \bibnamefont {Haley}}, \bibinfo {author}
  {\bibfnamefont {S.~F.}\ \bibnamefont {Weber}}, \bibinfo {author}
  {\bibfnamefont {A.}~\bibnamefont {Acharya}}, \bibinfo {author} {\bibfnamefont
  {E.}~\bibnamefont {Maniv}}, \bibinfo {author} {\bibfnamefont
  {Y.}~\bibnamefont {Qiu}}, \bibinfo {author} {\bibfnamefont {A.}~\bibnamefont
  {Aczel}}, \bibinfo {author} {\bibfnamefont {N.~S.}\ \bibnamefont
  {Settineri}}, \bibinfo {author} {\bibfnamefont {J.~B.}\ \bibnamefont
  {Neaton}}, \emph {et~al.},\ }\bibfield  {title} {\bibinfo {title} {Highly
  tunable magnetic phases in transition-metal dichalcogenide fe 1/3+ $\delta$
  nbs 2},\ }\href@noop {} {\bibfield  {journal} {\bibinfo  {journal} {Physical
  Review X}\ }\textbf {\bibinfo {volume} {12}},\ \bibinfo {pages} {021003}
  (\bibinfo {year} {2022})}\BibitemShut {NoStop}%
\bibitem [{\citenamefont {Wu}\ \emph {et~al.}(2023)\citenamefont {Wu},
  \citenamefont {Basak}, \citenamefont {Li}, \citenamefont {Kim}, \citenamefont
  {Ryan}, \citenamefont {Lu}, \citenamefont {Hashimoto}, \citenamefont
  {Nelson}, \citenamefont {Acevedo-Esteves}, \citenamefont {Haley} \emph
  {et~al.}}]{wu2023discovery}%
  \BibitemOpen
  \bibfield  {author} {\bibinfo {author} {\bibfnamefont {S.}~\bibnamefont
  {Wu}}, \bibinfo {author} {\bibfnamefont {R.}~\bibnamefont {Basak}}, \bibinfo
  {author} {\bibfnamefont {W.}~\bibnamefont {Li}}, \bibinfo {author}
  {\bibfnamefont {J.-W.}\ \bibnamefont {Kim}}, \bibinfo {author} {\bibfnamefont
  {P.~J.}\ \bibnamefont {Ryan}}, \bibinfo {author} {\bibfnamefont
  {D.}~\bibnamefont {Lu}}, \bibinfo {author} {\bibfnamefont {M.}~\bibnamefont
  {Hashimoto}}, \bibinfo {author} {\bibfnamefont {C.}~\bibnamefont {Nelson}},
  \bibinfo {author} {\bibfnamefont {R.}~\bibnamefont {Acevedo-Esteves}},
  \bibinfo {author} {\bibfnamefont {S.~C.}\ \bibnamefont {Haley}}, \emph
  {et~al.},\ }\bibfield  {title} {\bibinfo {title} {Discovery of charge order
  in the transition metal dichalcogenide fe x nbs 2},\ }\href@noop {}
  {\bibfield  {journal} {\bibinfo  {journal} {Physical Review Letters}\
  }\textbf {\bibinfo {volume} {131}},\ \bibinfo {pages} {186701} (\bibinfo
  {year} {2023})}\BibitemShut {NoStop}%
\end{thebibliography}%
\end{document}

\title{Supplementary Material for `Magnetoelastic coupling in intercalated transition metal dichalcogenides'}

\renewcommand{\thefigure}{S\arabic{figure}}

\renewcommand{\thetable}{S\arabic{table}}

\renewcommand{\thesection}{S\arabic{section}}

\renewcommand{\theequation}{S\arabic{equation}}

\author{A. Kar}
\thanks{These authors contributed equally to this work.}
%\email{phys.arunava20@gmail.com}
\affiliation{Donostia International Physics Center (DIPC), San Sebastián, Spain}

\author{R. Basak}
\thanks{These authors contributed equally to this work.}
\affiliation{Department of Physics, University of California San Diego, San Diego, California 92093, USA}

\author{Xue Li}
\thanks{These authors contributed equally to this work.}
\affiliation{Beijing National Laboratory for Condensed Matter Physics and Institute of Physics, Chinese Academy of Sciences, Beijing 100190, China}
\affiliation{University of Chinese Academy of Sciences, Beijing 100049, China}

\author{A. Korshunov}
\thanks{These authors contributed equally to this work.}
\affiliation{Donostia International Physics Center (DIPC), San Sebastián, Spain}

\author{D. Subires}
\affiliation{Donostia International Physics Center (DIPC), San Sebastián, Spain}
\affiliation{Departamento de Física Aplicada I, Universidad del País Vasco UPV/EHU, E-20018 San Sebastián, Spain}

\author{J. Phillips}
\affiliation{Departamento de Física Aplicada, Universidade de Santiago de Compostela, E-15782 Campus Sur s/n, Santiago de Compostela, Spain}
\affiliation{Instituto de Materiais iMATUS, Universidade de Santiago de Compostela, E-15782 Campus Sur s/n, Santiago de Compostela, Spain} 

\author{C.-Y. Lim}
\affiliation{Donostia International Physics Center (DIPC), San Sebastián, Spain}

\author{Feng Zhou}
\affiliation{Beijing National Laboratory for Condensed Matter Physics and Institute of Physics, Chinese Academy of Sciences, Beijing 100190, China}
\affiliation{School of Electronics and Information Engineering, Tiangong University, Tianjin 300387, China}

\author{Linxuan Song}
\affiliation{Beijing National Laboratory for Condensed Matter Physics and Institute of Physics, Chinese Academy of Sciences, Beijing 100190, China}
\affiliation{School of Electronics and Information Engineering, Tiangong University, Tianjin 300387, China}
\author{Wenhong Wang}
\affiliation{School of Electronics and Information Engineering, Tiangong University, Tianjin 300387, China}
\author{Yong-Chang Lau}
\affiliation{Beijing National Laboratory for Condensed Matter Physics and Institute of Physics, Chinese Academy of Sciences, Beijing 100190, China}
\affiliation{University of Chinese Academy of Sciences, Beijing 100049, China}

\author{G. Garbarino}
\affiliation{European Synchrotron Radiation Facility (ESRF), BP 220, F-38043 Grenoble Cedex 9, France}

\author{P. Gargiani}
\affiliation{ALBA Synchrotron Light Source, 08290 Barcelona, Spain}

\author{Y. Zhao}
\affiliation{Deutsches Elektronen-Synchrotron DESY, Notkestr. 85, 22607, Hamburg, Germany}

\author{C. Plueckthun}
\affiliation{Deutsches Elektronen-Synchrotron DESY, Notkestr. 85, 22607, Hamburg, Germany}

\author{S. Francoual}
\affiliation{Deutsches Elektronen-Synchrotron DESY, Notkestr. 85, 22607, Hamburg, Germany}

\author{A. Jana}
\affiliation{CNR-Istituto Officina
dei Materiali (CNR-IOM), Strada Statale 14, km 163.5, 34149 Trieste,
Italy}
\affiliation{International Center for Theoretical Physics (ICTP), 34151 Trieste,
Italy}

\author{I. Vobornik}
\affiliation{CNR-Istituto Officina
dei Materiali (CNR-IOM), Strada Statale 14, km 163.5, 34149 Trieste,
Italy}

\author{T. Valla}
\affiliation{Donostia International Physics Center (DIPC), San Sebastián, Spain}

\author{A. Rajapitamahuni}
\affiliation{National Synchrotron Light Source II, Brookhaven National Laboratory, Upton, New York 11973, USA}

\author{James G. Analytis}
\affiliation{Physics Department, University of California, Berkeley, California 94720, USA}
\affiliation{CIFAR Quantum Materials, CIFAR, Toronto, Ontario M5G 1M1, Canada}

\author{Robert J. Birgeneau}
\affiliation{Materials Science Division, Lawrence Berkeley National Lab, Berkeley, California 94720, USA}
\affiliation{Physics Department, University of California, Berkeley, California 94720, USA}
\affiliation{Department of Materials Science and Engineering, University of California, Berkeley, California 94720, USA}

\author{E. Vescovo}
\affiliation{National Synchrotron Light Source II, Brookhaven National Laboratory, Upton, New York 11973, USA}

\author{A. Bosak}
\affiliation{European Synchrotron Radiation Facility (ESRF), BP 220, F-38043 Grenoble Cedex 9, France}

\author{J. Dai}
\affiliation{ALBA Synchrotron Light Source, 08290 Barcelona, Spain}

\author{M. Tallarida}
\affiliation{ALBA Synchrotron Light Source, 08290 Barcelona, Spain}

\author{A. Frano}
\affiliation{Department of Physics, University of California San Diego, San Diego, California 92093, USA}

\author{V. Pardo}
\affiliation{Departamento de Física Aplicada, Universidade de Santiago de Compostela, E-15782 Campus Sur s/n, Santiago de Compostela, Spain}
\affiliation{Instituto de Materiais iMATUS, Universidade de Santiago de Compostela, E-15782 Campus Sur s/n, Santiago de Compostela, Spain} 

\author{S. Wu}
\affiliation{Department of Physics, Santa Clara University}

\author{S. Blanco-Canosa}
\email{sblanco@dipc.org}
\affiliation{Donostia International Physics Center (DIPC), San Sebastián, Spain}
\affiliation{IKERBASQUE, Basque Foundation for Science, 48013 Bilbao, Spain}

\maketitle

\section{Sample growth}
High-quality single crystals were synthesized using a chemical vapor transport method with polycrystalline precursors of Co(Fe), Nb(Ta), and S elements in the ratio of 0.33 : 1 : 2. The reagents were placed in an evacuated sealed silica tube using iodine as transport. The values of the resulting intercalation ratio x of individual single crystals were determined by energy dispersive x-ray spectroscopy (EDX) and correspond approximately to 1/3 concentration of the transition metal.
 
\section{Resistivity and magnetization measurement}
\begin{figure}
\begin{center}
\includegraphics[width=0.9\textwidth,draft=false]{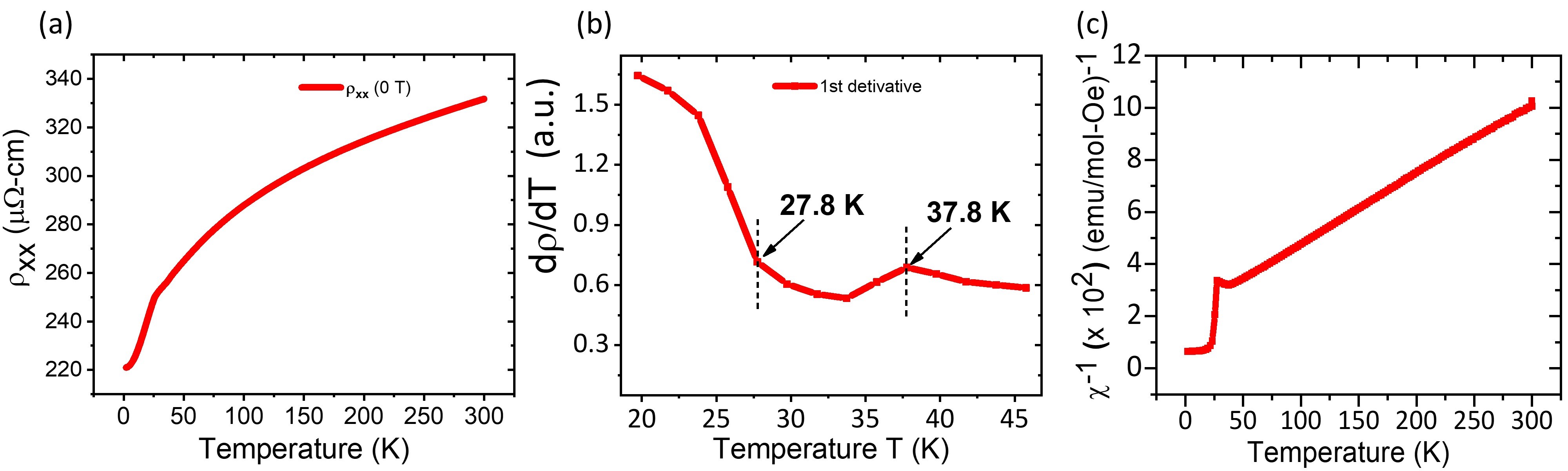}
\caption{(a) Temperature dependent Resistivity data of Co$_{1/3}$TaS$_2$. (b) First derivative representation of the resistivity data. (c) Inverse susceptibility plotting of Co$_{1/3}$TaS$_2$.}
\label{resistivity}
\end{center}
\end{figure} 
Consistent with the magnetization measurements shown in Fig. 1(e) of the main text, the temperature-dependent resistivity data (Fig. \ref{resistivity}) also reveal two magnetic phase transitions at T$_{N2}$=27.8 K and T$_{N1}$=37.8 K, with the major transition occurring below T$_{N2}$. A linear fit of the inverse susceptibility data (Fig. \ref{resistivity}(c)) yields a negative Curie-Weiss temperature ($\theta_{CW}$) of 75 K.
\begin{figure}
\begin{center}
\includegraphics[width=0.9\textwidth,draft=false]{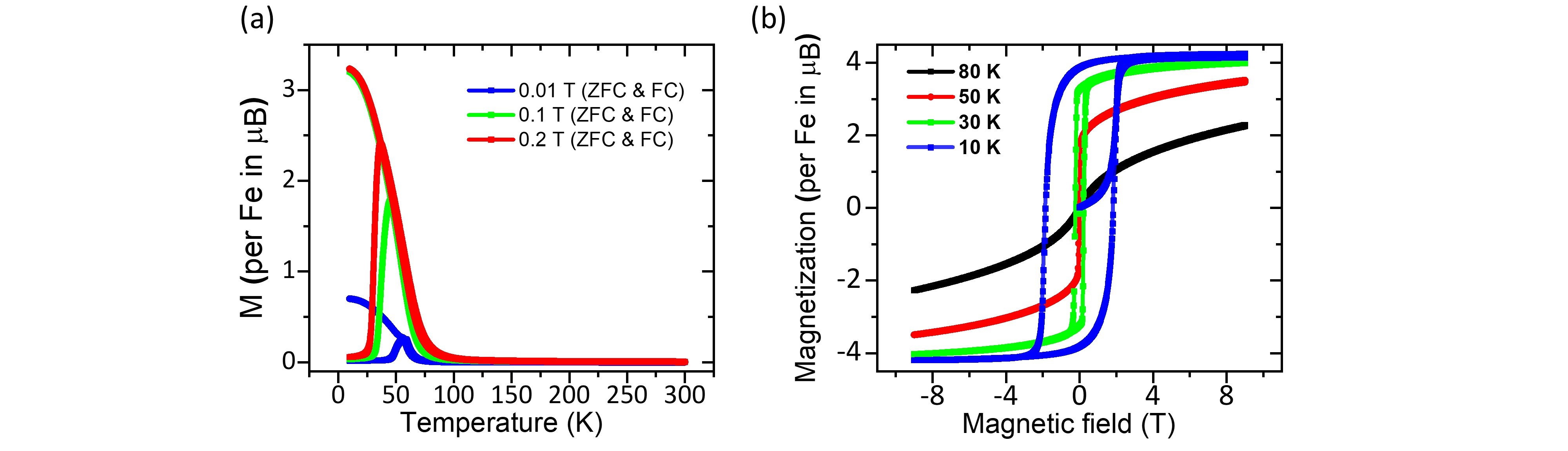}
\caption{Magnetization measurement spectra of Fe$_{1/3}$TaS$_2$. (a) Temperature-dependent DC Magnetic moment for different applied magnetic fields. (measurements were performed in both ZFC and FC conditions) (b) Magnetic field-dependent magnetic moment with different temperatures, above and below the magnetic transition temperature.}
\label{magnetization}
\end{center}
\end{figure}

Figure \ref{magnetization}(a) represents the temperature-dependent DC magnetization data per Fe ion measured parallel to the c-axis, magnetization easy axis of this sample, under applied fields of $\mu_0$H = 0.01, 0.1, and 0.2 T.  Around the Curie temperature, Tc = 58 K, the magnetization shows a significant increase at low field ($\mu_0$H = 0.01 T). In the applied magnetic field region, there is a clear separation between the zero-field-cooled (ZFC) and field-cooled (FC) branches. Irreversible magnetization processes that are typical of ferromagnetic activity are indicated by the observed bifurcation at lower fields. 

The field-dependent magnetization, shown in  Fig. \ref{magnetization}(b), at T = 10 K displays a nearly square-shaped hysteresis loop, with a sharp switching at the coercive field of $\mu_0$H = 2 T and a saturation magnetization of 4.2 $\mu_B$, which closely matches the expected moment for Fe$^{2+}$ ions in the high-spin state. As the temperature increases, the coercivity gradually decreases. By 50 K, the hysteresis loop disappears, although the magnetization reversal process remains sharp. Above the Curie temperature (T$_c$), the magnetic susceptibility data exhibit a characteristic curvature, indicative of field-induced polarization and the presence of short-range magnetic interactions within the system. 

\begin{figure}
\begin{center}
\includegraphics[width=0.9\textwidth,draft=false]{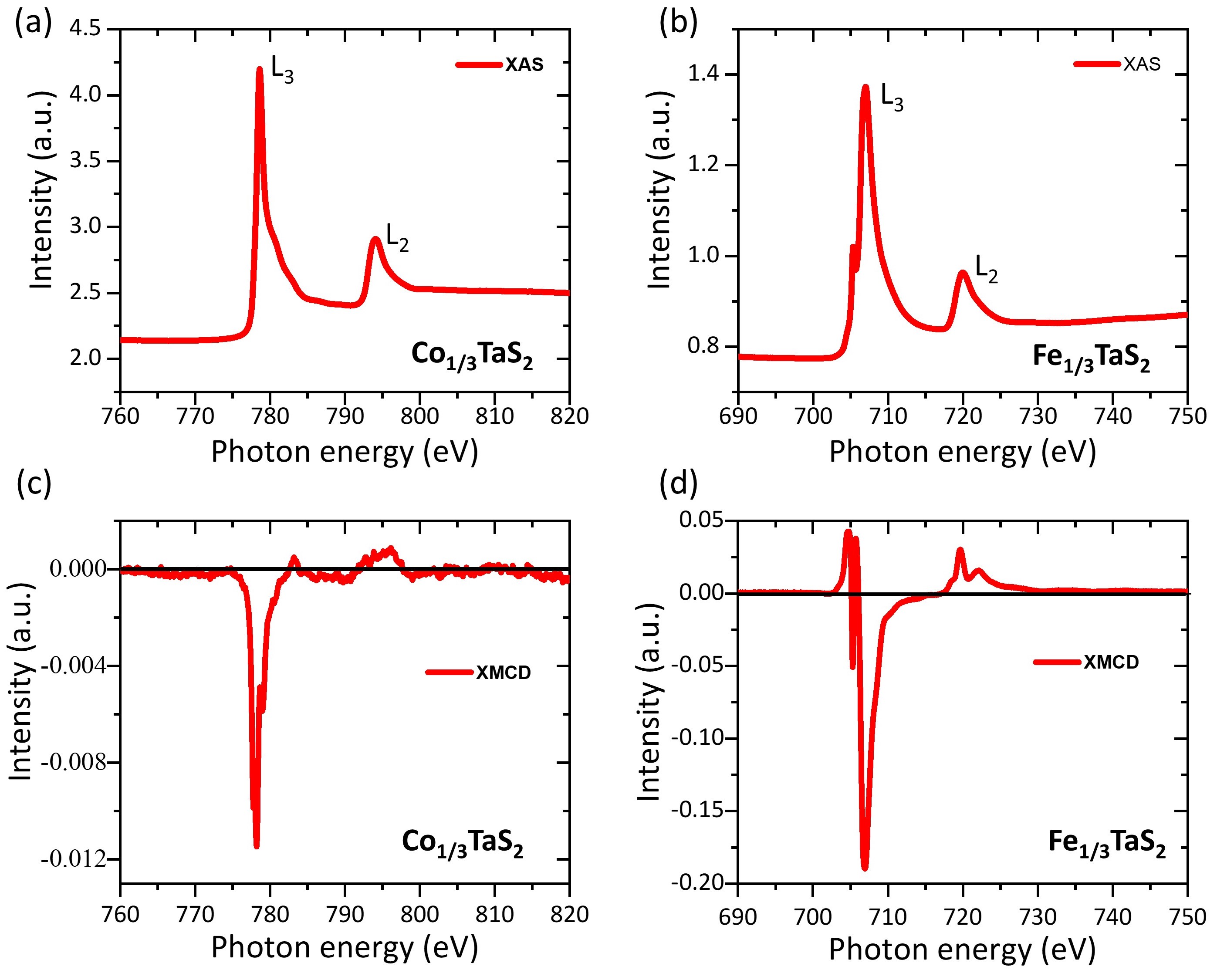}
\caption{Cobalt and Iron L-edge x-ray absorption spectra (XAS) and x-ray magnetic circular dichroism (XMCD) (2K temperature and 0T magnetic field) of (a)and (c) Co$_{1/3}$TaS$_2$;  (b) and (d) for Fe$_{1/3}$TaS$_2$ systems, respectively.}
\label{XMCD}
\end{center}
\end{figure} 

\begin{figure}
\begin{center}
\includegraphics[width=0.8\textwidth,draft=false]{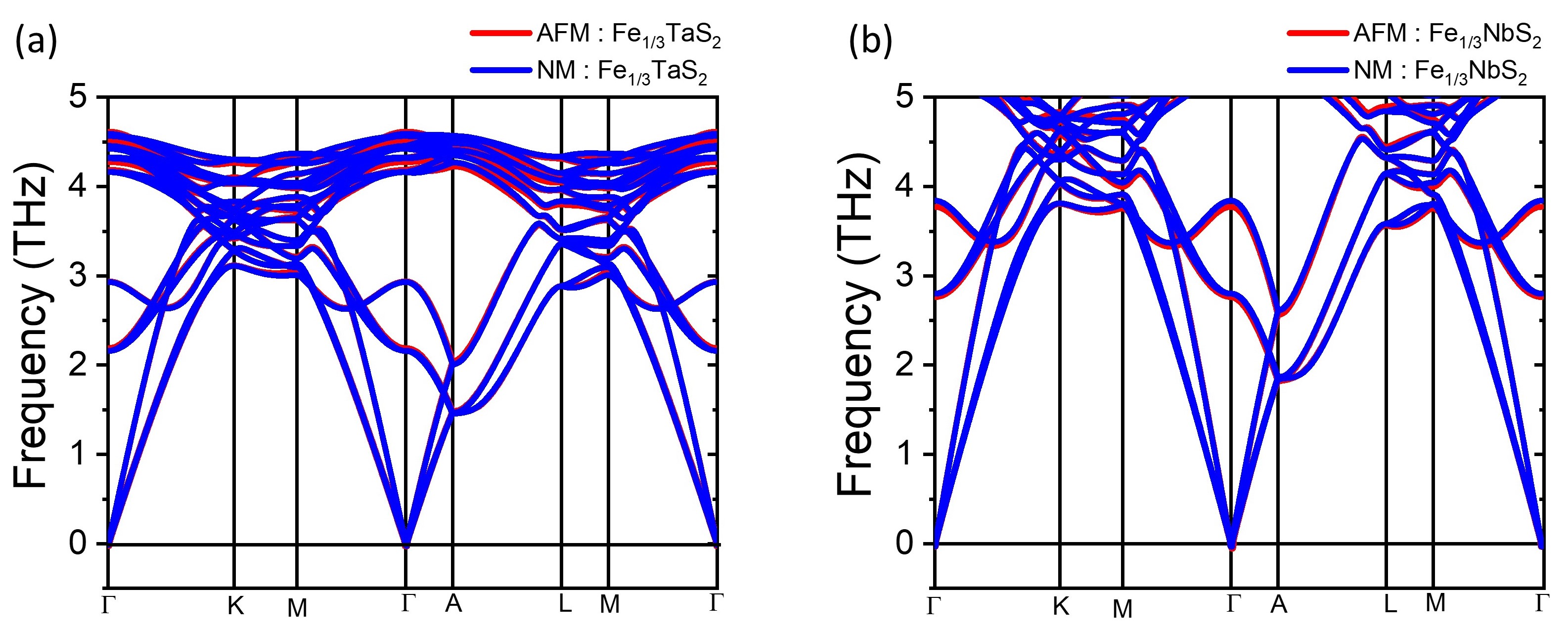}
\caption{DFT-calculated phonon dispersion spectra for (a) Fe$_{1/3}$TaS$_{2}$ and (b)Fe$_{1/3}$NbS$_{2}$. The spectra compare the phonon behavior of the magnetic and non-magnetic solutions for each system.}
\label{spin-phonon}
\end{center}
\end{figure}

\section{ARPES}

\begin{figure}
\begin{center}
\includegraphics[width=0.8\textwidth,draft=false]{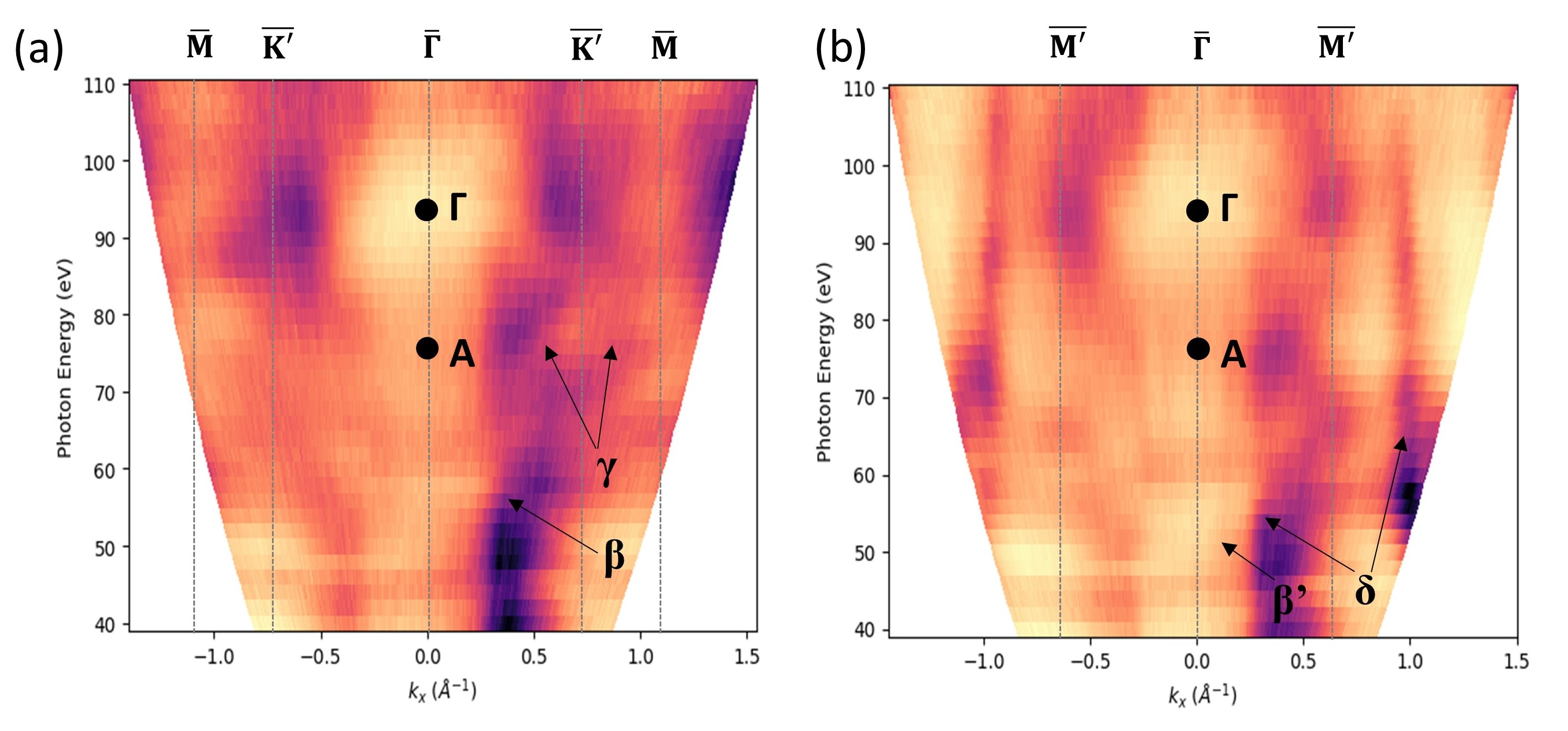}
\caption{Photon energy dependent behaviour of Co$_{1/3}$TaS$_2$  system (a) along $\bar{\Gamma}-\bar{K'}$ (b) along $\bar{\Gamma}-\bar{M'}$ high symmetry directions.}
\label{photon energy}
\end{center}
\end{figure}

To disclose the three-dimensional nature of the carrier pockets in Co$_{1/3}$TaS$_{2}$, photon energy-dependent photoemission intensity is measured. The band dispersion along $\bar{\Gamma}-\bar{\mathrm{K}'}$ and $\bar{\Gamma}-\bar{\mathrm{M}'}$ high symmetry directions with horizontal light polarization (h$\nu$ =40-120 eV) are represented in Fig \ref{photon energy}. The Fermi crossing of the $\beta$ band shows negligible k$_z$ dispersion, consistent with quasi 2D nature of the bulk 2H-TaS$_2$ system. The signal related to the $\beta'$ band is weak due to very low intensity near the E$_F$. Interestingly, the  $\gamma$ band shows a periodic variation with photon energy. The shallowness of this electron pocket becomes the maximum at k$_z$= $\pi$ plane and it becomes nearly flat at k$_z$= 0 plane. The periodic behavior of the electronic structure is also observed along k$_z$ and ${\bar{\Gamma}-\bar{M'}}$ direction. The corner touching point of these triangular electron pockets at $\mathrm{M}'$ becomes prominent when the shallowness is minimal.

\begin{figure}
\begin{center}
\includegraphics[width=1.0\textwidth,draft=false]{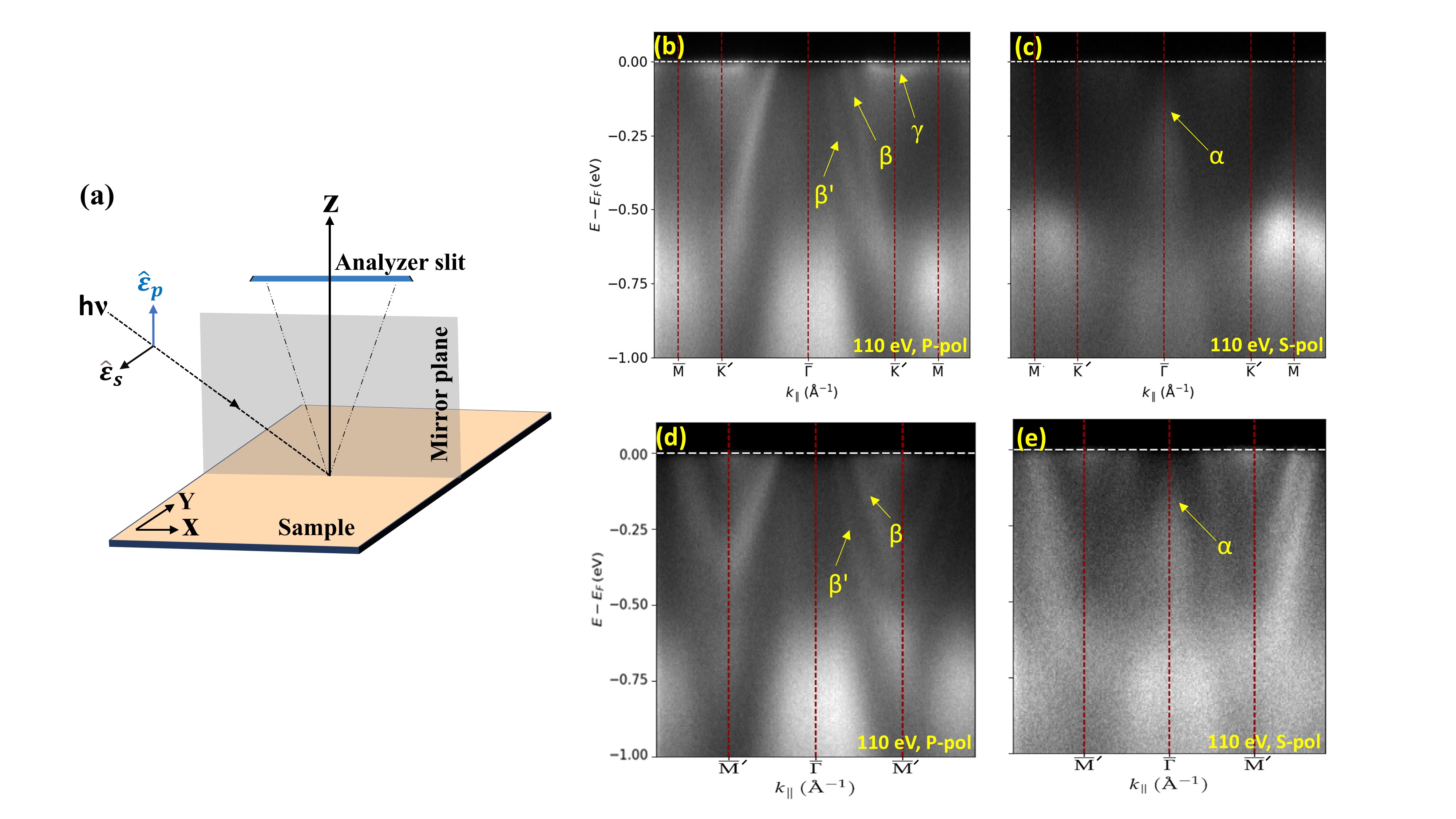}
\caption{Polarization dependent behavior of Co$_{1/3}$TaS$_2$ system (a) Schematic diagram of experimental geometry for ARPES experiment. (b) and (c) VB spectra obtained along $\bar{\Gamma}-\bar{K'}$ symmetry direction. (d) and (e) VB spectra obtained along $\bar{\Gamma}-\bar{M}'$ symmetry direction.  }
\label{polarization dependence}
\end{center}
\end{figure}

Furthermore, to resolve the possible multi-orbital nature of Co$_{1/3}$TaS$_{2}$ band structure, linear polarization-dependent ARPES measurements were performed as shown in Fig. \ref{polarization dependence}. Switching the light polarization from linear horizontal ($p$-pol) to linear vertical direction ($s$-pol), a strong intensity variation in the band structure is observed. The experimental geometry for polarization-dependent ARPES measurements is shown in Fig. \ref{polarization dependence}(a). However, the parent TMCD compound, 2H-TaS$_2$, space group P6$_3/$mmc, (194) contains in-plane and perpendicular mirror planes, whereas the 1/3 TM intercalated TMDC systems (P6$_3$22, \#182) do not exhibit any of the mirror-symmetric axes. Even in the absence of a symmetry plane, when Co$_{1/3}$TaS$_{2}$ compound is aligned along  $\Gamma-K'$ direction, the relative intensities of the ARPES spectra ($\alpha, \beta, \beta'$ and $\gamma$ bands) change upon switching the linear polarization from $p$-type type to $s$-type, but the BE of the photoemitted states remains constant. Because of the spatial symmetry of the $d$ orbitals, the photoemission intensity of a particular even (or odd) symmetry band can only be detected using $p$ (or $s$) polarised light, respectively. The $\alpha$ band at $\bar{\Gamma}$ point is visible only for the $s$-polarized light exhibiting its odd nature and can be attributed to the $d_{xy}$ and/or $d_{yz}$ orbital. Around the $\bar{K'}$ point, the electron-like band ${\gamma}$ only shows up for the $p$-polarized light, reflecting its even character and associating it with even Co $3d$ orbitals. The $\beta$ band is prominent under $p$-polarized light along both the $\bar{\Gamma}-\bar{K'}$ and $\bar{\Gamma}-\bar{M'}$ symmetry directions suggesting its orbital character as  $d_{z^2}$, or $d_{x^2-y^2}$. In contrast, the hole-like  $\beta$' band is not purely polarized, as it is observed under both $p$ and $s$ polarized light across both symmetry directions, indicating a likely mixture of various Co $3d$ orbitals. These observations align with the calculated orbital-projected band structure for the Co$_{1/3}$TaS$_{2}$ system, where the intermixing of Co $3d$ and Ta $5d$ orbitals shapes the observed band evolution. The periodic electrostatic potential from the intercalated ions induces static deformation, influencing the host material's band structure. Additionally, the overlap between the electronic wavefunctions of the intercalated ions and the host material plays a significant role in further modifying the conduction bands and potentially transforming the quasi-2D band structure of TMDs into a more three-dimensional topology.

\section{Density functional Theory calculation}

\begin{figure}[h!]
\begin{center}
\includegraphics[width=0.8\textwidth,draft=false]{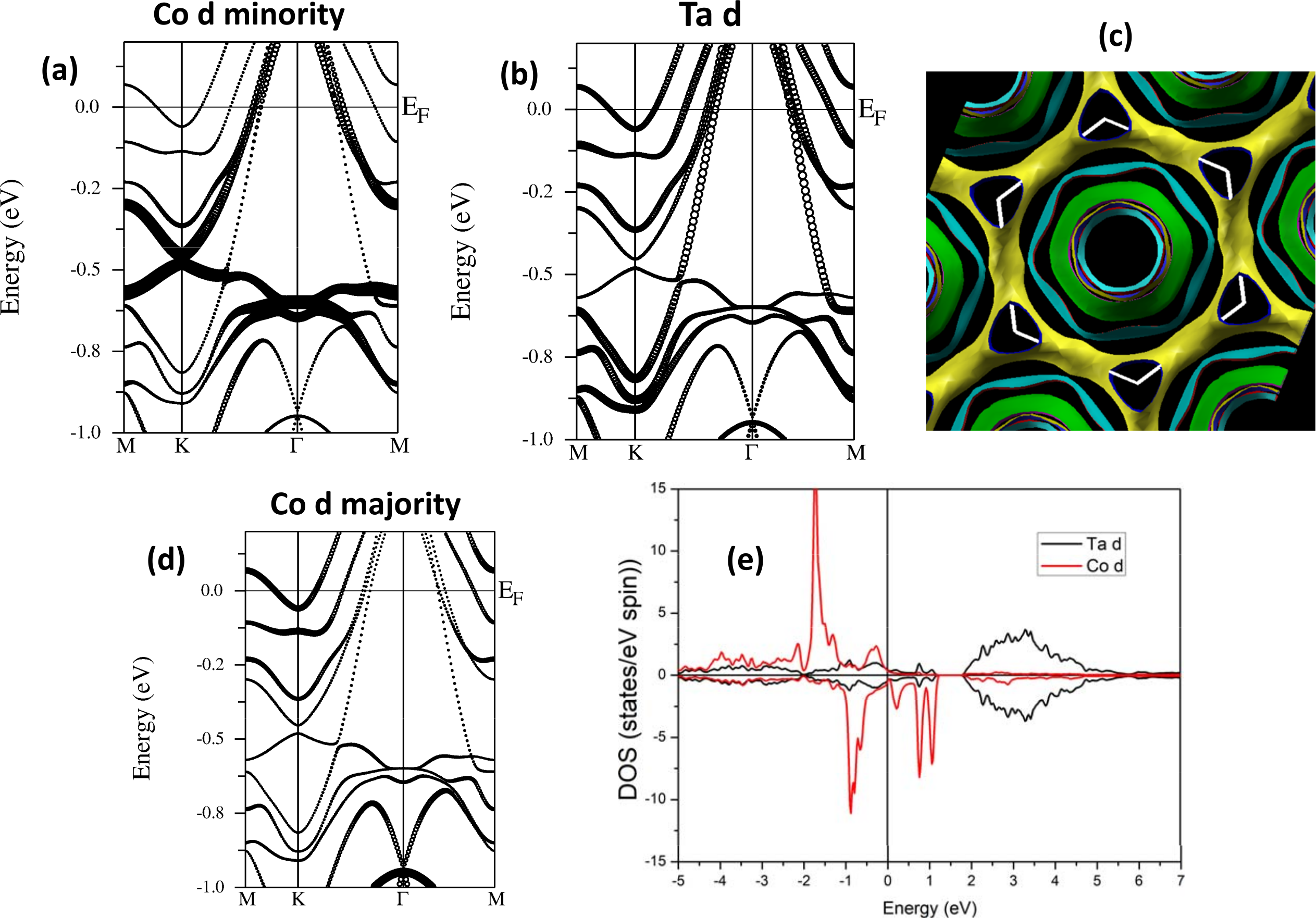}
\caption{DFT-calculated orbital and spin resolved dispersion spectra for Co$_{1/3}$TaS$_{2}$ (a) and (d) For Co $d$ minority and majority band structure, respectively. (b) Band structure obtained for Ta d bands. (c) Calculated Fermi surface projected along the c axis. (e) Atom resolved density of states (DOS) near Fermi level.}
\label{Co_DOS}
\end{center}
\end{figure}

Figure \ref{Co_DOS} shows a summary of the electronic structure of Co$_{1/3}$TaS$_2$. The GGA-PBE calculation shows the essential features found in the ARPES data. To match more closely the observed Fermi surfaces we lowered the chemical potential by 0.1 eV concerning the stoichiometric calculation. This creates an extra electron pocket at the $K$ point that would not be observed otherwise. The reason for this discrepancy could be small off-stoichiometries in either of the elements in the formula unit.
We can see that the Co $d$ states dominate the region close to the Fermi level. The Ta $d$ states are mostly located 2 eV above the Fermi level (see Fig. \ref{Co_DOS}d) but still substantial hybridizations make the Ta $d$ contribution close to the Fermi level significant. The Co $d$ states are full for the majority spin channel, being the partially-filled minority-spin states that dominate character at the Fermi level. The magnetic moment of the Co cations is close to 2 $\mu_B$ (1.8 $\mu_B$ for the GGA-PBE calculation), meaning an occupation close to the ionic d$^8$ configuration.
 
\begin{figure}[h!]
\begin{center}
\includegraphics[width=0.8\textwidth,draft=false]{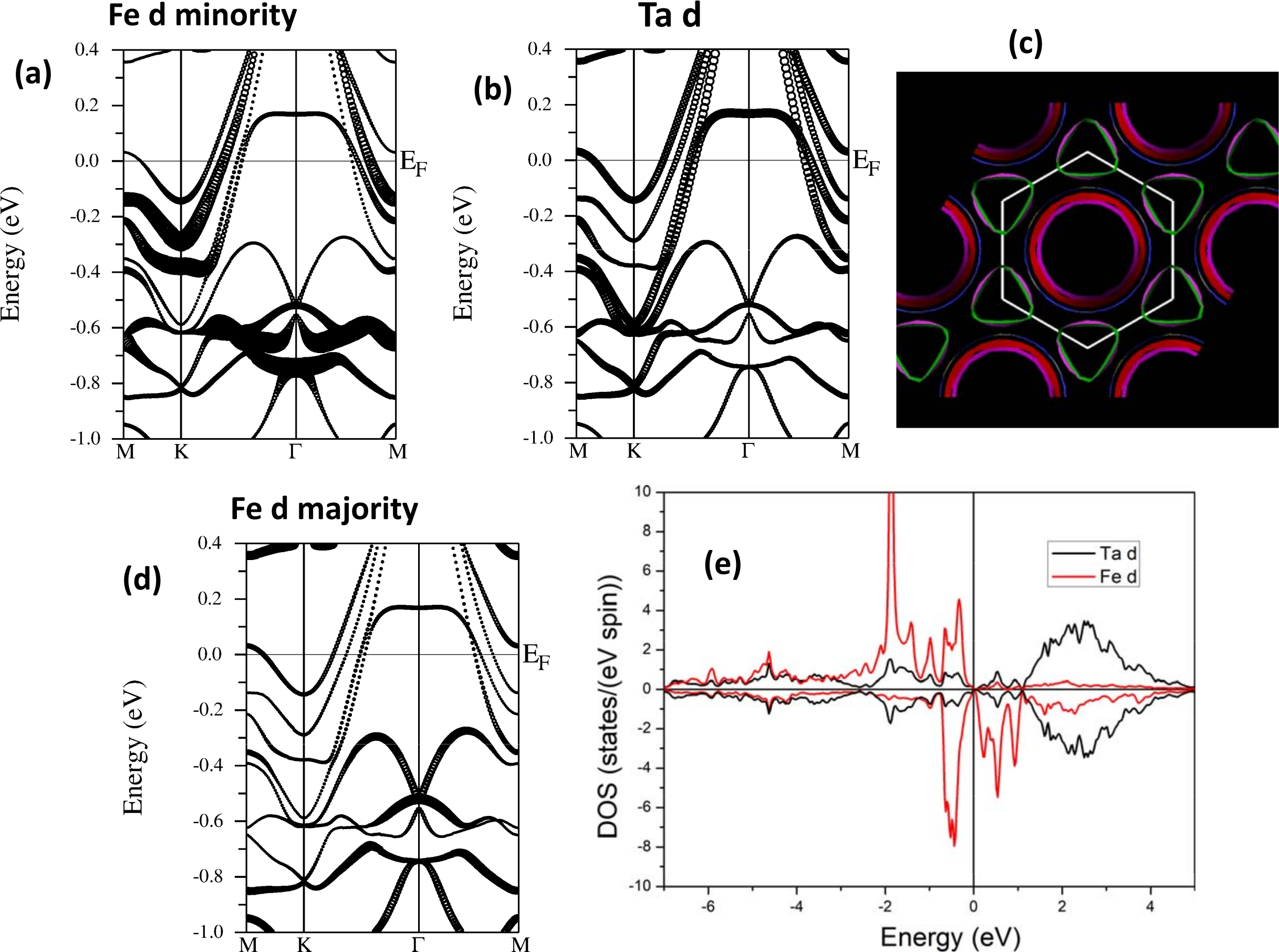}
\caption{DFT-calculated orbital and spin resolved dispersion spectra for Fe$_{1/3}$TaS$_{2}$ (a) and (d) For Fe  $d$ minority and majority band structure, respectively. (b) Band structure obtained for Ta d bands. (c) Calculated Fermi surface projected along the c axis. (e) Atom resolved density of states (DOS) near Fermi level.}
\label{Fe_DOS}
\end{center}
\end{figure}

Figure \ref{Fe_DOS} shows a summary of the electronic structure of Fe-TaS$_2$. In this case, the GGA-PBE calculation would be substantially different than the observed ARPES data. GGA solution is a low-spin configuration with a magnetic moment of 0.7 $\mu_B$ per Fe atom. However, when correlations are introduced through the LDA+U method [V. I. Anisimov et al., Phys. Rev. B, 48, 16929(1993)], the solution converged is high-spin with a magnetic moment of 2.2 $\mu_B$ per Fe atom for U= 5 eV. This is the solution presented in Fig. \ref{Fe_DOS}. To match more closely the experimental data, the chemical potential had to be shifted down by 0.44 eV. Counting the electrons in that region through the obtained density of states of the system implies an off-stoichiometry of about 0.53 electrons per formula unit (something of the sort Fe$_{0.25}$TaS$_2$ if stoichiometries were coming from Fe only). This Fermi level shifting moves the flat band around $\Gamma$ above the Fermi level (it would be otherwise below and it is not seen in the ARPES) and also creates the pocket around the $K$ point, that would be absent otherwise.
In this case, because the Fe $d$ are higher in energy than the Co d bands are, they dominate the character of the states at the Fermi level and make the Ta d states come closer in energy to the Fermi level (starting at only 1 eV above the Fermi level in this case). Again, being a high-spin state, all the majority-spin states are mostly occupied (closer to the Fermi level than the ones from Co are), sharing the same energy window with the minority states.

\section{Diffraction under magnetic field}
\begin{figure}[h!]
\begin{center}
\includegraphics[width=0.9\textwidth]{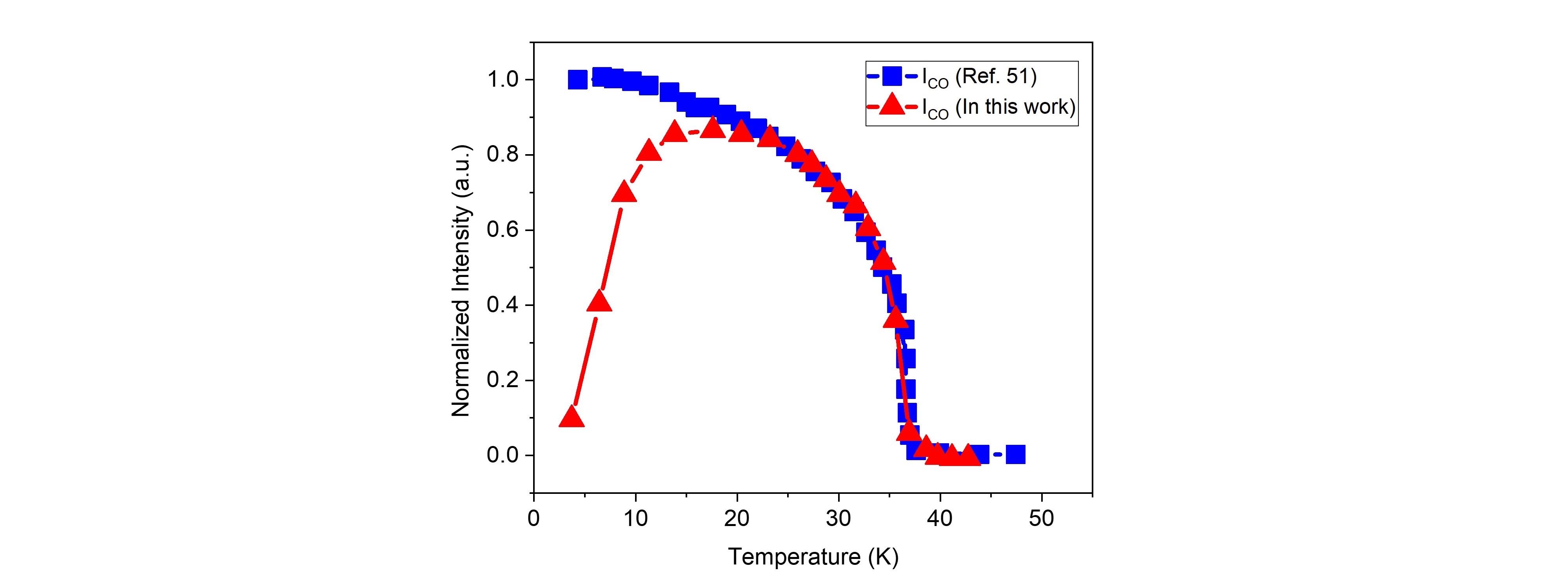}
\caption{Comparison of the charge order intensities of Fe$_{1/3}$NbS$_2$ measured at APS (blue squares) and DESY (red triangles). The different behavior of the intensity at low temperatures is due to the sample heating.}
\label{comparison}
\end{center}
\end{figure}

We note that temperature dependence of the charge order in Fe-NbS$_2$ we report here, Fig. 4 (b) in the main text and \ref{comparison}, shows an anomalous drop at low temperature, not observed in our previous studies \cite{wu2022highly,wu2023discovery}. The CDW signal was found to be homogeneous across the detected area, ruling out the possibility of sampling a specific inhomogeneous region. 
We attribute the differences to local beam heating, which was more pronounced in \cite{wu2023discovery}, given that 
we have used the same sample, hence eliminating sample variability concerns.

\begin{figure}
\begin{center}
\includegraphics[width=1.0\textwidth,draft=false]{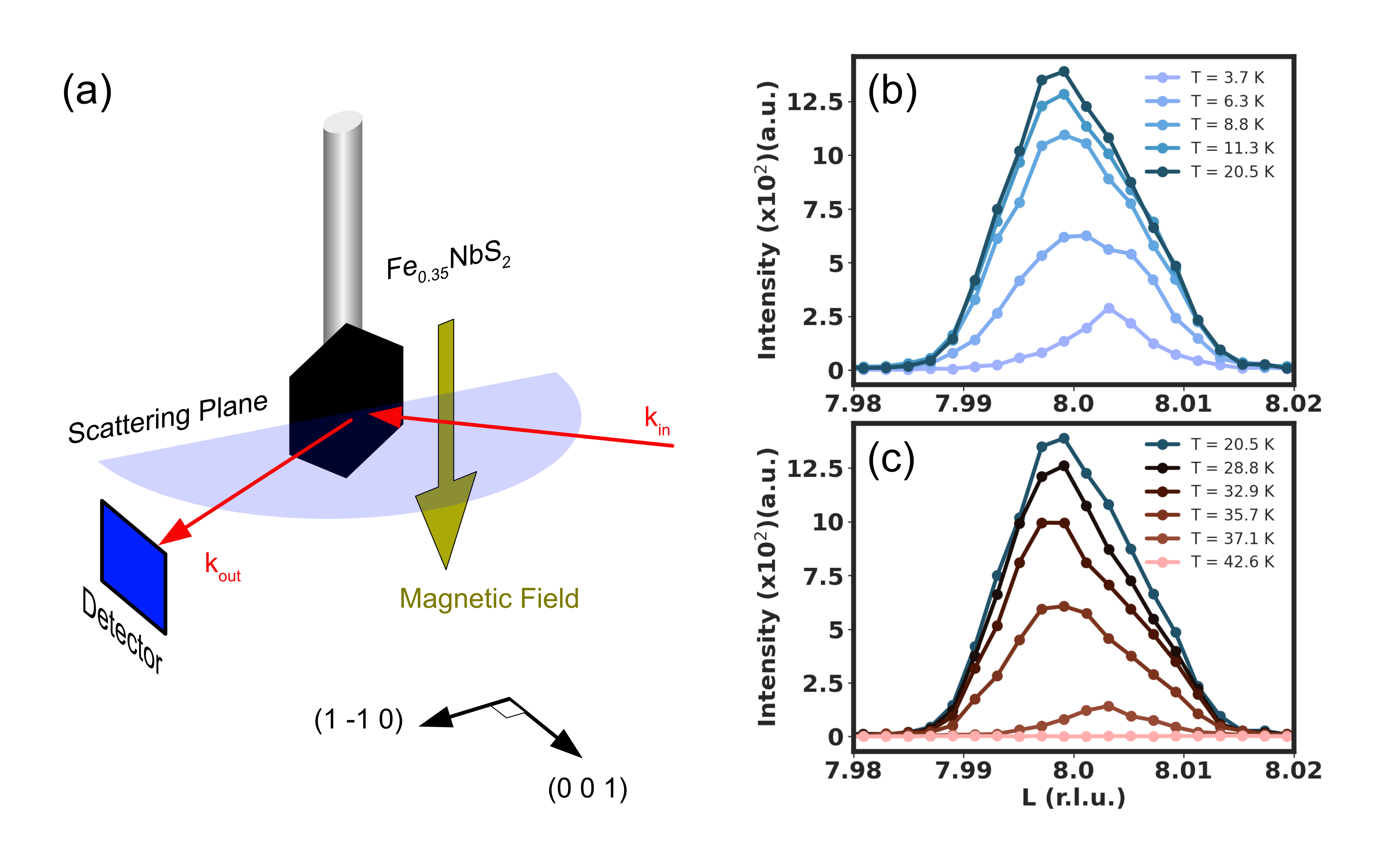}
\caption{X-ray scattering on Fe$_{0.35}$NbS$_2$ sample (a) Schematic representation of the experimental setup for hard X-ray scattering with an external magnetic field. The crystal orientation is shown along with the direction of the external magnetic field. (b) Selected scans showing low-temperature melting of the charge order peak at (0.5 -0.5 8) under zero-field cooling condition (c) Thermal melting of the peak with an onset temperature of 40 K. }
\label{scattering_image}
\end{center}
\end{figure}

The experimental geometry for the field-dependent scattering experiment is depicted in Fig. \ref{scattering_image}(a). First, we address the issue of temperature-dependent melting of the charge order (CO) peak below 15 K.  For each temperature we collected a th-2th scan (longitudinal scan) in the reciprocal space with a photo-diode as a detector.  A zero-field cooled measurement is shown in Fig. \ref{scattering_image}(b) where we can observe a clear dimming of the peak. We extracted the area under the curve after fitting a Gaussian to such peak intensity distribution which is depicted in Fig 4(b) and (c) in the main text. We have also shown in Fig. \ref{scattering_image} (c) the high-temperature melting of the CO peak. 
% Identical melting despite field cooled conditions are shown in Fig. 7 (c) and (d) for a field of 5 T and 10 T respectively. 

\begin{figure}[h!]
\begin{center}
\includegraphics[width=1.0\textwidth,trim={2cm 2cm 2cm 6cm}, clip,draft=false]{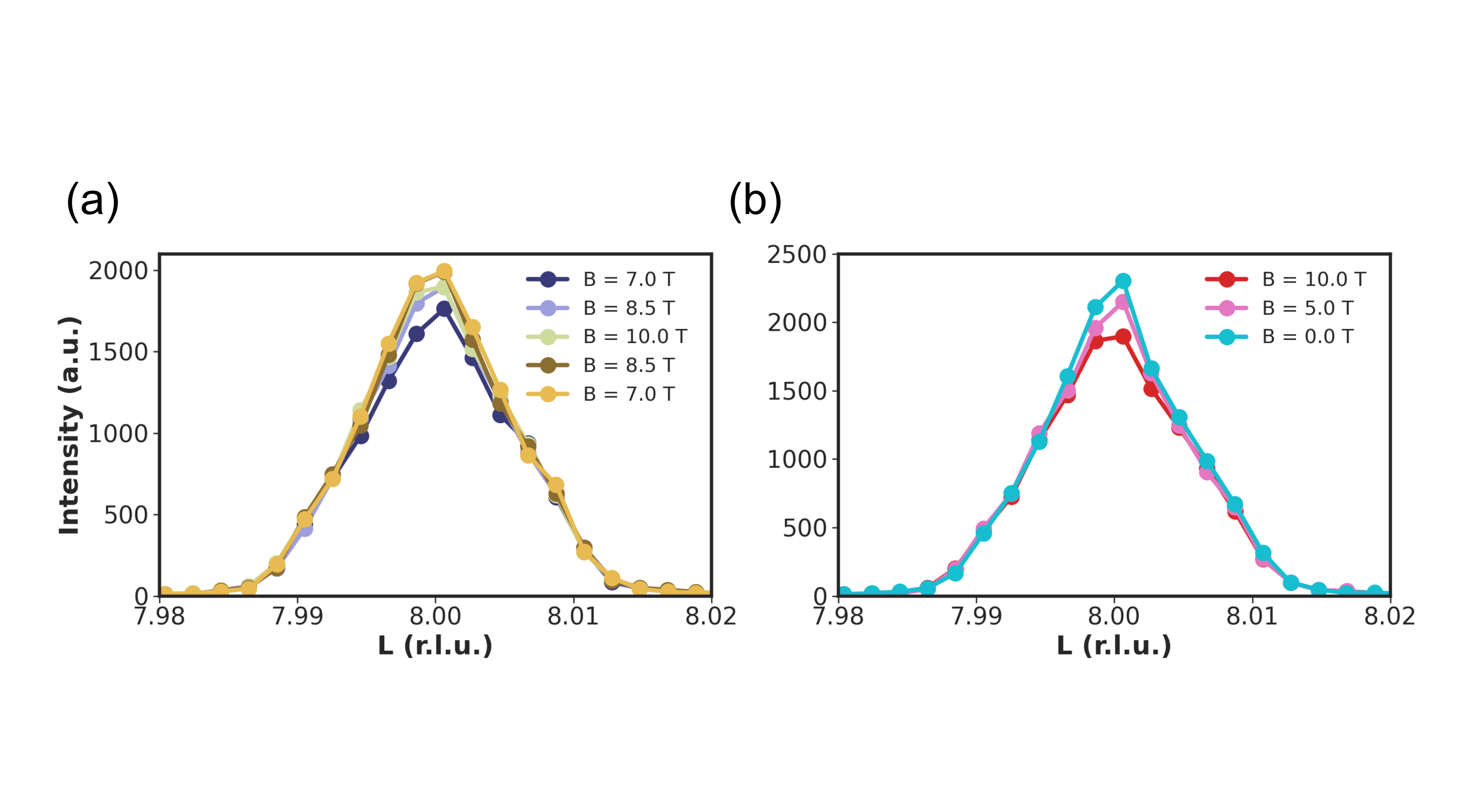}
\caption{X-ray scattering on Fe$_{0.35}$NbS$_2$ sample with external magnetic field: (a) Cycling magnetic field from 7 T to 10 T and back shows a change in peak intensity of the CO peak (b) As we keep lowering the magnetic field the peak intensity continues to increase.}
\label{zero field scattering}
\end{center}
\end{figure}

Next, we show illustrative raw data from the longitudinal scans taken on the CO peak at a temperature of 20 K while ramping the magnetic field. For instance, Fig. \ref{zero field scattering} (a) shows the effect of field hysteresis as we ramped the intensity from 7 T to 10 T and back. The peak shape and the intensity has clearly increased between the two 7 T scans. The trend continues as we decrease the field further to 0 T, as depicted in Fig. \ref{zero field scattering} (b). This peculiar behavior can be understood from a complex magneto-elastic energy landscape as discussed in the main text. 

\clearpage

\bibliography{reference}